\documentclass[11pt]{article}
\usepackage{old-arrows}
\usepackage[scaled=.9]{helvet}
\usepackage{charter}
\usepackage{XoohmG}
\omitBackreferences
 \newmdenv[linecolor=blue!60!green,
 linewidth=1,roundcorner=2pt,backgroundcolor=yellow!5!red!2,
 innerleftmargin=5pt,innerrightmargin=5pt,leftmargin=0pc,rightmargin=0pt,
 fontcolor=blue!60!black,font=\slshape,settings={\let\em=\bfseries}]{bBox}

\usepackage{cite}
\usepackage{thmtools}

\def\SI#1{\setbox9\hbox{$\SSS#1$}\5[-1pt]{\copy9\kern-.125\wd9}{\mathscr{I}}}

\newcommand{\snake}[6][]{\TikZ{[blue#1]\path[use as bounding box](0,0);
                          \draw[stealth-](0,#2)--++(-.1,0)arc(270:90:2*#3 and 1.5*#3)
                                         to++(#4+.1,0)arc(-90:90:2*#5 and 1.5*#5);
                          \path(#4+2*#5+.05,#2+3*#3+.5*#5)node
                                         {\makebox[0pt][l]{$\sss#6$}};}\ignorespaces}
\newcommand{\FF}[2][n]{F^{\sss(#1)}_{#2}}
\newcommand{\MF}[2][n]{{}^\wtd\mkern-5muF^{\sss(#1)}_{#2}}
\newcommand{\XX}[2][n-1]{X^{\sss(#1)}_{#2}}

\DeclareMathOperator*{\Flip}{Flip}
\newcommand{\pDs}[1]{{\pD^{^{\mkern-3mu_{\SSS\star}}}\mkern-7mu_{\smash{#1}}}}
\newcommand{\pDN}[1]{{\pD_{\smash{#1}}}}
\newcommand{\fan}[1]{{\S_{\smash{#1}}}}

\def\zZ#1#2{\big(\ZZ_{#1}{:}\,{#2}\big)}
\DeclareMathOperator*{\reg}{reg}

 \SfTitles
 \allowdisplaybreaks
 \numberwithin{equation}{section}
 \renewcommand{\baselinestretch}{1.1}
 
\begin{document}
\begin{center}
{\LARGE\sf\bfseries\boldmath
  Hirzebruch Surfaces, Tyurin Degenerations and Toric Mirrors:\\[1mm]
 \Large\sf\bfseries\boldmath
  Bridging 
  Generalized Calabi-Yau Constructions
}
\vspace{2mm}

\begin{tabular}{rcl}
\makebox[70mm][r]{\sf\bfseries Per Berglund$^*$} &and&
\makebox[70mm][l]{\sf\bfseries Tristan H\"{u}bsch$^\dag$}\\*[1mm]
\MC3c{\small\it
  $^*$%
      Department of Physics, University of New Hampshire, Durham, NH 03824, USA}\\[-1mm]
\MC3c{\small\it
  $^\dag$%
      Department of Physics \&\ Astronomy,
      Howard University, Washington, DC 20059, USA}\\[-1mm]
 {\tt per.berglund@unh.edu} &and& {\tt  thubsch@howard.edu}
\end{tabular}
\vspace{2mm}

{\sf\bfseries ABSTRACT}\\[3mm]
\parbox{150mm}{\addtolength{\baselineskip}{-1pt}\parindent=2pc\noindent
There is a large number of different ways of constructing Calabi-Yau manifolds, as well as related non-geometric formulations, relevant in string compactifications.
Showcasing this diversity, we discuss explicit deformation families of discretely distinct Hirzebruch hypersurfaces in $\IP^n\<\times\IP^1$ and identify their toric counterparts in detail. This precise isomorphism is then used to investigate some of their special divisors of interest, and in particular the secondary deformation family of their Calabi-Yau subspaces. Moreover, most of the above so called Hirzebruch scrolls are non-Fano, and their (regular) Calabi-Yau hypersurfaces are Tyurin-degenerate, but admit novel (Laurent) deformations by special rational sections as well as a sweeping generalization of the {\em\/transposition construction\/} of mirror models. This  bi-projective embedding also reveals a novel deformation connection between distinct toric spaces, and so also the various divisors of interest including their Calabi-Yau subspaces.
}

\begin{minipage}{150mm}
\vspace{\baselineskip}
\centering
\baselineskip=12pt\parskip=0pt
\tableofcontents
\end{minipage}
\end{center}

\section{Introduction, Rationale and Summary}
\label{s:IRS}
Constructing complex algebraic varieties as complete intersections of holomorphic hypersurfaces within a well-understood ``ambient'' space, $A$, has recently been generalized so as to include cases where some of those hypersurfaces have a negative degree over some factors in $A$~\cite{rgCICY1}.
 The diffeomorphism class and cohomology of such {\em\/generalized complete intersections\/} (gCI, gCICY if $c_1\<=0$) have been studied~\cite{rBH-Fm,rJL-gCIShv}, and they were soon provided with a rigorous scheme-theoretic definition~\cite{rGG-gCI}.
 Such constructions of immediate physics interest are anticanonical (Calabi-Yau) hypersurfaces in non-Fano varieties, their toric models and their Laurent deformations were further explored in~\cite{rBH-gB}, extending the already immense database~\cite{rKreSka00b} to include infinitely many, though not necessarily distinct, constructions.

 The purpose of this article, in part, is to provide a bridge between these different approaches, aiming to further explore the generalization~\cite{rBH-gB} of the transposition mirror model construction~\cite{rBH,rBH-LGO+EG,rMK-diss} and Batyrev's toric construction~\cite{rBaty01}; see also~\cite{rF+K-BHK} and references therein.
 To this end, we follow suit from the earlier work~\cite{rBH-Fm,rBH-gB} and continue to examine the generalized complete intersection Calabi-Yau models in the ``proof of concept'' showcasing the infinite sequence of deformation families $\ssK[{r||c|c}{\IP^n&1&n\\ \IP^1&m&2{-}m}]$ and their toric rendition. The bi-projective embedding is the ({\em\/generalized\/} when $m\<\geqslant3$) complete intersection of two hypersurfaces of bi-degrees $\pM{1\\m}$ and $\pM{n\\2{-}m}$, were the latter hypersurface is for $m\<\geqslant3$ well defined {\em\/only within\/} the former. This {\em\/ordered\/} approach reveals a detailed structure in this deformation family of Calabi-Yau models.

 In particular, we first focus on the deformation families
 $\ssK[{r||c}{\IP^n&1\\ \IP^1&m}]$ of Hirzebruch scrolls, and provide explicit, coordinate-level isomorphisms between such hypersurfaces~\cite{rH-Fm,rBH-Fm} and their toric rendition~\cite{rBH-gB}; see also~\cite{rF-TV,rGE-CCAG,rCLS-TV}. This naturally maps their cohomology data, as well as their special subspaces of interest, including the hallmark divisor of maximally negative self-intersection (dubbed {\em\/directrix\/}~\cite{rGrHa}), and then also their Calabi-Yau subspaces, as detailed in \SS\,\ref{s:HH}.
 The remainder of that section shows that the deformation family of Hirzebruch scrolls,
 $\ssK[{r||c}{\IP^n&1\\ \IP^1&m}]$, contains besides the {\em\/central\/} $\FF{m;0}$ also a hierarchy of its diffeomorphic but {\em\/discretely different\/} complex deformations, $\FF{m;\vec\e\,}$, each harboring less negative (sub-)directrices. This extends our comparisons across a detailed web of bi-projective and toric constructions --- both the infinite hierarchy of Hirzebruch scrolls, and then also their Calabi-Yau subspaces.

 Section~\ref{s:CYs} shows that the regular Calabi-Yau hypersurfaces,
 $\XX{m}\<\subset\FF{m;0}$, are for $m\<\geqslant3$ always Tyurin-degenerate, their codimension-1 singularity itself Calabi-Yau. While generic scrolls $\FF{m;\vec\e\,}$ admit smoothing such hypersurfaces by regular sections, those in the central $\FF{m;0}$ can only be desingularized by Laurent deformations~\cite{rBH-gB}. The latter require special attention to the putative pole singularities, detailed in \SS\,\ref{s:LdL'H}, but are shown in \SS\SS\,\ref{s:MTM}--\ref{s:MTMs} to admit a straightforward extension of the transposition mirror model construction~\cite{rBH,rBH-gB}.
 Finally, \SS\,\ref{s:Cox} discusses such Laurent deformations as {\em\/virtual varieties\/} (Weil divisors), as well as a recasting in terms of desingularized finite quotients of ramified multiple covers.
 
 The inclusion of these ideas and results in the gauged linear sigma models (GLSMs)~\cite{rPhases,rMP0} are discussed in \SS\,\ref{s:GLSM}, and our concluding remarks are collected in \SS\,\ref{s:Coda}. The technically more detailed material is deferred to the appendices. 
 As indicated throughout, the results presented herein indicate several avenues for further study, the pursuit of which is however beyond the scope of such a ``proof of concept'' article.

\section{Hirzebruch Scrolls}
\label{s:HH}
Following Hirzebruch's original definition~\cite{rH-Fm}, we identify the particular hypersurface
\begin{alignat}9
  \FF{m;0}&\coeq \big\{ p_0(x,y)=0\big\} ~\in\K[{r||c}{\IP^n&1\\\IP^1&m}],&\quad
   p_0(x,y)&\coeq x_0y_0\!^m\<+x_1y_1\!^m
 \label{e:bPnFm}
\iText{as the {\em\/central\/} member of the deformation family of degree-$\pM{1\\m}$ hypersurfaces in $\IP^n\<\times\IP^1$:}
  \FF{m;\e}&\coeq\{p_{\vec\e\,}(x,y)\<=0\} \in \K[{r||c}{\IP^n&1\\\IP^1&m}],&\quad
  p_{\vec\e\,}(x,y)&\coeq  p_0(x,y) +
     \sum_{a=0}^n\sum_{\ell=1}^{m-1} \e_{a\ell}\,x_a\,y_0\!^{m-\ell}y_1\!^\ell,
 \label{e:bPnFmE}
\end{alignat}
explicitly (and coarsely) parametrized by the $\e_{a\ell}\<\in\IC$.
 The gradient $\vd p_0\<=(y_0\!^m,y_1\!^m,\dots)$ of even the central model~\eqref{e:bPnFm} cannot vanish anywhere on $\IP^1$ since $y_0,y_1$ cannot both vanish: even $ p_0(x,y)$ is {\em\/transverse\/} (basepoint free), so $\FF{m;0}\<\coeq p_0^{-1}(0)$ is nonsingular, not just the generic $\FF{m;\e}\<\coeq p_\e^{-1}(0)$.
 Again following Hirzebruch~\cite{rH-Fm}, we identify $\FF{m;0}$ also with the $m$-twisted $\IP^{n-1}$-bundle over $\IP^1$ as well as the projectivization
 $\IP\big(\cO_{\IP^1}\oplus\cO_{\IP^1}(m)^{\oplus(n-1)}\big)$.

A key feature of deformation families such as~\eqref{e:bPnFmE} is that although the smooth hypersurfaces in the families with a fixed $m\<\simeq m\pMod{n}$ are all diffeomorphic to each other, they form a {\em\/discrete\/} collection of distinct complex manifolds --- and these distinctions also pertain to the Calabi-Yau hypersurfaces therein.

\subsection{Topological Characteristics}
\label{s:TopData}
As usual, $H^r(\FF{m},\ZZ)=H^r(\IP^n{\times}\IP^1,\ZZ)$ without torsion~\cite{rBH-Fm}, and
with $J_1^{\,n+1},\,J_2^{\,2},\,J_1^{\,n}J_2\<=0$,
\begin{equation}
 c(\FF{m})
  = \frac{(1+J_1)^{n+1}(1+J_2)^2}{1+J_1+mJ_2}
  =(1{+}J_1{-}mJ_2)(1{+}J_1)^{n-1}(1{+}J_2)^2. \label{e:CbP2T}
\end{equation}
The simplification owes to the identity
 $\frac{(1+J_1)^2}{1+J_1+mJ_2}\<=(1{+}J_1{-}mJ_2)$ insured by the nilpotence of $J_2$.
Standard (B\'ezout's theorem) computations then provide the $n$-tuple intersection numbers~\cite{rBeast}:
\begin{equation}
  [J_1\!^n]=
  \K[{r||c|c@{~}c@{~}c}{\IP^n&1&1&\cdots&1\\ \IP^1&m&0&\cdots&0}]=m,\qquad
  [J_1\!^{n-1}J_2]=
  \K[{r||c|c@{~}c@{~}c@{~}c}{\IP^n&1&1&\cdots&1&0\\ \IP^1&m&0&\cdots&0&1}]=1,
\end{equation}
and all other intersections vanish again owing to $J_2^{\,2}\<=0$. Also, powers of
 $(aJ_1{+}bJ_2)$ may be evaluated against complementary Chern classes to yield, e.g., for $n\<=4$:
\begin{subequations}
 \label{e:Fmod4c}\vspace*{-1mm}
\begin{gather}
  C_1^{~3}[aJ_1{+}bJ_2] =16[6a + (\2{4b{+}ma})],\qquad
  C_1{\cdot}C_2[aJ_1{+}bJ_2] =2[22a + 3(\2{4b{+}ma})],\\
  C_3[aJ_1{+}bJ_2] =12a + (\2{4b{+}ma}),\qquad
  C_1^{~2}[(aJ_1{+}bJ_2)^2] =8a[2a+(\2{4b{+}ma})],\\
  C_2[(aJ_1{+}bJ_2)^2] =a(8a+3(\2{4b{+}ma})],\qquad
  C_1[(aJ_1{+}bJ_2)^3] =a^2(2a +3(\2{4b{+}ma})].
\iText[-5pt]{Finally, the Chern {\em\/numbers\/} are $m$-independent:}
  C_1^{~4} =512,\qquad
  C_1^{~2}{\cdot}C_2 =224,\qquad
  C_1{\cdot}C_3 =56,\qquad
  C_2^{~2} =96,\qquad
  C_4 = \chi_{_E} =8.
\end{gather}
\end{subequations}
Jointly,~\eqref{e:Fmod4c} indicate an $[m\pMod{n}]$-dependence of these topological invariants for $n\<=4$, verified by the integral basis change, $\Tw{J}_1\<\coeq J_1{-}kJ_2$ and $\Tw{J}_2\<\coeq J_2$:
\begin{equation}
  [\Tw{J}_1\!^n]=(m{-}kn)\qquad\text{and}\qquad
  [\Tw{J}_1\!^{n-1}\Tw{J}_2]=1,\qquad k\<\in\ZZ,
\end{equation}
This implies that $\FF{m}\<{\approx_{\sss\IR}}\FF{m-kn}$ for integral $k$ are all diffeomorphic to each other~\cite{rWall}: they are the same real manifold, and so are then the Calabi-Yau hypersurfaces, $\XX{m}\<{\approx_{\sss\IR}}\XX{m-kn}\<\in\FF{m}[c_1]$; for details, see~\cite{rBH-Fm}.

The entire infinite $m$-sequence of deformation families of hypersurfaces~\eqref{e:bPnFmE} thus harbors precisely $n$ distinct real manifolds, distinguished only by $[m\pMod{n}]$. In particular, all transverse (and so smooth) scrolls in the deformation families $\ssK[{r||c}{\IP^n&1\\ \IP^1&m}]$ for any fixed $m\<\simeq m\pMod{n}$ are the same real manifold.

\subsection{Holomorphic Characteristics}
\label{s:DrX}
The $m$-sequence of deformation families of hypersurfaces~\eqref{e:bPnFmE} however admits infinitely many {\em\/complex\/} manifolds, distinguished by $m$, unreduced:
 The hallmark {\em\/holomorphic\/} characteristic of Hirzebruch's original~\cite{rH-Fm,rKC-Fm}, $\FF[2]{m}=\FF[2]m:=\IP\big(\cO\<\oplus\cO(m)\big)$, is its exceptional irreducible curve, $S_m$, a holomorphic hypersurface of self-intersection $-m$, the {\em\/directrix\/}~\cite[p.\,525]{rGrHa}. Correspondingly, each Hirzebruch $n$-fold $\FF{m}$ contains an {\em\/exceptional\/} irreducible (holomorphic) hypersurface $S_m\<{\subset_\IC}\FF{m}$ of self-intersection $-(n{-}1)m$. Additional relevant holomorphic distinctions are discussed in Appendix~\ref{s:holoD}, including the result
\begin{subequations}
 \label{e:H*nFmT}
\begin{equation}
   \dim H^0(\FF{m},T) =n^2{+}2 + \D^{\sss(n)}_m
   \qquad\text{and}\qquad
   \dim H^1(\FF{m},T) =\D^{\sss(n)}_m,
\end{equation}
where the number of exceptional contributions is, using the step-function
 $\vq_a^b\<\coeq\{1~\text{if}~a\<\leqslant b,~0~\text{otherwise}\}$:
\begin{alignat}9
   \D^{\sss(n)}_{m;0}&=\vq_1^m(n{-}1)(m{-}1),&\quad&\text{for}~~
   \FF{m,0}=\{x_0\,y_0^{\,m}{+}x_1\,y_1^{\,m}\<=0\}
    \in\ssK[{r||c}{\IP^n&1\\\IP^1&m}], \label{e:H*nFmT0}\\
   \D^{\sss(n)}_{m;\e\neq0}&<\D^{\sss(n)}_{m;0};&\quad&
    \text{for generic cases, }\D^{\sss(n)}_{m;\e\neq0}=0.
\end{alignat}
\end{subequations}
As there always exist more local reparametrizations than local deformations of the complex structure, $\dim H^0(\FF{m},T)\<>\dim H^1(\FF{m},T)$, the scrolls $\FF{m}$ are effectively rigid: their space of complex structure deformations modulo reparametrizations is discrete~\cite{rBH-Fm}.

 This ``jumping''~\eqref{e:H*nFmT} in the dimensions of $H^*(\FF{m;\e},T)$ depending on the concrete choice of the defining equation~\eqref{e:bPnFm}--\eqref{e:bPnFmE} again illustrates the variability of complex manifolds provided by even a simple deformation family such as $\ssK[{r||c}{\IP^n&1\\ \IP^1&m}]$.
 Even the simplest ($\FF[2]2\<\leadsto\FF[2]0$, see~\cite{rGHSAR} and~\cite[\SS\,3.1.2]{rBeast}) of such discrete deformations has been known to affect string compactifications~\cite{Morrison:1996na,rPhasesMF}. Another, phenomenologically relevant effect of such discrete deformations was explored in~\cite{rBHL1,rBHL2,Gray:2019tzn}.

\paragraph{The Directrix:}
The homology class of the directrix is easy to represent as $[S_m]=[J_1]{-}[mJ_2]$, so indeed
\begin{equation}
 [S_m]^n=
  \K[{r||c|c@{~}c@{~}c}{\IP^n&1&~~1&\cdots&~~1\\ \IP^1&m&-m&\cdots&-m}]
   = m+n(-m) = -(n{-}1)m.
\end{equation}
An {\em\/irreducible\/} holomorphic submanifold representative of $[S_m]$ must be the zero-locus of a degree-$\pM{\3-1\\-m}$ global holomorphic section. No such section exist on $A\<=\IP^n{\times}\IP^1$, but there {\em\/does\/} exist a unique such section on $\FF{m}\<=\FF{m;0}$ and is easily constructed following the techniques introduced in~\cite{rgCICY1,rBH-Fm,rGG-gCI}. To highlight the novelty and more general uses of this explicit construction, we adapt from~\cite{rBH-Fm}: The key point is to identify sections $\Fs(x,y)$ on the zero-locus $\{ p_{\vec\e\,}\<=0\}\subset A$ with the restriction of the {\em\/equivalence class\/} of sections\ftn{In physics, gauge potentials are a prime example, being defined only up to gauge transformations: $A_\m\<\simeq A_\m\<+\vd_\m\l$. This enables the Wu-Yang construction of a magnetic monopole~\cite{rWY-MM}.},
 $[\Fs(x,y)\pMod{ p_{\vec\e\,}}]$, on all of $A$.
For example, a total degree-$\pM{\3-1\\-m}$ multiple of $ p_0(x,y)$ is of the form
\begin{equation}
 \frac{ p_0(x,y)}{(y_0\,y_1)^m}
 =\Big(\frac{x_0}{y_1\!^m} + \frac{x_1}{y_0\!^m}\Big),
  \qquad\deg=\pM{\3-1\\-m\\},
 \label{e:lam0p}
\end{equation}
which serves as the $r_0\<=r_1\<=m$ case of the more general:
\begin{cons}\label{C:Fs}
Given a degree-$\pM{1\\m}$ hypersurface $\{p_{\vec\e\,}(x,y)\<0\}\subset\IP^n{\times}\IP^1$ as in~\eqref{e:bPnFmE}, construct
\begin{equation}\mkern-60mu
 \deg\<=\pM{\3-1\\m{-}r_0{-}r_1}{:}\quad
  \Fs_{\vec\e\,}(x,y;\l)\coeq
   \Flip_{y_0}\Big[\frac1{y_0\!^{r_0}\,y_1\!^{r_1}}p_{\vec{\e}\,}(x,y)\Big]
    \pMod{p_{\vec{\e}\,}(x,y)},
 \label{e:gDrX}
\end{equation}
progressively decreasing $r_0{+}r_1\<=2m,2m{-}1,\<\cdots$, and keeping only those Laurent polynomials that contain both $y_0$- and $y_1$-denominators but no $y_0,y_1$-mixed ones. The ``\/$\Flip_{y_i}$'' operator changes the relative sign of the rational monomials with $y_i$-denominators.
 For algebraically independent such sections, restrict to a subset with maximally negative degrees that are not overall $(y_0,y_1)$-multiples of each other.
\end{cons}
\noindent
In particular, the $r_0\<=r_1\<=m$ and $ p_0(x,y)\<=\lim_{\vec\e\to0}p_{\vec\e\,}(x,y)$ case produces the degree-$\pM{~~1\\-m}$ directrix:
\begin{equation}
 \Fs(x,y)\<=\Fs_0(x,y)
  \<=\Big[\Big(\frac{x_0}{y_1\!^m}-\frac{x_1}{y_0\!^m}\Big)
               +\frac{\l}{(y_0\,y_1)^m} p_0(x,y)\Big]
    =\bigg\{\begin{array}{@{}l@{~~\text{if}~~}ll}
           +2\frac{x_0}{y_1\!^m} &y_1\neq0, &\l=+1,\\[2pt]
           -2\frac{x_1}{y_0\!^m} &y_0\neq0, &\l=-1.\\
           \end{array}
 \label{e:DrX}
\end{equation}
Designed to generalize this patch-wise feature, the mod-$p_{\vec\e}$ equivalence class of sections has a well-defined and holomorphic local representative everywhere on $A$. Since the difference
 $\Fs_{\vec\e\,}(x,y;\l){-}\Fs_{\vec\e\,}(x,y;\l')$ vanishes where $p_{\vec\e\,}(x,y)\<=0$, the two local representatives such as~\eqref{e:DrX} define a single well-defined holomorphic section~\eqref{e:gDrX} on the zero-locus $\FF{m;\vec\e}\<\coeq\{p_{\vec\e\,}(x,y)\<=0\}$.
 Moreover, $\vd\Fs_0\<=\big(\frac1{y_1\!^m},{-}\frac1{y_0\!^m},\dots\big)$ cannot vanish anywhere on $\IP^1$ since $y_0,y_1\<<\infty$; the analogous is true of $\Fs_{\vec\e\,}(x,y)$ for generic $\vec\e$. The section $\Fs_{\vec\e\,}(x,y)|_{\smash{\FF{m;\vec\e}}}$ is thereby transverse (basepoint free) and the holomorphic hypersurface $\big(S_{m;\vec\e}\<\coeq\Fs^{-1}_{\vec\e\,}(0)\big)\<\subset\FF{m;\vec\e}$ is nonsingular and so irreducible. Away from $\{p_{\vec\e\,}(x,y)\<=0\}$, $\Fs_{\vec\e\,}(x,y)$ can only define an equivalence class of subspaces corresponding to $[S_{m;\vec\e}\,]\<\in H_*(A)$.
 If $r_0\<=r_1$ and $p_{\vec\e\,}(x,y)$ is $y_0\<\iff y_1$ symmetric, $\Flip_{y_0}$ evidently flips the sign in $p_{\vec\e\,}(x,y)$ itself, but not so more generally; see \SS\,\ref{s:DscDef} for examples.

 The Czech cohomology framework was explicitly shown to provide such constructions with a rigorous scheme-theoretic definition~\cite{rGG-gCI}. Technically, the putative poles in~\eqref{e:DrX} are evaded by clearing the denominators and connecting the patch-wise defined sections by the Mayer-Vietoris sequence. Reassured by the existence of this formal-foundational framework, here we continue the analysis following~\cite{rBH-Fm,rBH-gB}. Also, the toric framework reached in the next subsection will reveal that these technical complexities are not intrinsic, but a property of the embedding.

 It should be clear that mod-$p_{\vec\e}$ equivalence classes of sections more negative than~\eqref{e:gDrX} cannot have a well-defined holomorphic representative everywhere on $A$. In turn, any multiple of $\Fs_{\vec\e\,}(x,y)$ by a regular $x,y$-polynomial is also a holomorphic mod-$p_{\vec\e}$ equivalence class of sections, of a correspondingly less negative degree, but is clearly not algebraically independent.

\subsection{Toric Rendition}
\label{s:bP>T}
Let's focus first on the central, $\vec\e\<=0$ case, where the explicitly complementary form of~\eqref{e:bPnFm} and~\eqref{e:DrX} suggests the reparametrization
\begin{subequations}
 \label{e:bP=T}
\begin{equation}
  (x_0,x_1,x_2,\cdots;y_0,y_1) \to (\, p_0\,,\,\Fs\,,x_2,\cdots;y_0,y_1),\qquad
  \det\Big[\frac{\vd(\, p_0\,,\,\Fs\,,\,x_2,\cdots;y_0,y_1)}
                {\vd(x_0,x_1,x_2,\cdots;y_0,y_1)}\Big]\<={-}2,
 \label{e:0psxy}
\end{equation}
which leaves the $ p_0\<=0$ hypersurface parametrized by $(\Fs,x_2,\cdots;y_0,y_1)$. The new variables inherit the $\IP^n{\times}\IP^1$ degrees, and are identified with the Cox variables of the toric rendition of $\FF{m}$ as given in~\cite{rBH-gB}:
\begin{equation}
 \begin{array}{@{}c@{~}c@{~}c@{~}c@{~}c@{~}c@{~~}c@{~}c@{}}
      &x_0&x_1&x_2&\cdots&x_n&y_0&y_1\\ \toprule\nGlu{-2pt}
 \IP^n& 1 & 1 & 1 &\cdots& 1 & 0 & 0 \\[-1mm]
 \IP^1& 0 & 0 & 0 &\cdots& 0 & 1 & 1 \\
 \end{array}
~~\begin{array}{c}
   \too{\sss\eqref{e:0psxy}}\\ ~\\
  \end{array}~~
 \begin{array}{@{}c@{~}c@{~}c@{~}c@{~}c@{~~}c@{~}c@{}}
   p_0&\3-\Fs&x_2&\cdots&x_n&y_0&y_1\\ \toprule\nGlu{-2pt}
   1 &\3-1& 1 &\cdots& 1 & 0 & 0 \\[-1mm]
   m &-m & 0 &\cdots& 0 & 1 & 1 \\
 \end{array}
~~\begin{array}{c}
   \too{\, p_0=0\,}\\ ~\\
  \end{array}~~
 \begin{array}{@{}c@{~}c@{~}c@{~}c@{~}c@{~}c@{~}c@{~}c@{}}
 \3-X_1&X_2&\cdots&X_n&X_{n+1}&X_{n+2}\\ \toprule\nGlu{-2pt}
 \3-1& 1 &\cdots& 1 & 0 & 0 \\[-1mm]
 -m & 0 &\cdots& 0 & 1 & 1 \\
 \end{array}
 \label{e:bP>T}
\end{equation}
\end{subequations}
The $\IP^n{\times}\IP^1$-inherited degrees form the Mori vectors (given in the rows of the right-hand side tabulation), i.e., the GLSM gauge charges~\cite{rPhases,rMP0}, $Q^a_i\<\coeq Q^a(X_i)$: $Q^1=(1,1,\cdots,1,0,0)$ and $Q^2=(-m,0,\cdots,0,1,1)$.

The $n$-space orthogonal to these two $(n{+}2)$-vectors is spanned by $n$ integral $(n{+}2)$-vectors $\n^\k_i$ (with $\k\<=1,\dots,n$ and $i\<=1,\dots,n{+}2$), a choice of which is shown in the upper $n$ rows:
\begin{equation}
 \vC{\begin{picture}(42,37)(0,2)
   \put(22.3,17.7){\TikZ{\path[use as bounding box](0,0);
               \draw[blue,thick,-stealth](0,0)--++(2.1,-.35);
                \path[blue](2,-.1)node{$\hat{e}_1$};
               \draw[blue,thick,-stealth](0,0)--++(-.02,2.2);
                \path[blue](.3,2.1)node{$\hat{e}_3$};
               \draw[blue,thick,-stealth](0,0)--++(-1.6,-.3);
                \path[blue](-1.7,-.4)node{$\hat{e}_2$};
              }}
   \put(0,0){\includegraphics[width=40mm]{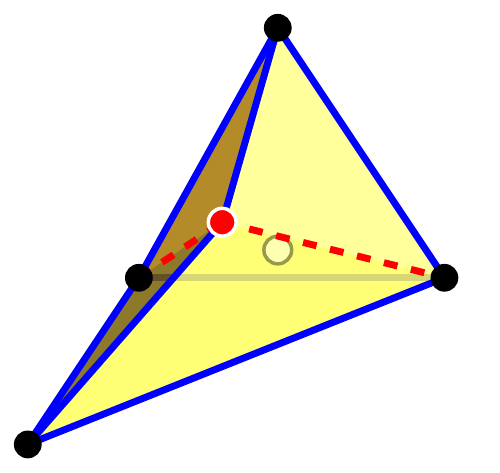}}
   \put(22.3,17.7){\TikZ{\path[use as bounding box](0,0);
                \path(-1.3,.8)node{$\pDs{\FF[3]4}$};
                \path(-1.3,1.6)node{$n\<=3$};
                \path(1.3,1.6)node{$m\<=4$};
                \path[red](-.15,.35)node{$\n_1$};
                \path(1.4,-.5)node{$\n_2$};
                \path(-1.4,0)node{$\n_3$};
                \path(-.3,1.8)node{$\n_4$};
                \path(-2.1,-1.25)node{$\n_5$};
              }}
 \end{picture}}
 \qquad
  \begin{array}{@{}c@{~}|c@{~}c@{~}c@{~}c@{~}c@{\,}|@{\,}c@{~}c@{}}
   & X_1  & X_2  & X_3  &\cdots& X_n  &X_{n+1}& X_{n+2} \\ \toprule\nGlu{-2pt}
   &\!-1  &  1   &  0   &\cdots&  0   &   0   &-m \\[-3pt] 
\MR3*{\rotatebox{90}{$\fan{\FF{m}}$}}
   &\!-1  &  0   &  1   &\cdots&  0   &   0   &-m \\[-3pt]
 \TikZ{\path[use as bounding box](0,0);
        \path(-.35,.25)node{$\left.\rule{0pt}{12mm}\right\{$};
        \draw[thick,->](-.45,.225)--++(-.4,0);
            }%
   &~~~\vdots&\vdots&\vdots&\ddots&\vdots&\vdots &~~\vdots\\[-3pt]
   &\!-1  &  0   &  0   &\cdots&  1   &   0   &-m \\[-3pt]
   &   0  &  0   &  0   &\cdots&  0   &   1   &-1 \\[-1pt] \midrule\nGlu{-2pt}
Q^1&   1  &  1   &  1   &\cdots&  1   &   0   &\3-0
 \TikZ{\path[use as bounding box](0,0);
        \path(.2,-.1)node{$\Big\}$};
        \draw[thick,->](.3,-.1)to[out=0,in=235]++(1,.5);
            }%
                                                  \\[-3pt]
Q^2&\!-m  &  0   &  0   &\cdots&  0   &   1   &\3-1\\
  \end{array}
  \qquad
 \vC{\TikZ{[scale=.5]
           \path[use as bounding box](-3,-4)--(5.5,2);
            \corner{(0,0)}{0}{90}{1}{green};
            \corner{(0,0)}{0}{-77}{1}{yellow};
            \corner{(0,0)}{-77}{-215}{1}{blue};
            \corner{(0,0)}{90}{145}{1}{red};
           \draw[thick,-stealth](0,0)--++(1,0);
            \path[right](1,0)node
                 {\scriptsize$\vec{Q}_2,{\cdots},\vec{Q}_n=\pM{1\\0}$};
           \draw[thick,-stealth](0,0)--++(0,1);
            \path[right](-.6,1.4)node
                 {\scriptsize$\vec{Q}_{n+1},\,\vec{Q}_{n+2}=\pM{0\\1}$};
           \draw[thick,-stealth](0,0)--++(1,-4);
            \path[right](1,-3.7)node{\scriptsize$\vec{Q}_1=\pM{\3-1\\-m}$};
           \draw[blue,very thick,-stealth](0,0)--++(-3,2);
            \path[blue,right](-3.8,2.5)node
                 {\scriptsize$\vec{Q}_0=\pM{-n\\m-2}\<=-\sum_i\vec{Q}_i$};
           \filldraw[thick,fill=white](0,0)circle(1mm);
           \path(.6,-2)node{$n=3$~~~~$m=4$};
            }}
 \label{e:nFmQnu}
\end{equation}
In turn, the {\em\/columns\/} in the tabulation~\eqref{e:nFmQnu} specify:
 ({\small\bf1})~Below the divide are the 2-vector generators, $\vec{Q}_i$, of the secondary fan, shown at right in~\eqref{e:nFmQnu} for $n\<=3$ and $m\<=4$.
 ({\small\bf2})~Above the horizontal divide are the $n$-vector generators,
 $\vec\n_i\smt\fan{\FF{m}}$, of the (primary) fan defining the 
 toric variety $\FF{m}$.
 The so-called {\em\/spanning polytope\/}~\cite{rGE-CCAG},
 $\pDs{\FF{m}}$, is shown at left: its faces are the bases of the cones in $\fan{\FF{m}}$, denoted ``$\pDs{\FF{m}}\<\lat\fan{\FF{m}}$.'' For $m\<\geqslant3$, $\pDs{\FF{m}}$ is non-convex ($\n_1$ is a saddle-point), reflecting that $\FF{m}$ not Fano.
 This fan also shows $\FF{m}$ to be a $\IP^{n-1}$-fibration (encoded by $\n_1,\cdots,\n_n$) over the base-$\IP^1$ (encoded by $\n_{n+1},\n_{n+2}$).

 The mutually defining relation,
\begin{equation}
  \sum_{i=1}^{n+2}Q^a_i\,\n^\k_i=0,\qquad
  \Big\{\begin{array}{@{\,}r@{\,=\,}l}
    a & 1,2; \\ 
    \k& 1,{\cdots},n. \\ 
  \end{array}
 \label{e:Qnu=0}
\end{equation}
specifies the integral components $\n^\k_i$ only up to linear combinations
 $\n^\k_i\simeq \sum_\l c^\k{}_\l\,\n^\l_i$, so
the integral $n$-vectors $\vec\n_i$ are defined only up to overall $\GL(n;\ZZ)$ transformations.
 Analogously, the GLSM gauge charges $Q^a_i$ are defined only up to linear combinations
 $Q^a_i\simeq \sum_b C^a{}_b\,Q^b_i$, so the secondary fan generators $\vec{Q}_i$ are defined only up to $\GL(2;\ZZ)$ transformations. Verifying certain additional conditions~\cite{rBKK-tvMirr}, the $Q^a$-vectors are identified as the Mori vectors for the toric variety specified in~\eqref{e:nFmQnu}.

The toric specification~\eqref{e:nFmQnu} is detailed~\cite{rKKMS-TE1,rD-TV,rF-TV,rGE-CCAG,rCLS-TV}:
 Each top-dimensional cone $\s_{\sss I}\<\in\fan{\FF{m}}$ (over a facet of $\pDs{\FF{m}}$) encodes a $\IC^n$-like chart of $\FF{m}$, glued together as per their intersection, $\s_{\sss I}\<\cap\s_{\sss J}\<\subset\fan{\FF{m}}$. The complete hierarchy of these mutual intersections, down to the 1-cones $\n_i$, fully specifies not only the space $\FF{m}$ but also its toric holomorphic submanifolds and {\em\/their\/} mutual intersections~\cite{rD-TV}. In particular, each of these 1-cones specifies a divisor defined as the zero locus of a (Cox) variable, such as the $\n_i\<\mapsto X_i$ in~\eqref{e:nFmQnu}, which in turn generate the homogeneous coordinate ring of the toric space~\cite{rCox}.

\subsection{The Anticanonical System}
\label{s:K*}
Another key holomorphic characteristic of the ambient space in which we seek Calabi-Yau hypersurfaces,
\begin{equation}
 \Big(\XX{m}\subset\big(\FF{m}\<\coeq\{ p_0(x,y)\<=0\}\big)\Big)
  \in \K[{r||c|c}{\IP^n&1&n\\ \IP^1&m&2{-}m}],\qquad
   p_0(x,y)\<=x_0y_0\!^m{+}x_1y_1\!^m,
 \label{e:bPnFmX}
\end{equation}
are the anticanonical sections
 $\G\big(\cK^*_{\smash{\FF{m}}}\<=\cO\pM{n\\2{-}m}\big|_{\smash{\FF{m}}}\big)$, i.e.,
degree-$\pM{n\\2{-}m}$ defining equations $q(x,y)\<=0$.
 For $m\<=0$ and $1$, all $3\binom{2n-1}n$ such sections are regular $(x,y)$-polynomials on $A\<=\IP^n{\times}\IP^1$, while for $m\<=2$ only $2\binom{2n-1}n\<=\binom{2n}n$ are regular, global polynomials. The remaining $\binom{2n-1}n$ sections are {\em\/non-polynomial\/}~\cite{rGHPD,rBeast}, and stem from certain 1-forms on $A\<=\IP^n{\times}\IP^1$. In the general, $m\<\geqslant3$ cases, those 1-forms are the sole source of $\cK^*_{\smash{\FF{m}}}$-sections, as evident from the Koszul resolution of the restriction
 $\cO_A\pM{n\\2{-}m}\<\to\cO\pM{n\\2{-}m}|_{\smash{\FF{m}}}$~\cite{rgCICY1,rBH-Fm}.

For the required total degree-$\pM{n\\2-m}$ and $m\<\geqslant3$, we list products of non-negative powers of the variables~\eqref{e:bP=T} except $ p_0$, all of which must have at least one $\Fs(x,y)$-factor:
\begin{equation}
  q(x,y;\l)\<\coeq
   \sum_{k=0}^{n-1} ~~\sum_{\ell=0}^{\makebox[0pt][c]{$\SSS km+2$}}
    \underbrace{c^{\sss(n-k-1)}_\ell(x_2,\cdots,x_n)}_{\deg\,=~\sss\pM{n-k-1\\0}}\;
     \underbrace{\big(y_0\!^{km+2-\ell}y_1\!^\ell\big)}_{+~\pM{0\\2+km}}\;
      \underbrace{\Fs(x,y;\l)^{k+1}}_{+~(k+1)\pM{~~1\\-m}}.
 \label{e:sK*}
\end{equation}
where the $c^{\sss(n-k-1)}_\ell$ are regular polynomials of their arguments, and the mod-$ p_0$ equivalence is inherited from the $\Fs$-factor. This may be seen as a generalization of Construction~\ref{C:Fs} and~\eqref{e:gDrX}.

The toric rendition encodes the anticanonical sections by the {\em\/polar\/}~\cite{rF-TV,rGE-CCAG,rCLS-TV} of the spanning polytope, $\pDs{\FF{m}}\<\lat\fan{\FF{m}}$ (such as in~\eqref{e:nFmQnu}, left and middle):
\begin{equation}
  (\pDs{\FF{m}})^\circ
  \define \{\, u{:}~~ \vev{v,u}{+}1\<\geqslant0,~~ v\<\in\pDs{\FF{m}}\,\}.
  \label{e:StdP}
\end{equation}
These {\em\/regular\/} anticanonical sections are then all of the form~\cite{rBaty01}:
\begin{alignat}9
  H^0(\FF{m},\cK^*) &\ni \sum_{\m\in M\cap(\pDs{\FF{m}})^\circ}
                  a_\mu \Big(\prod_{\n_i\smt\pDs{\FF{m}}} X_i^{~\vev{\n_i,\m}+1}\Big)
 \label{e:CsK*}
\iText{where $\n_i\<\smt\pDs{X}$ are the vertices of $\pDs{X}$, i.e., the 1-cone generators of $\S_X$, with $N$-lattice co-prime coordinates specifying the Cox variables $X_i$, and $M$ is the lattice dual to $N$. This yields}
  H^0(\FF[2]m,\cK^*)&\ni~
  X_1X_2 \big(c_0^1\,X_3\!^2 +c_1^1\,X_3X_4 +c_2^1\,X_4\!^2\big)\nn\\*
 &\quad~~ +X_1\!^2 \big(c_0^0\,X_3\!^{m+2} +c_1^0\,X_3\!^{m+1}X_4 +\cdots
                              +c_{m+1}^0\,X_3X_4\!^{m+1} +c_{m+2}^0\,X_4\!^{m+2}\big),
 \label{e:TsK*}
\end{alignat}
exactly matching the $n\<=2$ case of~\eqref{e:sK*} after renaming the variables as in~\eqref{e:bP>T} and having simplified, e.g., $c_0^1(X_2)=c_0^1\,X_2$ and $c_0^0(X_2)=c_0^0$, so the coefficients $c_i^{n-k-1}$ in~\eqref{e:TsK*} are plain constants. These regular polynomials indeed all have an overall factor of $X_1\<\iff\Fs(x,y)$, and so fully agree with~\eqref{e:sK*}.

The ``tuning'' of $q(x,y;\l)$ in~\eqref{e:sK*} to~\eqref{e:bPnFm} builds the moduli space of generalized complete intersections such as~\eqref{e:bPnFmX} over the deformation space (even if discrete) of the general type ambient spaces such as~\eqref{e:bPnFm}. While we defer a detailed study of this hierarchy, let us consider a few examples.

\subsection{Discrete Deformations}
\label{s:DscDef}
Consider the Hirzebruch scroll
\begin{figure}[htbp]
$$
  \begin{array}{@{}r@{~~}r@{~~}r@{~~}r@{~}|@{~}r@{~~}r@{}}
   & X_1  & X_2 & X_3 & X_4 & X_5 \\ \toprule\nGlu{-2pt}
\MR3*{\rotatebox{90}{$\fan{\FF[3]5}$}}
   & -1  &  1  &  0  &  0  & -5 \\[-3pt] 
   & -1  &  0  &  1  &  0  & -5 
 \TikZ{\path[use as bounding box](0,0);
        \path(.2,.115)node{$\left.\rule{0pt}{7mm}\right\}$};
        \draw[thick,->](.3,.1)--++(.4,0);
            }%
                                \\[-3pt] 
   &  0  &  0  &  0  &  1  & -1 \\[-1pt] \midrule\nGlu{-2pt}
Q^1&  1  &  1  &  1  &  0  &  0 \\[-1pt]
Q^2& -5  &  0  &  0  &  1  &  1 \\
  \end{array}
\qquad\quad
 \vC{\begin{picture}(50,40)
  \put(0,0){\includegraphics[viewport=80mm 0mm 250mm 130mm, clip, height=40mm]
            {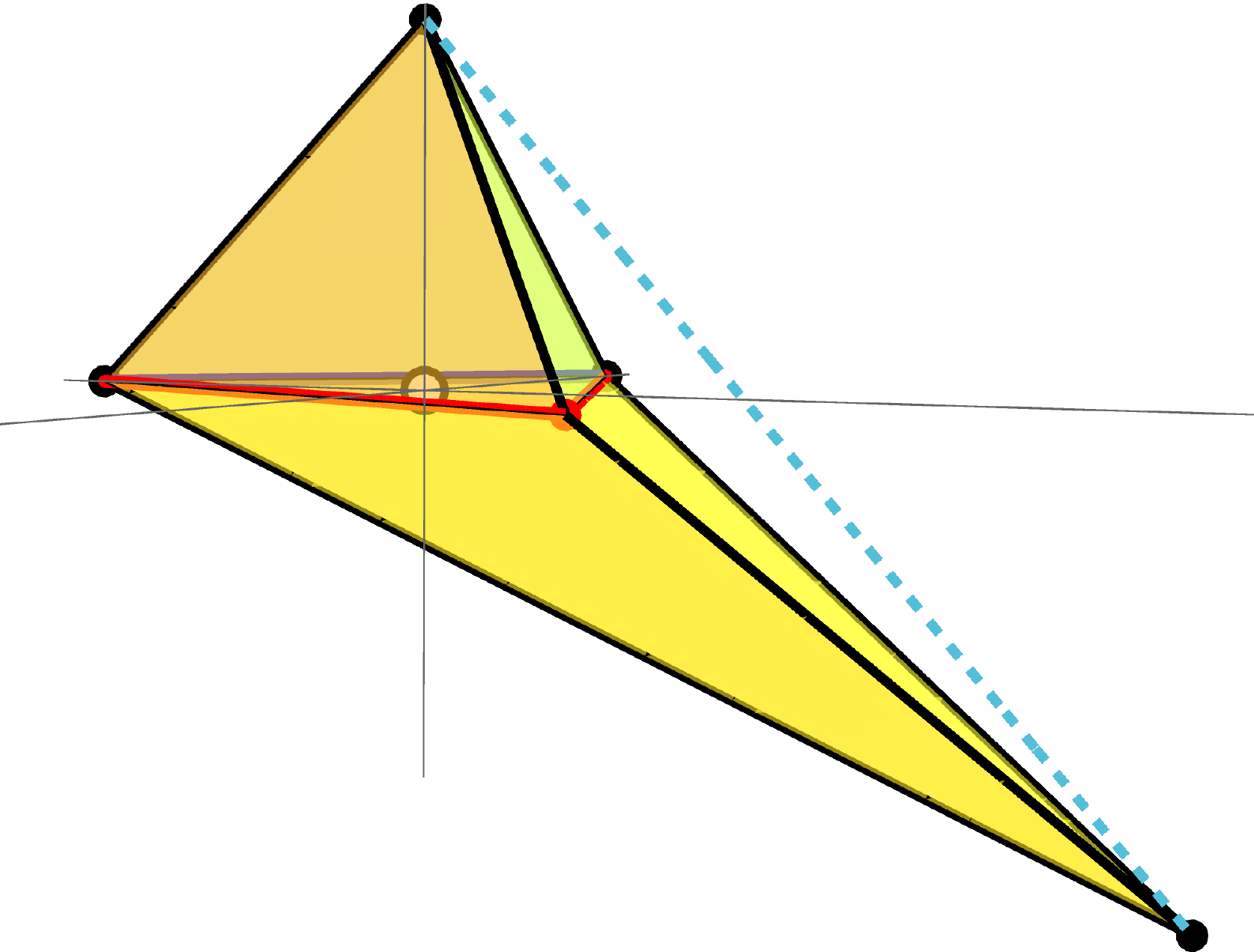}}
  \put(22,19){\footnotesize\C1{$\bS{\n_1}$}}
  \put(26,25){\footnotesize$\n_2$}
  \put(1.5,21){\footnotesize$\n_3$}
  \put(20,38){\footnotesize$\n_4$}
  \put(39,33){\Large$\pDs{\FF[3]{\!\sss(5,0,0)}}$}
  \put(52,0){\footnotesize$\n_5$}
 \end{picture}}
\qquad
 \vC{\begin{picture}(30,40)
  \put(3,0){\includegraphics[height=40mm]
            {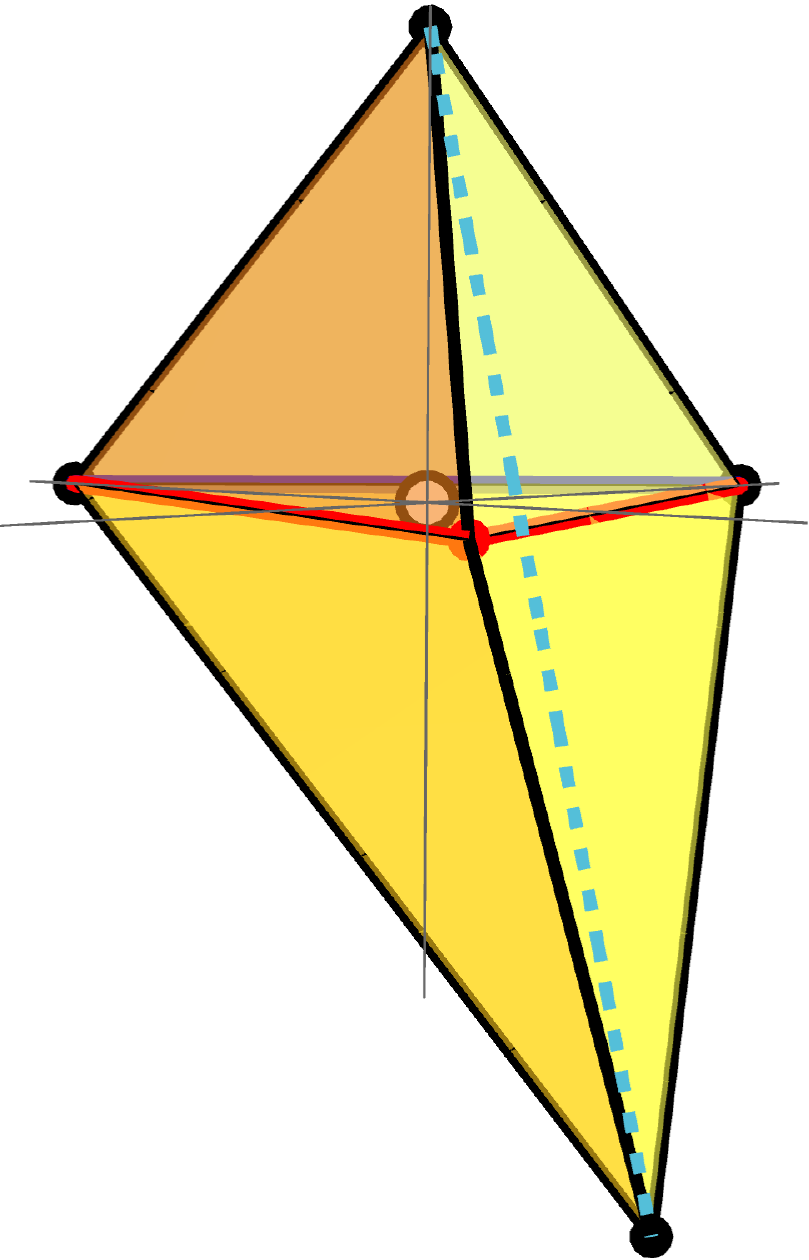}}
  \put(15,19){\footnotesize\C1{$\bS{\n_1}$}}
  \put(28,25){\footnotesize$\n_2$}
  \put(1.5,25){\footnotesize$\n_3$}
  \put(12,38){\footnotesize$\n_4$}
  \put(19,0){\footnotesize$\n_5$}
 \end{picture}}
$$
 \caption{The toric specification of $\FF[3]5$ (left) and its spanning polytope (middle and right)}
 \label{f:3F5}
\end{figure}
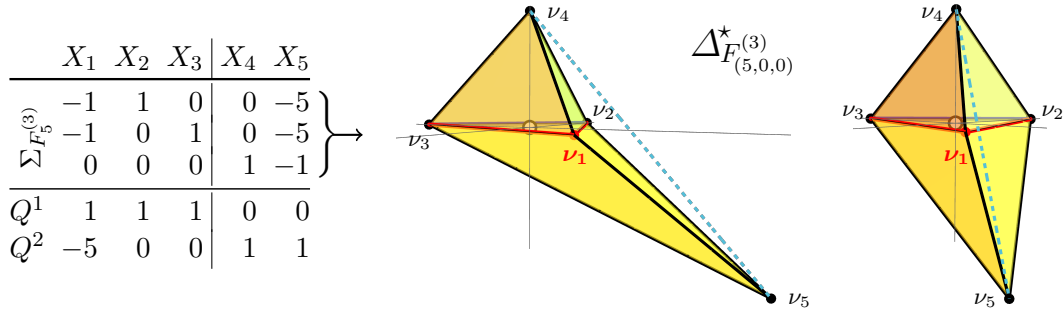
 $\FF[3]{5;0}\<=\{x_0\,y_0\!^5{+}x_1\,y_1\!^5\}\in\ssK[{r||c}{\IP^3&1\\ \IP^1&5}]$,
with its unique degree-$\pM{\3-1\\-5}$ directrix, $\Fs(x,y)$; see~\eqref{e:DrX}. The reparametrization~\eqref{e:0psxy} leads to the toric rendition in Figure~\ref{f:3F5},
its spanning polytope, $\pDs{\FF[3]{(5,0,0)}}$, depicted to the right of the tabulation from two vantage points for clarity. It is non-convex at the saddle-point, $\n_1$.
 The horizontal polygon spans the fan of the fibre-$\IP^2$ and
 $-\!\sum_{i=1}^3Q^2(X_i)\<=5$ is the total twist in this $\IP^2$-bundle over $\IP^1$.

\paragraph{A Simple Deformation:}
Consider deforming the $n\<=3$, $m\<=5$ central case~\eqref{e:bPnFm} in the
 $\ssK[{r||c}{\IP^3&1\\ \IP^1&5}]$ deformation family, which corresponds to the toric specification of $\FF[3]5$ in Figure~\ref{f:3F5}:
\begin{subequations}
 \label{e:5>41a}
\begin{alignat}9
 && p_1(x,y)&= x_0\,y_0\!^5 +x_1\,y_1\!^5 +x_2\,y_1\!^4 y_0\!^1.
 \label{e:p1}
\iText{It admits two algebraically independent directrices:}
 \pM{\3-1\\-4}{:}&\quad&
 \Fs_{1,1}(x,y)
         &=\frac{x_0\,y_0}{y_1^5} +\frac{x_2}{y_1^4} -\frac{x_1}{y_0^4}~\pMod{p_1},\\
 \pM{\3-1\\-1}{:}&\quad&
 \Fs_{1,2}(x,y)
         &=\frac{x_0}{y_1} -\frac{x_2}{y_0} -\frac{x_1\,y_1^4}{y_0^5}~\pMod{p_1}.
\end{alignat}
\end{subequations}
As above, the reparametrization
\begin{equation}
 (x_0,x_1,x_2,{\cdots};y_0,y_1)\to(p_1,\Fs_{1,1},\Fs_{1,2},{\cdots};y_0,y_1),\qquad
\det\big[\frac{\vd(p_1,\Fs_{1,1},\Fs_{1,2},{\cdots};y_0,y_1)}{\vd(x_0,x_1,x_2,{\cdots};y_0,y_1)}\big]=4
 \label{e:41psxy}
\end{equation}
again has a constant Jacobian, and produces the toric rendition:
\begin{equation}
  \begin{array}{@{}r@{~~}r@{~~}r@{~~}r@{~}|@{~}r@{~~}r@{}}
   & X_1  & X_2 & X_3 & X_4 & X_5 \\ \toprule\nGlu{-2pt}
\MR3*{\rotatebox{90}{$\fan{\FF[3]{5;3}}$}}
   & -1  &  1  &  0  &  0  & -3 \\[-3pt] 
   & -1  &  0  &  1  &  0  & -4 
 \TikZ{\path[use as bounding box](0,0);
        \path(.2,.115)node{$\left.\rule{0pt}{7mm}\right\}$};
        \draw[thick,->](.3,.1)--++(.4,0);
            }%
                                \\[-3pt] 
   &  0  &  0  &  0  &  1  & -1 \\[-1pt] \midrule\nGlu{-2pt}
Q^1&  1  &  1  &  1  &  0  &  0\\[-1pt]
Q^2& -4  & -1  &  0  &  1  &  1\\
  \end{array}
\qquad\qquad
 \vC{\begin{picture}(45,40)
  \put(0,0){\includegraphics[viewport=50mm 0mm 220mm 150mm, clip, height=41mm]
            {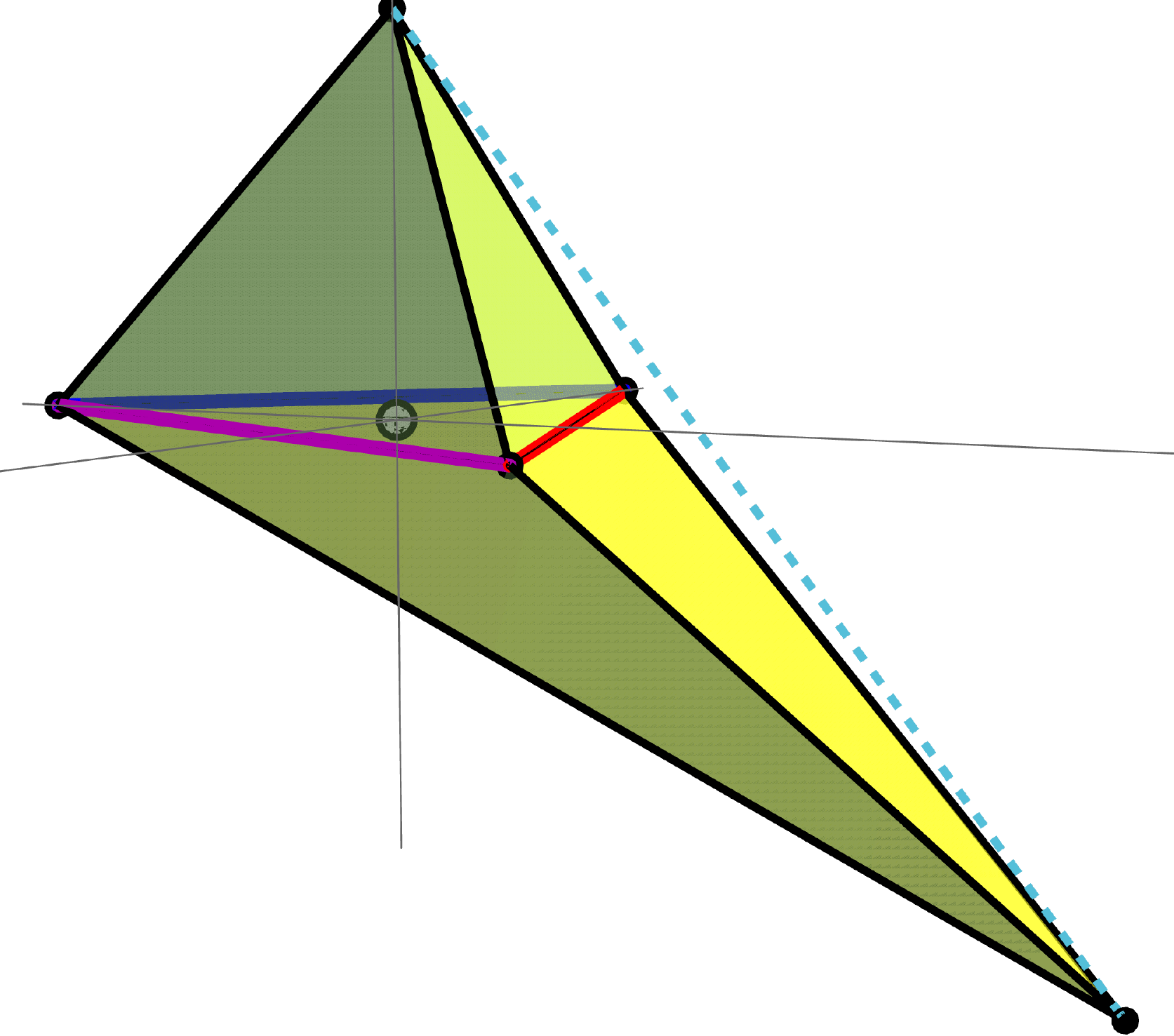}}
  \put(18,19){\footnotesize\C1{$\bS{\n_1}$}}
  \put(25,27){\footnotesize$\n_2$}
  \put(-1,22.5){\footnotesize$\n_3$}
  \put(17,40){\footnotesize$\n_4$}
  \put(46,0){\footnotesize$\n_5$}
  \put(20,13){\TikZ{\path[use as bounding box](0,0);
                    \path[Sage,left](0,0)node{flat};
                    \path[Sage,left](.6,-.4)node{non-convex};
                    \path[Sage,left](1.2,-.8)node{rectangle};
                    \draw[Sage,thick,->](-.7,.1)to[out=120,in=225]++(0,.8);
              }}
  \put(33,33){\Large$\FF[3]{\!\sss(4,1,0)}$}
 \end{picture}}
\qquad
 \vC{\begin{picture}(27,40)
  \put(0,0){\includegraphics[viewport=128mm 0mm 240mm 165mm, clip, height=42mm]
            {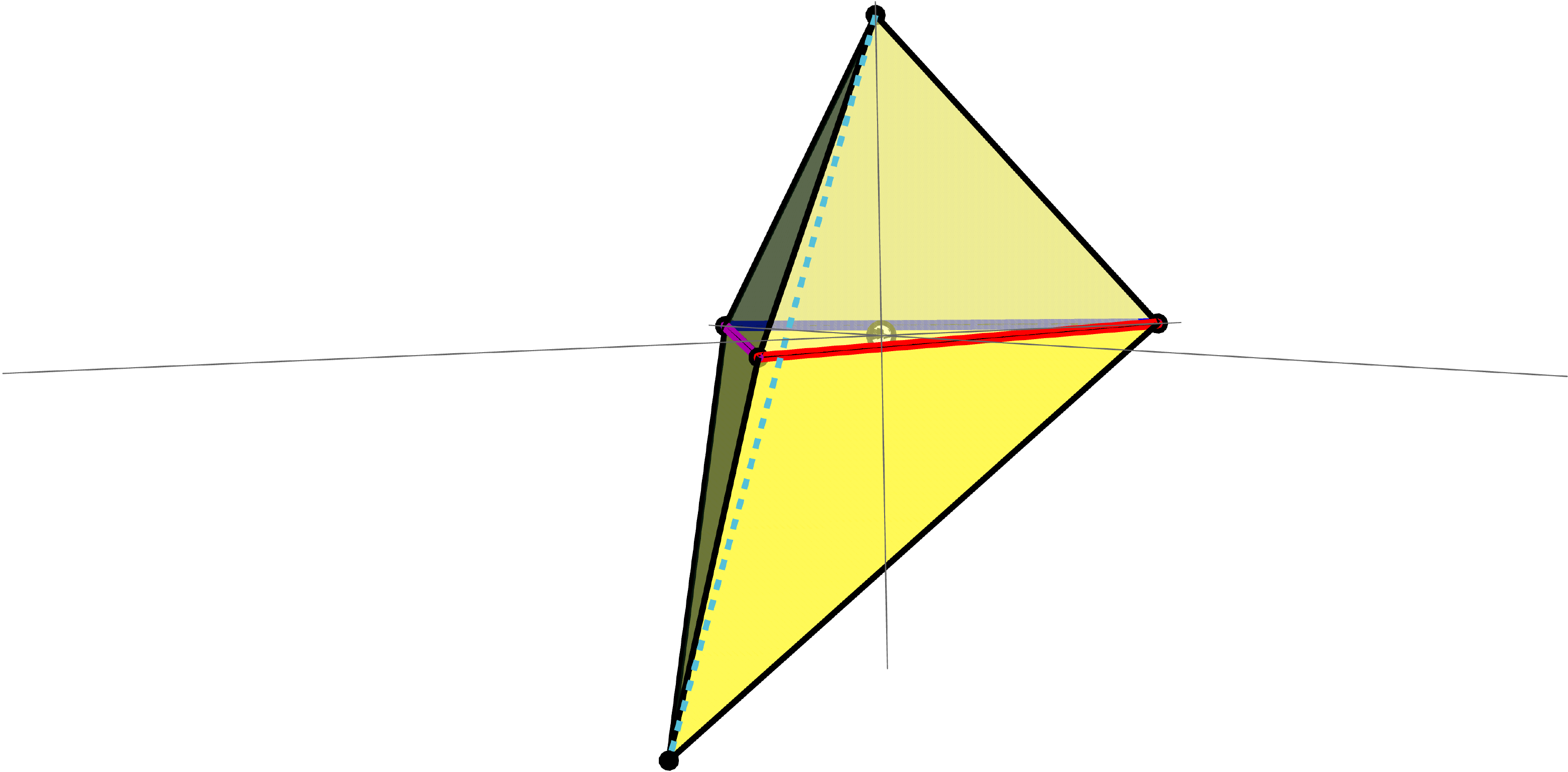}}
  \put(7,18){\footnotesize\C1{$\bS{\n_1}$}}
  \put(27,25){\footnotesize$\n_2$}
  \put(0,22.5){\footnotesize$\n_3$}
  \put(8,39){\footnotesize$\n_4$}
  \put(4,0){\footnotesize$\n_5$}
 \end{picture}}
 \label{e:3F41s}
\end{equation}
where the Cox variables are $X_1\<=\Fs_{1,1}$, $X_2\<=\Fs_{1,2}$, $X_3\<=x_3$, $X_4\<=y_0$ and $X_5\<=y_1$.

\paragraph{Another Simple Deformation:}
Another simple deformation within the $\ssK[{r||c}{\IP^3&1\\ \IP^1&5}]$ deformation family,
\begin{subequations}
 \label{e:5>32a}
\begin{alignat}9
 \pM{1\\5}{:}&\quad&
 p_2(x,y)&= x_0\,y_0\!^5 +x_1\,y_1\!^5 +x_2\,y_1\!^3\,y_0\!^2
 \label{e:p2}
\iText{admits two algebraically independent directrices:}
 \pM{\3-1\\-3}{:}&\quad&
 \Fs_{2,1}(x,y)
         &=\frac{x_0\,y_0^2}{y_1^5} +\frac{x_2}{y_1^3} -\frac{x_1}{y_0^3}~\pMod{p_2},\\
 \pM{\3-1\\-2}{:}&\quad&
 \Fs_{2,2}(x,y)
         &=\frac{x_0}{y_1^2} -\frac{x_2}{y_0^2} -\frac{x_1 y_1^3}{y_0^5}~\pMod{p_2}.
\end{alignat}
\end{subequations}
As above, the reparametrization
\begin{equation}
 (x_0,x_1,x_2,{\cdots};y_0,y_1)\to(p_2,\Fs_{2,1},\Fs_{2,2},{\cdots};y_0,y_1),\qquad
\det\big[\frac{\vd(p_2,\Fs_{2,1},\Fs_{2,2},{\cdots};y_0,y_1)}{\vd(x_0,x_1,x_2,{\cdots};y_0,y_1)}\big]=4
 \label{e:32psxy}
\end{equation}
again has a constant Jacobian and produces the toric rendition:
\begin{equation}
  \begin{array}{@{}r@{~~}r@{~~}r@{~~}r@{~}|@{~}r@{~~}r@{}}
   & X_1  & X_2 & X_3 & X_4 & X_5 \\ \toprule\nGlu{-2pt}
\MR3*{\rotatebox{90}{$\fan{\FF[3]{5;2}}$}}
   & -1  &  1  &  0  &  0  & -1 \\[-3pt] 
   & -1  &  0  &  1  &  0  & -3 
 \TikZ{\path[use as bounding box](0,0);
        \path(.2,.115)node{$\left.\rule{0pt}{7mm}\right\}$};
        \draw[thick,->](.3,.1)--++(.4,0);
            }%
                                \\[-3pt] 
   &  0  &  0  &  0  &  1  & -1 \\[-1pt] \midrule\nGlu{-2pt}
Q^1&  1  &  1  &  1  &  0  &  0\\[-1pt]
Q^2& -3  & -2  &  0  &  1  &  1\\
  \end{array}
\qquad\quad
 \vC{\begin{picture}(52,45)
  \put(0,0){\includegraphics[viewport=25mm 0mm 155mm 105mm, clip, height=43mm]
            {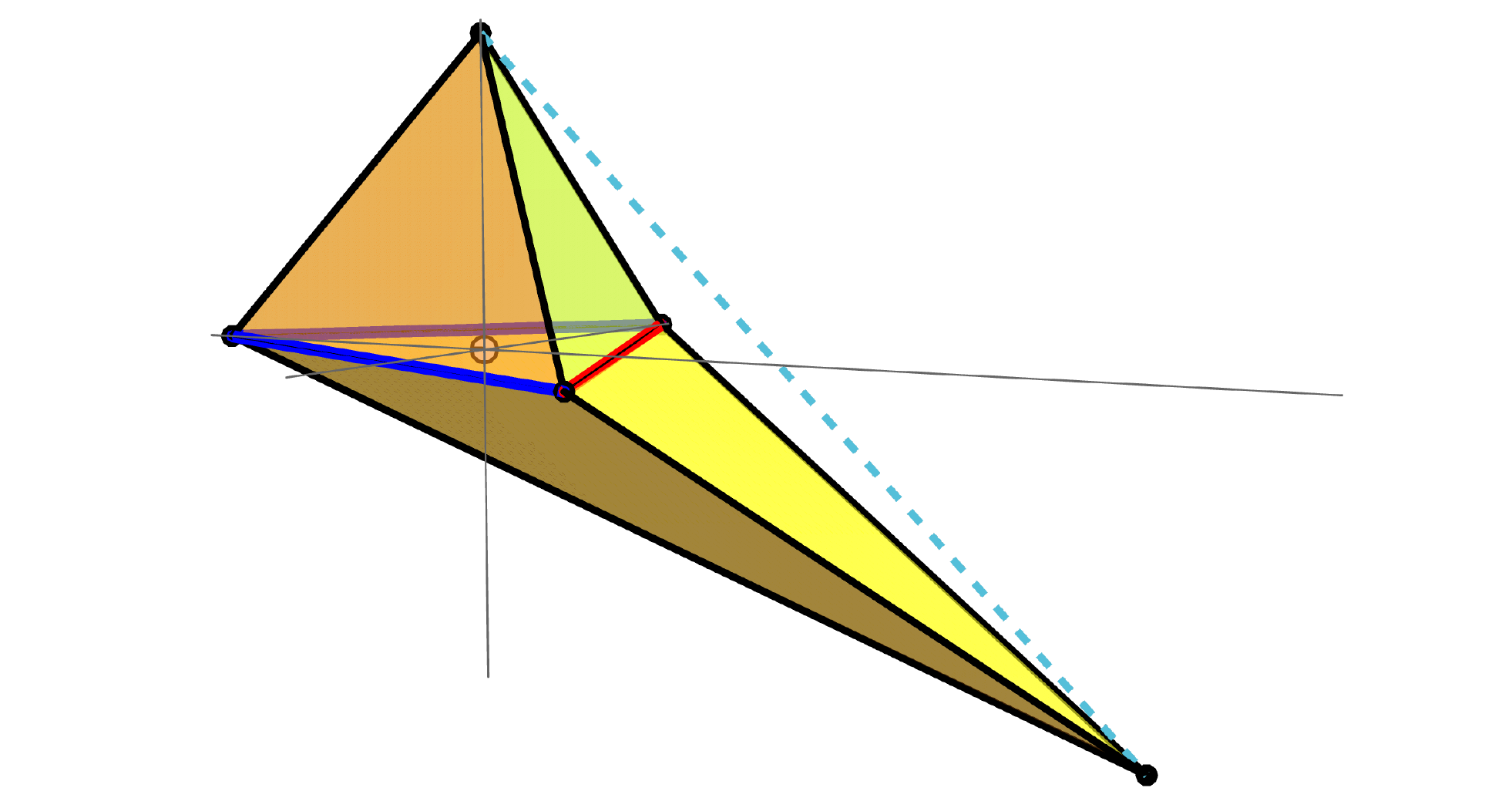}}
  \put(23,21){\footnotesize\C1{$\bS{\n_1}$}}
  \put(27,26){\footnotesize$\n_2$}
  \put(-1,22){\footnotesize$\n_3$}
  \put(17,41){\footnotesize$\n_4$}
  \put(51,4){\footnotesize$\n_5$}
  \put(37,33){\Large$\FF[3]{\!\sss(3,2,0)}$}
 \end{picture}}
\qquad
 \vC{\begin{picture}(35,40)
  \put(0,0){\includegraphics[viewport=35mm 0mm 130mm 110mm, clip, height=40mm]
            {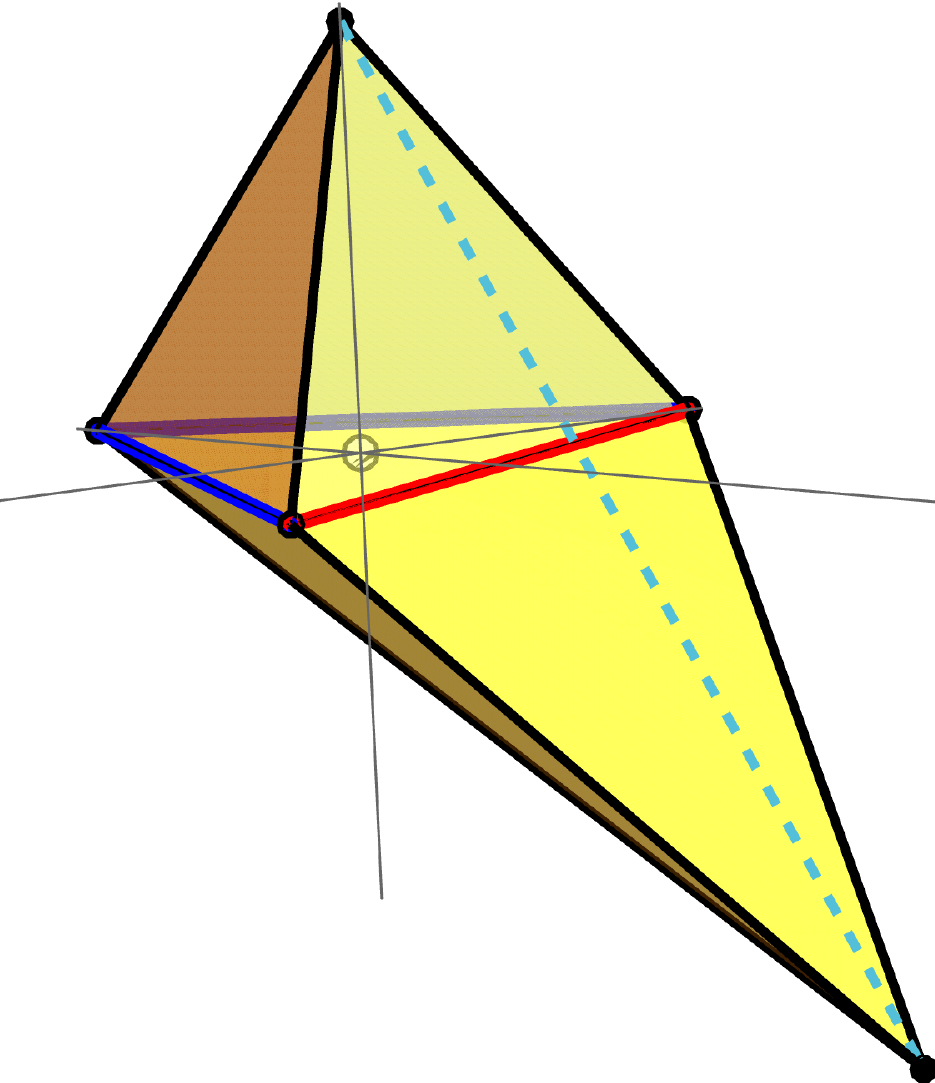}}
  \put(8,17){\footnotesize\C1{$\bS{\n_1}$}}
  \put(27,25){\footnotesize$\n_2$}
  \put(0,22){\footnotesize$\n_3$}
  \put(14,39){\footnotesize$\n_4$}
  \put(29,0){\footnotesize$\n_5$}
 \end{picture}}
 \label{e:3F32s}
\end{equation}
where the Cox variables are $X_1\<=\Fs_{2,1}$, $X_2\<=\Fs_{2,2}$, $X_3\<=x_3$, $X_4\<=y_0$ and $X_5\<=y_1$.

\paragraph{A Double Deformation:}
Consider a further, $\IP^1$-symmetrizing deformation of~\eqref{e:p2}:
\begin{subequations}
 \label{e:5>32s}
\begin{alignat}9
 \deg&=\pM{1\\5}{:}&\quad
 p_3(x,y)&= x_0\,y_0\!^5 +x_1\,y_1\!^5 +x_2\,y_1\!^3\,y_0\!^2 +x_3\,y_1\!^2\,y_0\!^3
  \label{e:p3}
\iText{and admits three algebraically independent directrices:}
 \deg&=\pM{\3-1\\-2}{:}&\quad
 \Fs_{3,1}(x,y)
         &=\frac{x_0}{y_1\!^2} -\frac{x_2}{y_0\!^2}
           -\frac{x_3\,y_1}{y_0\!^3} -\frac{x_1\,y_1\!^3}{y_0\!^5}~\pMod{p_3},
 \label{e:5>32s1}\\
 \deg&=\pM{\3-1\\-2}{:}&\quad
 \Fs_{3,2}(x,y)
         &=\frac{x_0\,y_0^3}{y_1^5} +\frac{x_2\,y_0}{y_1^3}
           +\frac{x_3}{y_1^2} -\frac{x_1}{y_0^2}~\pMod{p_3},
 \label{e:5>32s2}\\
 \deg&=\pM{\3-1\\-1}{:}&\quad
 \Fs_{3,3}(x,y)
         &=\frac{x_0\,y_0^2}{y_1^3} +\frac{x_2}{y_1}
            -\frac{x_3}{y_0} -\frac{x_1\,y_1^2}{y_0^3}~\pMod{p_3}.
 \label{e:5>32s3}
\end{alignat}
\end{subequations}
These $\Fs_{3,i}(x,y)$ have four monomials instead of just two in~\eqref{e:0psxy}.
 As before, the reparametrization
\begin{equation}
  (x_0,x_1,x_2,x_3,{\cdots};y_0,y_1)\to
 (p_3,\Fs_{3,1},\Fs_{3,2},\Fs_{3,3},{\cdots};y_0,y_1),
 \label{e:221psxy}
\end{equation}
has a constant Jacobian,
 $\det\big[\frac{\vd(p_3,\Fs_{3,1},\Fs_{3,2},\Fs_{3,3},{\cdots};y_0,y_1)}
               {\vd(x_0,x_1,x_2,x_3,{\cdots};y_0,y_1)}\big]\<=8$.
 The 3-dimensional hypersurface $p_3(x,y)\<=0$ has the straightforward toric rendition with the Cox variables $X_i\<=\Fs_{3,i}(x,y)$, $X_4\<=y_0$ and $X_5\<=y_1$:
\begin{equation}
  \begin{array}{@{}r@{~~}r@{~~}r@{~~}r@{~}|@{~}r@{~~}r@{}}
   & X_1  & X_2 & X_3 & X_4 & X_5 \\ \toprule\nGlu{-2pt}
\MR3*{\rotatebox{90}{$\fan{\FF[3]{5;\e_3}}$}}
   & -1  &  1  &  0  &  0  &  0 \\[-3pt] 
   & -1  &  0  &  1  &  0  & -1 \\[-3pt] 
   &  0  &  0  &  0  &  1  & -1 \\[-1pt] \midrule\nGlu{-2pt}
Q^1&  1  &  1  &  1  &  0  &  0\\[-1pt]
\Tw{Q}^2& -2  & -2  & -1  &  1  &  1\\
  \end{array}
~~\approx_{\sss\IR}~~
  \begin{array}{@{}r@{~~}r@{~~}r@{~~}r@{~}|@{~}r@{~~}r@{}}
   & X_1  & X_2 & X_3 & X_4 & X_5 \\ \toprule\nGlu{-2pt}
\MR3*{\rotatebox{90}{$\fan{\FF[3]{5;\e_3}}$}}
   & -1  &  1  &  0  &  0  &  0 \\[-3pt] 
   & -1  &  0  &  1  &  0  & -1 
 \TikZ{\path[use as bounding box](0,0);
        \path(.2,.115)node{$\left.\rule{0pt}{7mm}\right\}$};
        \draw[thick,->](.3,.1)--++(.4,0);
            }%
                                \\[-3pt] 
   &  0  &  0  &  0  &  1  & -1 \\[-1pt] \midrule\nGlu{-2pt}
Q^1&  1  &  1  &  1  &  0  &  0\\[-1pt]
Q^2& -1  & -1  & 0  &  1  &  1\\
  \end{array}
\qquad\qquad
 \vC{\begin{picture}(30,40)
  \put(0,0){\includegraphics[height=40mm]{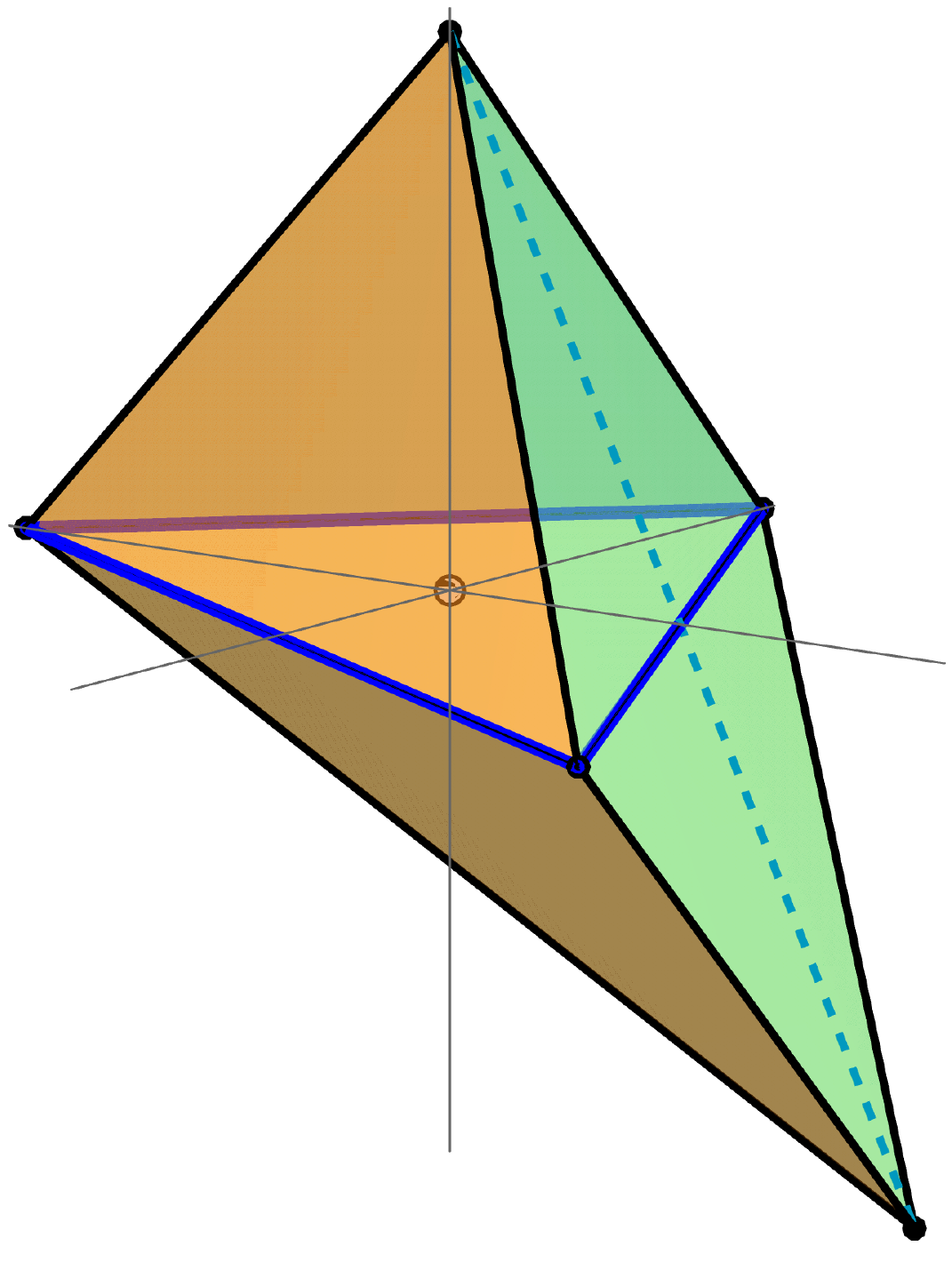}}
  \put(20,14){\footnotesize\C1{$\bS{\n_1}$}}
  \put(25,23){\footnotesize$\n_2$}
  \put(0,20){\footnotesize$\n_3$}
  \put(9,38){\footnotesize$\n_4$}
  \put(30,1){\footnotesize$\n_5$}
  \put(18,37){\TikZ{\path[use as bounding box](0,0);
                    \path[Sage,right](0,0)node{flat};
                    \path[Sage,right](.2,-.4)node{convex};
                    \path[Sage,right](.5,-.8)node{rectangle};
                    \draw[Sage,thick,->](1.4,-1)to[out=-90,in=0]++(-.9,-.7);
              }}
  \put(0,10){\Large$\FF[3]{\!\sss(1,1,0)}$}
 \end{picture}}
 \label{e:221}
\end{equation}
where the choice of $Q$-charges on the far left (bottom two rows) follows from the change of variables~\eqref{e:221psxy} with~\eqref{e:5>32s}, which simplifies to $Q^2=\Tw{Q}^2{-}Q^1$, reflecting the $\FF[3]5\<{\approx_{\sss\IR}}\FF[3]2$ diffeomorphism of Hirzebruch scrolls. In turn, the $\fan{\FF[3]{5;\e_3}}\<\smt\pDs{\FF[3]{(1,1,0)}}$ specification~\eqref{e:221} unambiguously specifies this latter choice of 5-vectors, $(Q^1,Q^2)$, as the correct Mori vectors~\cite{rBKK-tvMirr}, consistent with a star-triangulation of the spanning polytope and the corresponding simplicial unit subdivision of the fan.
 This type of discrete deformations $\FF{m}\<\leadsto\FF{m\pMod{n}}$ have been seen to affect string compactifications since early on, notably in the simplest form, $\FF[2]2\<\leadsto\FF[2]0$~\cite{Morrison:1996na,rPhasesMF}.
 By effectively reducing the negativity of $X_1,X_2$ and the total twist from 5 to 2, the resulting toric specification $\FF[3]{\sss(1,1,0)}$ in~\eqref{e:221} no longer features directrices as negative as~\eqref{e:5>32s1}--\eqref{e:5>32s2}, and deforms the non-Fano hypersurface~\eqref{e:5>32s3} into the almost Fano $\FF[3]{\sss(1,1,0)}$.
 
\paragraph{A Comparison:}
Two rather distinct-looking members of this deformation family of 3-folds, the $y_0\<\iff y_1$-symmetrized versions of~\eqref{e:p1}, and an asymmetric deformation of~\eqref{e:p2}:
\begin{equation}
  \ARR.{@{}r@{\;}c@{\;}l@{}}{~~\\[1mm]
  \K[{r||c}{\IP^3&1\\ \IP^1&5}] &\ni&
  \ARR.{@{}l@{~}r@{~}l@{}}
        {x_0\,y_0\!^5 \<+x_1\,y_1\!^5&+x_2\,y_0\!^4\,y_1 \<+x_3\,y_0\,y_1\!^4&=0\\[2mm]
         x_0\,y_0\!^5 \<+x_1\,y_1\!^5&+x_2\,y_0\!^4\,y_1 \<+x_3\,y_0\!^3\,y_1\!^2&=0}.\\
  \midrule\nGlu{-1pt}
  &\To& \bigg\{\ARR.{@{}r@{\;=\;}l@{}}
             {\Tw{Q}^1(x_0,\cdots,y_1)&(1,1,1,0,0)\\
              \Tw{Q}^2(x_0,\cdots,y_1)&(-3,-1,-1,1,1)}.}.
  ~~\To~~
  \begin{array}{@{}r@{~~}r@{~~}r@{~~}r@{~}|@{~}r@{~~}r@{}}
   & X_1  & X_2 & X_3 & X_4 & X_5 \\ \toprule\nGlu{-2pt}
\MR3*{\rotatebox{90}{$\fan{\FF[3]{(2,0,0)}}$}}
   & -1  &  1  &  0  &  0  & -2 \\[-3pt] 
   & -1  &  0  &  1  &  0  & -2 \\[-3pt] 
   &  0  &  0  &  0  &  1  & -1 \\[-1pt] \midrule\nGlu{-2pt}
Q^1&  1  &  1  &  1  &  0  &  0 \\[-1pt]
Q^2& -2  &  0  &  0  &  1  &  1 \\
  \end{array}
 \label{e:41=32}
\end{equation}
Each of them admits a (different) collection of one degree-$\pM{~~1\\-3}$ and {\em\/two\/} independent degree-$\pM{~~1\\-1}$ directrices. Via analogous constant-Jacobian changes of variables, they both lead to the same toric $\FF[3]{(3,1,1)}\<{\approx_{\sss\IR}}\FF[3]{(2,0,0)}$, where the last equivalence is again the toric rendition of Wall's diffeomorphism~\cite{rWall}. This shows that there exist rather nontrivial identifications within the coarse parameter space of $\ssK[{r||c}{\IP^3&1\\ \IP^1&5}]$. For each $n\<\geqslant2$,
 $\FF{(2,0,\cdots)}$ is almost Fano: both its spanning and its Newton polytope is convex and reflexive, although $\pDs{\FF{(2,0\cdots)}}$ has a degree-2 edge, which is polar to a {\em\/double\/} $(n{-}2)$-face in $\pDN{\FF{(2,0\cdots)}}$. 

By modifying the spanning polytope and its central fan,
 $\pDs{\FF{\ora{\bS{m}}}}\<\lat\fan{\FF{\ora{\bS{m}}}}$,
these and other deformations also modify the Newton polytope, both its regular part and the extension, and thereby also the entire anticanonical system.

\begin{figure}[htb]
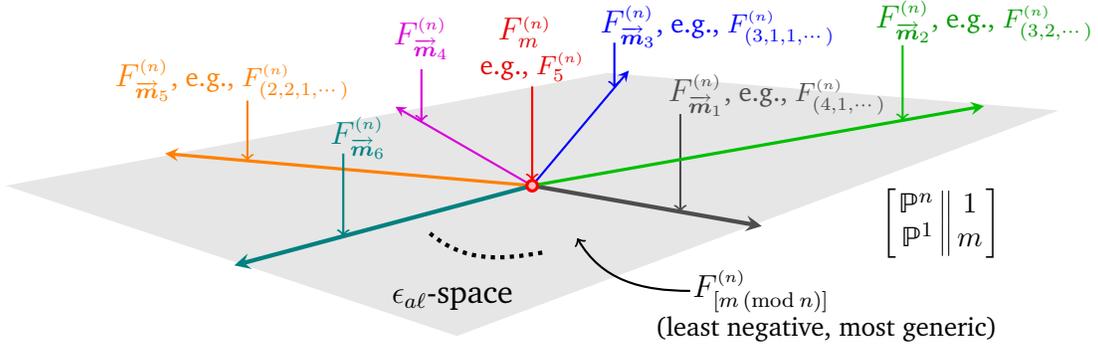

 \begin{center}
  \TikZ{\path[use as bounding box](-7,-1.5)--(7.5,2.5);
        \fill[gray!20](-7,0)--(-1,-2)--(7,1)--(1,1.5);
         \path(-2,-1.5)node[right]{\large$\e_{a\ell}$-space};
        \draw[black!70,ultra thick,-stealth](0,0)--++(-10:3.1);
         \path[black!70](-10:1.7)+(0,1.5)node[right]
              {{\large$\FF{\ora{\bS{m}}_1}$}\small, e.g., $\FF{(4,1,\cdots)}$};
         \draw[black!70,<-,thick](-10:2)--++(0,1.3);
        \draw[very thick,Green,-stealth](0,0)--++(10:6.1);
         \path[green!60!black](10:4.5)+(0,1.4)node[right]
              {{\large$\FF{\ora{\bS{m}}_2}$}\small, e.g., $\FF{(3,2,\cdots)}$};
         \draw[Green,<-,thick](10:5)--++(0,1);
        \draw[thick,blue,-stealth](0,0)--++(50:2);
         \path[blue](50:1.2)+(0,1.2)node[right]
              {{\large$\FF{\ora{\bS{m}}_3}$}\small, e.g., $\FF{(3,1,1,\cdots)}$};
         \draw[blue,<-,thick](50:1.7)--++(0,.6);
        \draw[thick,Magenta,-stealth](0,0)--++(150:2.1);
         \path[Magenta](150:1.7)+(0,.7)node[above]{\large$\FF{\ora{\bS{m}}_4}$};
         \draw[Magenta,<-,thick](150:1.7)--++(0,.7);
        \draw[very thick,orange,-stealth](0,0)--++(175:4.9);
         \path[orange](175:2.3)+(0,1.2)node[left]
              {{\large$\FF{\ora{\bS{m}}_5}$}\small, e.g., $\FF{(2,2,1,\cdots)}$};
         \draw[orange,<-,thick](175:3.8)--++(0,.8);
        \draw[ultra thick,teal,-stealth](0,0)--++(195:4.1);
         \path[teal](195:2.6)+(.2,.9)node[above]{\large$\FF{\ora{\bS{m}}_6}$};
         \draw[teal,<-,thick](195:2.6)--++(0,1.1);
        \draw[ultra thick, dotted](205:1.5)to[out=-40,in=190](280:.9);
        \filldraw[red,fill=pink,very thick](0,0)circle(.7mm);
         \draw[red,<-,thick](0,.07)--++(0,1.25);
         \path[red!90!black](0,1.25)node[above]{\small e.g., $\FF5$};
         \path[red!90!black](-.1,1.7)node[above]{\large$\FF{m}$};
         \draw[<-,thick](.6,-.7)to[out=-60,in=180]++(1.5,-.7);
        \path(2,-1.4)node[right]{\large$\FF{[m\pMod{n}]}$};
        \path(1.5,-1.9)node[right]{\small(least negative, most generic)};
        \path(4.5,-.5)node[right]{$\K[{r||c}{\IP^n&1\\ \IP^1&m}]$};
        }
 \end{center}
 \caption{A rough sketch of the full deformation family of degree-$\pM{1\\m}$ hypersurfaces in $\IP^n\<\times\IP^1$}
 \label{f:disDef}
\end{figure}

\paragraph{The General Picture:}
The hypersurfaces~\eqref{e:p1}, \eqref{e:p2} and~\eqref{e:p3} are evidently deformations of the $n\<=3$, $m\<=5$ case of~\eqref{e:bPnFm}.
 Consequently, $\FF[3]{\!\sss(4,1,0)}$, $\FF[3]{\!\sss(3,2,0)}$ and
 $\FF[3]{\!\sss(2,2,1)}\<{\approx_{\sss\IR}}\FF[3]{\!\sss(1,1,0)}$
  are all explicit (discrete) deformations of $\FF[3]5$. 
The evident generalizations of these explicit constructions suggest:
\begin{clam}
For any integral $n$-tuple $\ora{\bS{m}}$ with $\sum_{i=1}^n\bS{m}_i\<=m$, the $\ora{\bS{m}}$-twisted $\IP^{n-1}$-bundle over $\IP^1$ may be identified with toric variety
 $\FF{\ora{\bS{m}}}$, specified by the two Mori $(n{+}2)$-vectors, $Q^1\<=(1,\cdots,1;0,0)$ and
 $Q^2\<=({-}\ora{\bS{m}};1,1)$. They are all (discrete) deformations of $\FF{m}$, specified by
 $Q^2\<=(-m,0,\cdots,0;1,1)$, and may all be located in specific regions of
 $\ssK[{r||c}{\IP^n&1\\ \IP^1&m}]$, the full deformation family of simple
 degree-$\pM{1\\m}$ hypersurfaces in $\IP^n\<\times\IP^1$.
 Equivalently,
 $\FF{\ora{\bS{m}}}\approx\IP\big(\cO_{\IP^1}\oplus\bigoplus_i\cO_{\IP^1}(m_i)\big)$.
\end{clam}
\noindent
Recall that $h^1(\FF{m},T)\<<h^0(\FF{m},T)$ implies that all $\FF{m}$ are effectively rigid, so that their space of complex structures is discrete, which suggests the general situation illustrated in Figure~\ref{f:disDef}.
The concrete examples~\eqref{e:p1}, \eqref{e:p2}, \eqref{e:p3} and~\eqref{e:41=32} do not have any explicit coefficients shown since those are easily absorbed by $(x,y)$-rescaling. However, writing them out explicitly shows that for each parameter, only $\e_{a\ell}\<\neq0$ vs.\ $\e_{a\ell}\<=0$ is distinguished: all these models are infinitesimally near each other. For a related but differently constructed explicit deformation family containing $\FF[2]2$ and $\FF[2]0$ see~\cite{rGHSAR,rBeast}. Either way, the result of~\eqref{e:41=32} shows that the coarse $\e_{a\ell}$-parameter space in~\eqref{e:bPnFmE} is subject to highly nontrivial identifications.

\section{Calabi-Yau Subspaces}
\label{s:CYs}
We now turn to Calabi-Yau hypersurfaces in the {\em\/central\/} hypersurface in the deformation family $\ssK[{r||c}{\IP^n&1\\ \IP^1&m}]$; see Figure~\ref{f:disDef}. Less special cases then include deformations such as discussed in the previous section, and the deformation space of Calabi-Yau hypersurfaces therein builds atop the effectively discrete one in Figure~\ref{f:disDef}.

\subsection{Tyurin Degeneration: Calabi-Yau Matryoshke}
\label{s:CY-CY}
The explicit expansions~$\eqref{e:sK*}\<\approx\eqref{e:TsK*}$ show that for $m\<\geqslant3$, every anticanonical section of $\FF{m}$ factorizes:
\begin{alignat}9
  H^0(\FF{m},\cK^*) &\ni q(x,y) = \Fc(x,y){\cdot}\Fs(x,y),\qquad
  \deg(\Fs)\<=\pM{1\\-m},\quad \deg(\Fc)\<=\pM{n-1\\2},
 \label{e:q=cs}\\
  \Fc(x,y)&\coeq
   \sum_{k=0}^{n-1} ~~\sum_{\ell=0}^{\makebox[0pt][c]{$\SSS km+2$}}
     c^{\sss(n-k-1)}_\ell(x_2,\cdots,x_n)\,
                     \big(y_0\!^{km+2-\ell}y_1\!^\ell\big)\,\Fs^k(x,y).
 \label{e:CpX}
\end{alignat}
 Thus, all anticanonical (Calabi-Yau) hypersurfaces {\em\/reduce\/} to a union of two components:
\begin{equation}
  m\<\geqslant3,~~
   \FF{m} \supset \big(\,\XX{m}\<\coeq q^{-1}(0)\,\big)
   =\big(\,C_m\<\coeq\Fc^{-1}(0)\,\big)\cup \big(\,S_m\<\coeq\Fs^{-1}(0)\,\big).
 \label{e:nXm}
\end{equation}
As divisors in $\FF{m}$, $[\XX{m}]=[C_m]\<+[S_m]$.
With generic coefficient polynomials $c^{\sss(n-k-1)}_i(x_2,\cdots,x_n)$, the component
 $\Fc^{-1}(0)\<\subset\FF{m}$ is non-singular and holomorphic, and so irreducible.
 With $\Fs^{-1}(0)$ having been named the {\em\/directrix\/}~\cite{rGrHa}, we call $\Fc^{-1}(0)$ the {\em\/complementrix.}

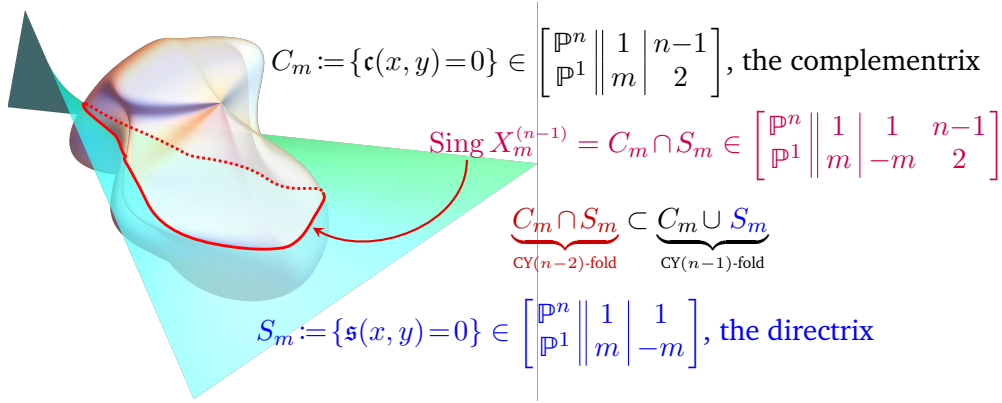
\begin{figure}[htbp]
 \begin{center}
 \setbox8=\hbox{\tiny CY$(n{-}2)$-fold}
 \setbox9=\hbox{\tiny CY$(n{-}1)$-fold}
  \begin{picture}(130,45)
   \put(-2,-7){\includegraphics[viewport=.5mm 30mm 266mm 230mm, clip, height=53mm]
              {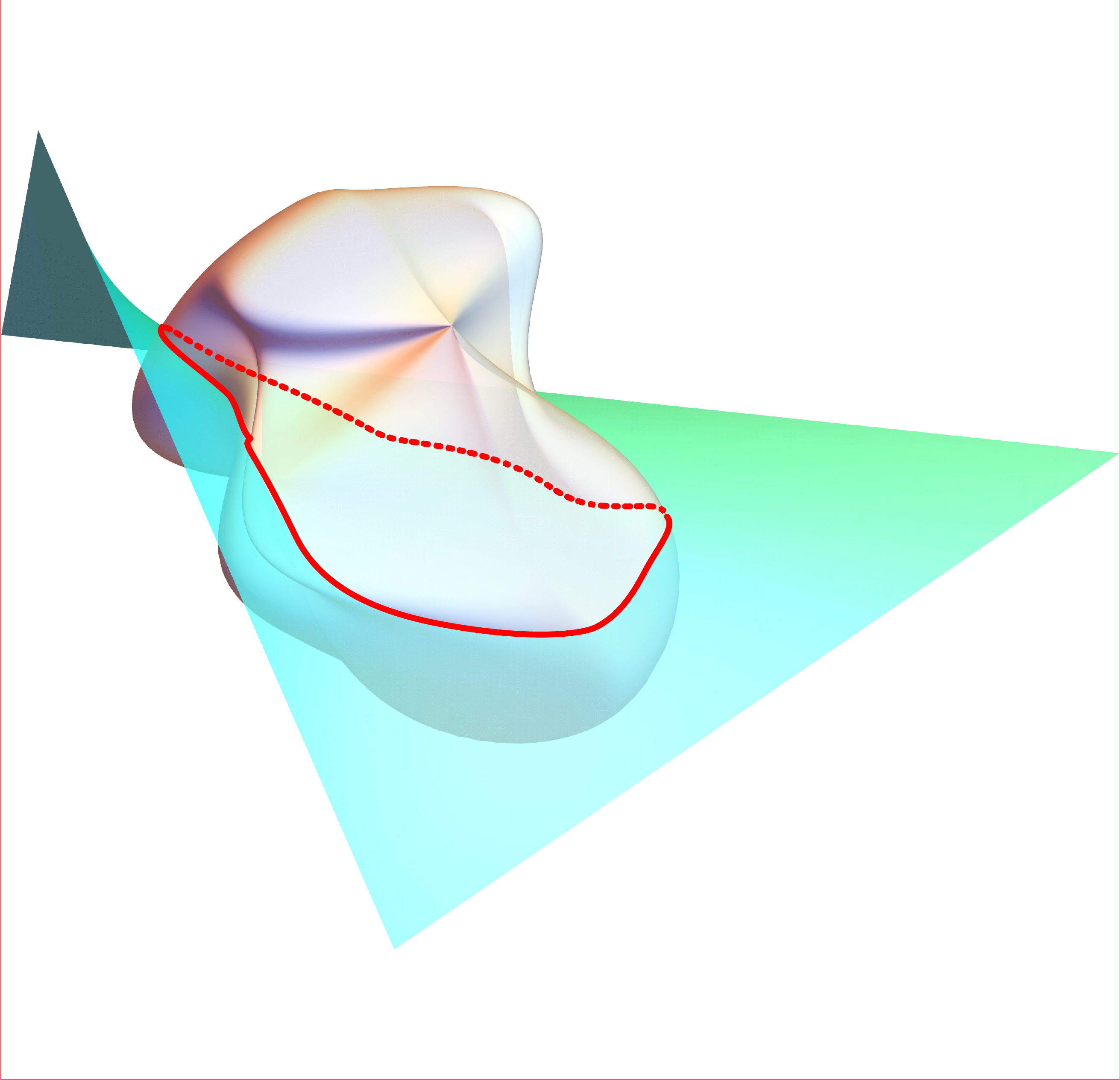}}
   \lput(31,1){\color{Blue}$S_m\<\coeq\{\Fs(x,y)\<=0\}
                             \in \K[{r||c|c}{\IP^n&1&1\\ \IP^1&m&-m}]$, the directrix}
   \lput(33,37){$C_m\<\coeq\{\Fc(x,y)\<=0\}\in\K[{r||c|c}{\IP^n&1&n{-}1\\ \IP^1&m&2}]$,
                the complementrix}
   \lput(54,26){\color{Rouge}$\Sing\XX{m} = C_m\<\cap S_m
                              \in \K[{r||c|cc}{\IP^n&1&1&n{-}1\\ \IP^1&m&-m&2}]$}
   \put(65,26){\TikZ{\path[use as bounding box](0,0);
                     \draw[red!90!black,thick,-stealth]
                          (-.6,-.1)to[out=-90,in=-30]++(-2.05,-.9);
               }}
   \lput(65,16){${\color[rgb]{.8,0,0}\underbrace{C_m\<\cap S_m}_{\copy8}}\subset
                  \underbrace{C_m\<\cup\,\C3{S_m}}_{\copy9}$}
  \end{picture}
 \end{center}
 \caption{The generic Calabi-Yau hypersurface $\XX{m}\subset\FF{m}$ for $m\<\geqslant3$
          and its codimension-1 singularity}
 \label{f:nXm}
\end{figure}

\paragraph{Singularity:}
Being {\em\/reducible\/} for $m\<\geqslant3$, the generic Calabi-Yau $(n{-}1)$-fold $\XX{m}\subset\FF{m}$ is singular precisely at the intersection of its components:
\begin{equation}
   \Sing\XX{m} = C_m\cap S_m \in \K[{r||c|cc}{\IP^n&1&\3-1&n{-}1\\ \IP^1&m&-m&2}].
 \label{e:c2Xm}
\end{equation}
The row-wise sum of degrees shows that $\Sing\XX{m}$ is itself a Calabi-Yau subspace, now of codimension-2 in $\FF{m}$ --- a Calabi-Yau {\em\/matryoshka}\ftn{We have recently learned that C.~Doran has been independently using the same term and metaphor for such iteratively nested chains of Calabi-Yau subspaces in presentations.}, a Calabi-Yau-within-Calabi-Yau.
 The reducible Calabi-Yau hypersurface~\eqref{e:nXm} and its codimension-1 singular set are sketched in Figure~\ref{f:nXm}; it fits within the framework of ``Constructive Calabi-Yau manifolds''~\cite{Lee:2006wf} and exhibits the so-called Tyurin degeneration~\cite{tyurin2003fano}.

Owing to the simple forms of $ p_0(x,y)$ and $\Fs(x,y)$ and the reparametrization~\eqref{e:bP=T}, it follows that
\begin{equation}
  (x_0\,y_0\!^m{+}x_1\,y_1\!^m)\<\eqco p_0(x,y) =0=
  \Fs(x,y)\<\coeq\Big(\frac{x_0}{y_1\!^m}{-}\frac{x_1}{y_0\!^m}\Big)
  \quad\Iff\quad  x_0\<=0\<=x_1,
 \label{e:x0x1}
\end{equation}
which is not surprising, given the constant-Jacobian reparametrization equivalence $( p_0,\Fs,\dots)\approx(x_0,x_1,\dots)$ found in~\eqref{e:0psxy}.
This in turn leaves
\begin{equation}
  {}^\sharp\!\XX[n-2]m=\Sing\XX{m}
  =\Big\{ \sum_{\ell=0}^2 c^{\sss(n-1)}_\ell(x_2,\cdots,x_n)(y_0\!^{2-\ell}y_1\!^\ell)=0 \Big\}
  \in\K[{r||c}{\IP^{n-2}&n{-}1\\ \IP^1&2}],
 \label{e:CpXred}
\end{equation}
which is nonsingular for generic choices of the coefficient functions $c^{\sss(n-1)}_\ell(x_2,\cdots,x_n)$ --- and is a regular anticanonical (Calabi-Yau) hypersurface.

\paragraph{Order Matters:}
The left-to-right ordering of the hypersurfaces in
 $\ssK[{r||c|cc}{\IP^n&1&\3-1&n{-}1\\ \IP^1&m&-m&2}]$~\eqref{e:c2Xm} {\em\/is relevant\/} within the framework of {\em\/generalized complete intersections\/}~\cite{rgCICY1,rBH-Fm,rGG-gCI} and the $|$-separation signifies this:
  The 2nd, degree-$\pM{\3-1\\-m}$ degree hypersurface $\Fs^{-1}(0)$ is well defined only within the 1st, degree-$\pM{1\\m}$ hypersurface $\FF{m}\<\coeq p_0^{-1}(0)$. Away from
 $p_0^{-1}(0)$, the {\em\/equivalence class\/}~\eqref{e:DrX} is holomorphic, but its zero-locus $\Fs^{-1}(0)$ is not well defined. In the toric rendition~\eqref{e:nFmQnu}, $p_0(x,y)\<=0$ and this ``tuning'' hold by the definition of $\FF{m}$; see~\eqref{e:bP>T}.
 In contrast, the 3rd, degree-$\pM{n{-}1\\2}$ defining polynomial~\eqref{e:CpX} of the complementrix is regular on all of $A\<=\IP^n{\times}\IP^1$, and the zero-locus $\Fc^{-1}(0)$ is well defined everywhere on $A$, including $p_0^{-1}(0)$ and $\Fs^{-1}(0)$.
 
\begin{figure}[htb]
 \begin{center}
  \begin{picture}(140,40)(2,0)
   \put(0,-2){\includegraphics[height=40mm]{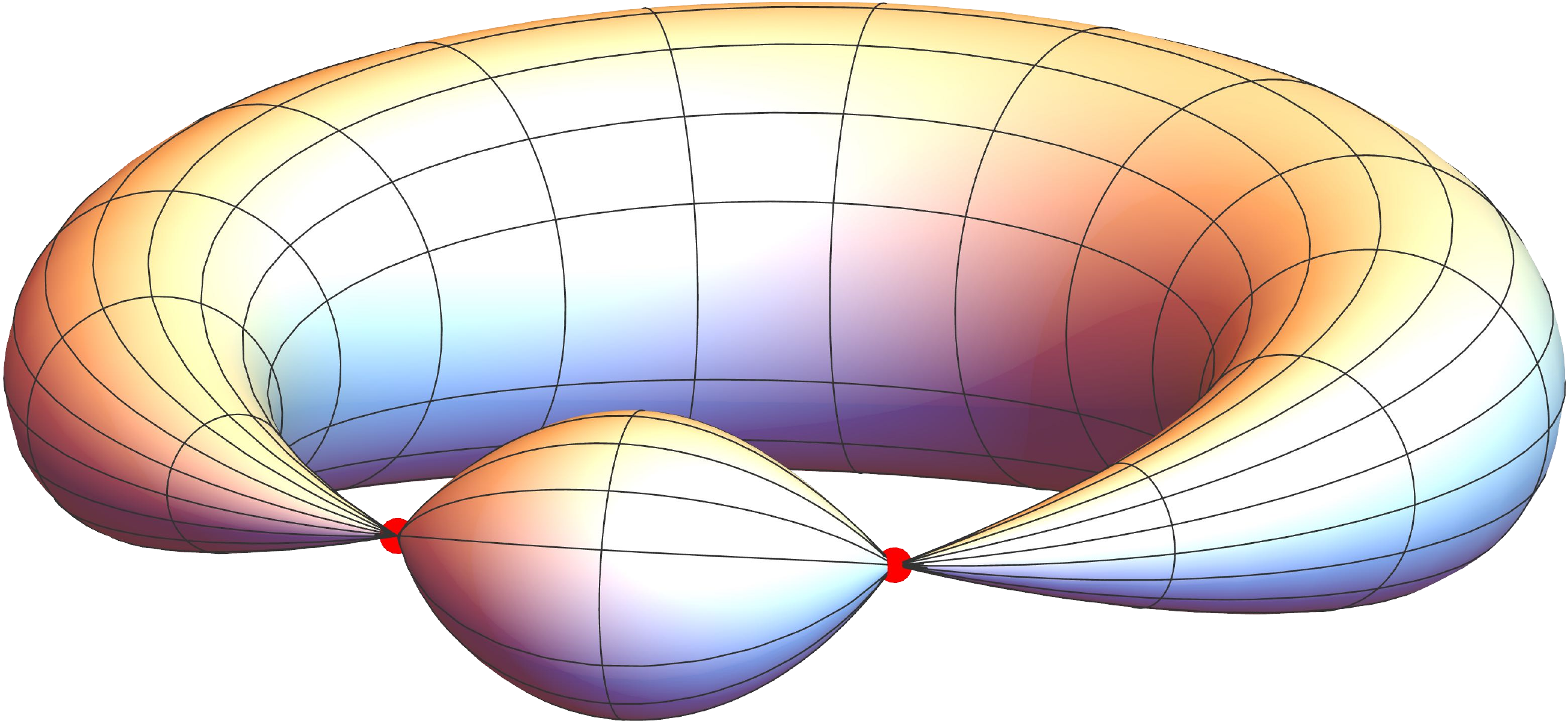}}
   \put(85,28){$C_m\<\in\ssK[{r||cc}{\IP^2&1&1\\ \IP^1&m&2}]$}
   \put(-6,0){\C3{$\ssK[{r||c|c}{\IP^2&1&~~1\\ \IP^1&m&-m}]\<\ni S_m$}}
   \put(65,0){\C1{$\Sing\XX[1]m\<\in\ssK[{r||c|cc}{\IP^2&1&~~1&1\\ \IP^1&m&-m&2}]$}}
   \put(110,14){$\left.\rule{0pt}{10ex}\right\}
                 \XX[1]m\<\in\ssK[{r||c|c}{\IP^2&1&2\\ \IP^1&m&2{-}m}]$}
   \put(117,8){$=C_m\cup S_m$}
   \put(50,20){\TikZ{\path[use as bounding box](0,0);
                   \draw[red,thick,->](1.4,-1.9)to[out=180,in=-35]++(-1.35,.45);
                   \draw[red,thick,->](1.4,-1.9)to[out=180,in=0]++(-2.9,-.45)
                                                to[out=180,in=-90]++(-1.3,1);
                   \draw[blue,thick,->](-3.1,-1.8)to[out=30,in=200]++(1,.5);
                   \draw[thick,->](3.4,.9)to[out=180,in=60]++(-.5,-.3);
                   }}
  \end{picture}
 \end{center}
 \caption{The general, $m\<\geqslant3$, case of the Calabi-Yau subspaces in a Hirzebruch surface, $\FF[2]m$}
 \label{f:p2Torus}
\end{figure}
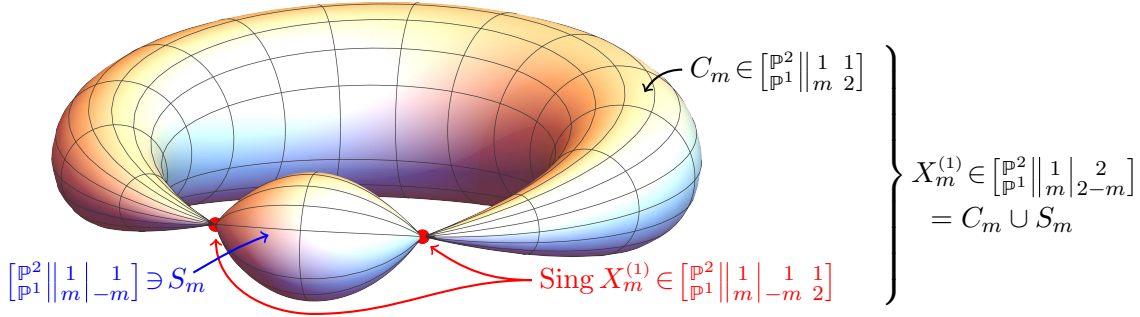

 Owing to~\eqref{e:x0x1} and the reduction~\eqref{e:c2Xm}$\leadsto$\eqref{e:CpXred}, the common zero-locus $p_0^{-1}(0)\<\cap\Fs^{-1}(0)$ is equivalent to the $x_0\<=0\<=x_1$
subspace, $\IP^{n-2}\<\times\IP^1\<\subset A$. Indeed, for the $n\<=2$ case, the original Hirzebruch {\em\/surface,} the directrix is equivalent to $\IP^0\<\times\IP^1$ --- the simple (algebraic) line of self-intersection $-m$ within $\FF[2]{m}$~\cite{rH-Fm,rGrHa}.
 Therein, the complementrix is the anticanonical hypersurface, which is indeed two points --- the Calabi-Yau 0-fold:
\begin{equation}
  \K[{r||c|cc}{\IP^2&1&\3-1&1\\ \IP^1&m&-m&2}] ~\6[1pt]{\sss\eqref{e:x0x1}}\approx~
  \K[{l||c}{\IP^0&1\\ \IP^1&2}]{:}\quad
  \Fc(x,y)\<=x_2\,y_0\,y_1\quad \text{where}~x_2\simeq\l x_2\<\neq0,
\end{equation}
which is the singularity, $\Sing\XX[1]{m}$. Thus,
 $\XX[1]{m}\in\ssK[{r||c|c}{\IP^2&1&2\\ \IP^1&m&2{-}m}]$
is a twice-pinched torus; see Figure~\ref{f:p2Torus}.

\paragraph{Being Exceptional:}
The foregoing constructions explicitly depend on the central choice~\eqref{e:bPnFm} within the deformation family $\ssK[{r||c}{\IP^n&1\\ \IP^1&m}]$,
and cannot be completed for the deformed choices~\eqref{e:bPnFmE}. In particular, the total degree-$\pM{\3-1\\-m}$ multiple of $p_{\vec\e\,}(x,y)$ is
\begin{equation}
  \frac{p_{\vec\e\,}(x,y)}{(y_0\,y_1)^m} = \Big(\frac{x_0}{y_1\!^m} + \frac{x_1}{y_0\!^m}\Big)
  +\sum_{a=0}^n\sum_{\ell=1}^{m-1} \e_{a\ell}\,x_a\,y_0\!^{-\ell}y_1\!^{\ell-m}.
\end{equation}
The $\e_{a\ell}$-expansion contains rational monomials with mixed denominators and a $y_0,y_1$-independent numerator when $\e_{a\ell}\<\neq0$, each of which has a pole in both coordinate charts in $\IP^1$ and so cannot be holomorphic in the manner of~\eqref{e:DrX}\ftn{With a total degree $\pM{\3-1\\-m}$ and poles of order $\ell$ at $y_0\<=0$ and $m{-}\ell$ at $y_1\<=0$ with $\ell\<\in[1,m{-}1]$, such monomials cannot be split into sums of two partial fractions (nor any change of variables), each with a single pole.}.

This proves the central hypersurface~\eqref{e:bPnFm} to be the unique one in the deformation family $\ssK[{r||c}{\IP^n&1\\ \IP^1&m}]$ that has an irreducible directrix with the maximally negative self-intersection, $[S_m]^n=-(n{-}1)m$. Consequently, the anticanonical hypersurfaces $\XX{m}\<\subset\FF{m}$ are necessarily singular only for this central case. The smooth Calabi-Yau 3-folds of the form~\eqref{e:bPnFmX} reported in~\cite{rgCICY1} refer\ftn{We thank James Gray for confirming this detail.} to non-central cases~\eqref{e:bPnFmE}, some of which were discussed in \SS\,\ref{s:DscDef}.

The above facts add to the connectivity among (generalized) complete intersection Calabi-Yau $n$-folds. Suffice it here to provide an example, deferring a more detailed analysis to a separate effort:
\begin{enumerate}[itemsep=-1pt, topsep=-1pt]

 \item The {\em\/generic\/} Calabi-Yau 3-folds in the deformation family $\ssK[{r||c|c}{\IP^4&1&~~4\\ \IP^1&5&-3}]$ are smooth and are diffeomorphic to generic members in $\ssK[{r||c|c}{\IP^4&1&4\\ \IP^1&1&1}]$ of {\em\/regular\/} complete intersections. This then connects at least some gCICYs into the ``web of Calabi-Yau 3-folds''~\cite{rGHC,Candelas:1989ug,Avram:1995pu,Avram:1997rs}.

 \item The {\em\/special\/} Calabi-Yau 3-folds in the deformation family $\ssK[{r||c|c}{\IP^4&1&~~4\\ \IP^1&5&-3}]$ using the {\em\/central,} Hirzebruch-like defining equation~\eqref{e:bPnFm} are all singular (Tyurin-degenerate, see below), but their singular set is itself a smooth Calabi-Yau (K3) 2-fold, thus connecting to the web of Calabi-Yau 2-folds.
\end{enumerate}

\paragraph{Tyurin Degeneration:}
By reducing for $m\<\geqslant3$ to a union $\XX{m}\<=(C_m\<\cup S_m)$~\eqref{e:nXm} where
 $\fX\<\coeq(C_m\<\cap S_m)$ is a codimension-2 Calabi-Yau space~\eqref{e:c2Xm}, each
 Calabi-Yau $\XX{m}$ hypersurface in the {\em\/central\/} Hirzebruch scroll~\eqref{e:bPnFm},
 $\big(\FF{m}\<\coeq p_0^{-1}(0)\big)$, is {\em\/Tyurin degenerate\/}~\cite{tyurin2003fano}.  In the bi-projective embedding and restricting to
 $\fX\<\coeq{}^\sharp\!\XX[n-2]m\in\ssK[{r||c|cc}{\IP^n&1&~~1&n{-}1\\ \IP^1&m&{-}m&2}]$, the adjunction relations
\begin{equation}
   T_\fX\into T_{C_m}\big|_\fX \onto \cO_A\pM{n-1\\2}\big|_\fX,
    \qquad\text{and}\qquad
   T_\fX\into T_{S_m}\big|_\fX \onto \cO_A\pM{1\\-m}\big|_\fX
\end{equation}
identify the two rightmost sheaves as the normal sheaves of $\fX\<\subset C_m$ and
 $\fX\<\subset S_m$, respectively. Then,
\begin{equation}
   \cO_A\pM{n-1\\2}\big|_\fX \otimes \cO_A\pM{1\\-m}\big|_\fX
   =\cO_A\pM{n\\2-m}\big|_\fX =\cK^*_{\smash{\FF{m}}}\big|_{\fX\subset\XX{m}\subset\FF{m}}
\end{equation}
is the restriction to $\fX\<\coeq{}^\sharp\!\XX[n-2]m\<\subset\XX{m}\<\subset\FF{m}$ of the anticanonical sheaf of $\FF{m}$, a section of which defines $\XX{m}\<\subset\FF{m}$, so $\cK^*_{\smash{\XX{m}}}\<=\cO_{\smash{\XX{m}}}$.
 In fact, both $S_m$ and $C_m$ are {\em\/quasi-Fano\/} by the $n$-dimensional generalization of definition~\cite[Def.\,2.2]{tyurin2003fano}: They both contain the smooth codimension-2
 $\,^\sharp\!\XX[n-2]m$, and their structure sheaf cohomology vanishes except $H^0\<\approx\IC$, so $h^q(C_m,\cO)\<=\d_{q,0}\<=h^q(S_m,\cO)$, reproducing the defining property, $\c(\cO_{C_m})\<=1\<=\c(\cO_{S_m})$; see Appendix~\ref{s:qFCS}.
 The mirror-pair constructions below, in \SS\,\ref{s:MTM}, should then provide a testing ground for the so-called {\em\/DHT~conjecture\/}~\cite{doran2016mirror,Kanazawa:2017wp,Doran:2020vo,Doran:2018um,Barrott:2021wx,doran2021mirror,Doran:2021ud}.

\subsection{Laurent Deformations and Intrinsic Limit}
\label{s:LdL'H}
Since the entire anticanonical system~\eqref{e:sK*}, i.e.,~\eqref{e:TsK*} factorizes for $m\<\geqslant3$, the necessarily Tyurin-degenerate Calabi-Yau hypersurfaces
 $\XX{m}\<\subset\FF{m}$ in the {\em\/central\/} Hirzebruch scroll~\eqref{e:bPnFm}, i.e.,~\eqref{e:nFmQnu} cannot be smoothed by regular deformations.
 However, the rational sections encoded by the {\em\/extended\/} Newton polytope~\cite{rBH-gB} make the anticanonical system of $\FF{m}$ transverse, and so afford a Laurent smoothing of $\XX{m}\<\subset\FF{m}$.
 The simple, $n\<=2$ case in Figure~\ref{f:p2Torus} certainly suggests that the singularity {\em\/should have\/} a {\em\/crepant\/} smoothing, i.e., without changing the (vanishing) canonical class.

\paragraph{Laurent Deformations:}
The Calabi-Yau models in~\cite{rBH-gB} deform the reducible hypersurface~\eqref{e:nXm} by including certain very specific rational monomials in the defining section~\eqref{e:sK*}, i.e.,~\eqref{e:TsK*}. Suffice it here to showcase the $\XX[1]3\<\subset\FF[2]3$ example in its toric rendition, and focus on the cornerstone (extremal) polynomial
\begin{equation}
   f(x;a)= a_1 x_1\!^2x_3\!^5 +a_2x_1\!^2x_4\!^5
        +a_3\frac{x_2\!^2}{x_4} +a_4\frac{x_2\!^2}{x_3}\quad
   \in~\G(\cK^*_{\smash{\FF[2]3}}).
 \label{e:csp2F3}
\end{equation}
This particular choice of rational monomials will be explained below; see~\eqref{e:2F3SN}.
 Identifying $\FF[2]3$ with the MPCP-desingularization of $\IP^2_{(3{:}1{:}1)}$ prepends $x_1\<\simeq\l^0x_1$ to $(x_2,x_3,x_4)\<\simeq(\l^3x_2,\l x_3,\l x_4)$. Thus, $(x_1,x_2)$ are the homogeneous coordinates of the exceptional $\IP^1$, so identified by the
 $(x_1,x_2,x_3,x_4)\<\simeq(\tw\l^1x_1,\tw\l^1x_2,x_3,x_4)$ symmetry.
 These two $\IC^*$-rescalings are equivalent to the $n\<=2$, $m\<=3$ case of~\eqref{e:nFmQnu}.

\begin{figure}[htb]
 \begin{center}
  \begin{picture}(172,32)(0,2)
   \put(0,0){\includegraphics[width=56mm]{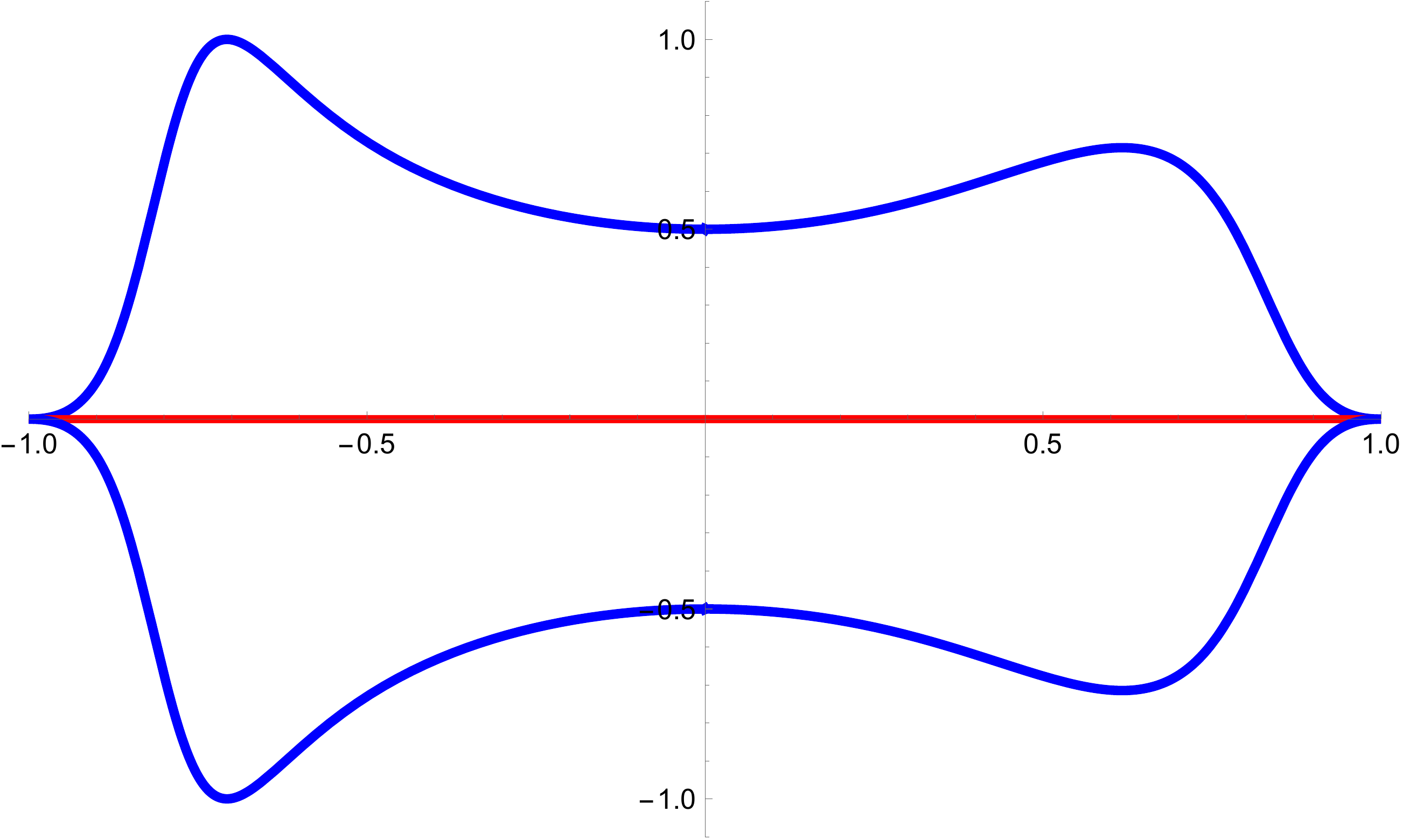}}
    \put(0,0){(a)}
    \put(13,0){$\e\<=0$}
    \cput(28.15,17.5){\C1{\small the directrix, $S_3$}}
    \cput(28.15,10.2){\C3{\small the complementrix, $C_3$}}
   \put(58,0){\includegraphics[width=56mm]{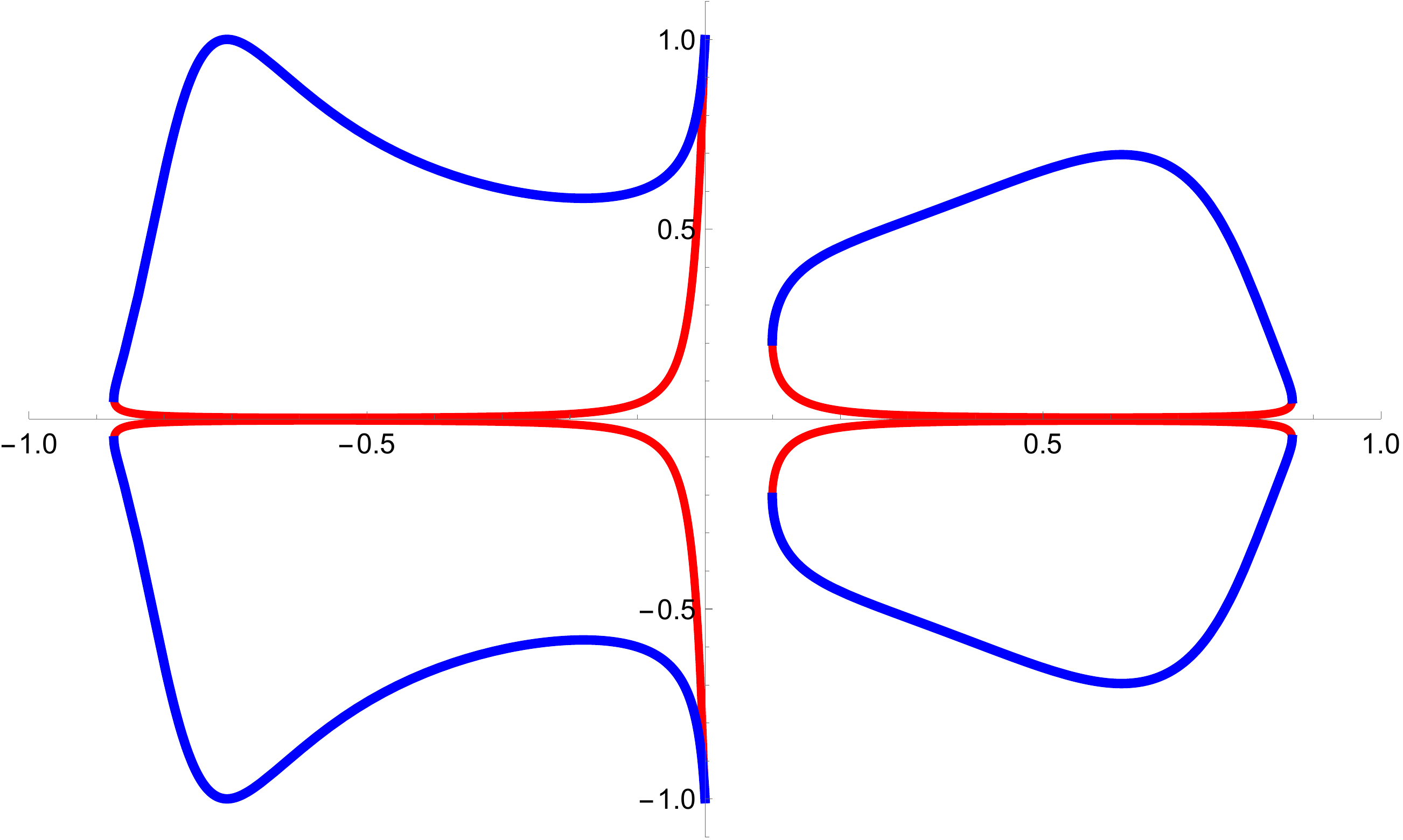}}
    \put(58,0){(b)}
    \put(71,0){$\e\<=\fRc1{40}$}
    \put(84.77,31.15){$\ostar$}
    \put(84.77,0.5){$\ostar$}
    \cput(85.5,16){\TikZ{\path[use as bounding box](0,0);
                   \draw[thick,densely dashed,<->](.15,1.45)to[out=-65,in=65]++(0,-2.75);
                   }}
   \put(116,0){\includegraphics[width=56mm]{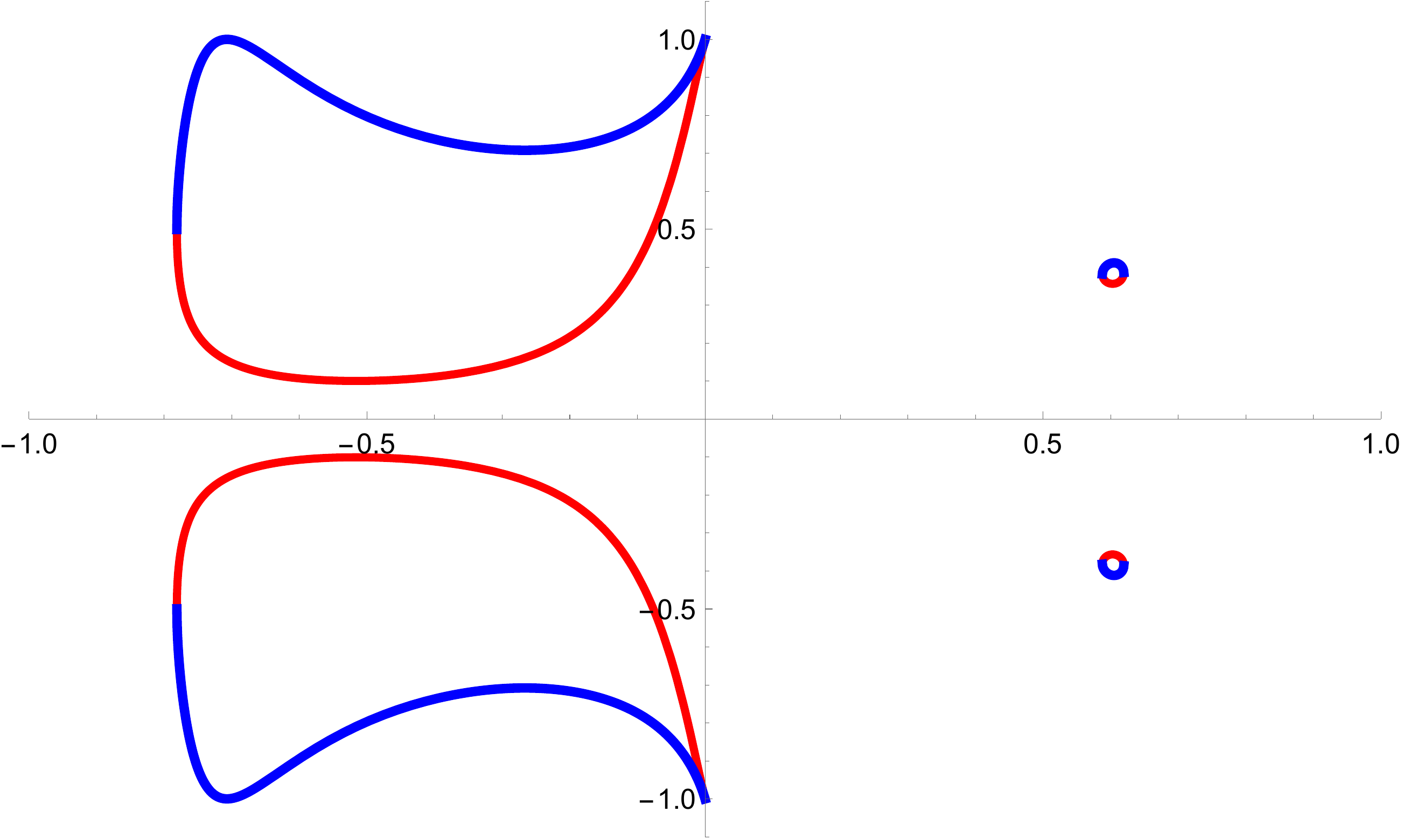}}
    \put(116,0){(c)}
    \put(129,0){$\e\<=\fRc1{6}$}
    \put(142.8,31){$\ostar$}
    \put(142.8,0.5){$\ostar$}
    \cput(143.5,16){\TikZ{\path[use as bounding box](0,0);
                   \draw[thick,densely dashed,<->](.15,1.45)to[out=-65,in=65]++(0,-2.75);
                   }}
  \end{picture}
 \end{center}
 \caption{Several plots of $x_1\!^2x_3\!^5\<+x_1\!^2x_4\!^5\<-x_1x_2x_3\!^2\<+\e\frac{x_2^2}{x_4}\<=0$, restricted to the real circles $x_1\<\to y\<\in[-1,1{\simeq}{-}1]$, $x_2\<\to\sqrt{1{-}y^2}$, $x_3\<\to\sqrt{1{-}x^2}$ and $x_4\<\to x\<\in[-1,1{\simeq}{-}1]$; $\{y\<=0\}$ is the directrix, $\{x\<=0\}$ is the pole-locus; the putative pole-in-zero-locus at $(x,y)=(0,{\pm}1)$, filled by the intrinsic limit (below), is marked by the $\ostar$ symbol}
 \label{f:smDrX}
\end{figure}
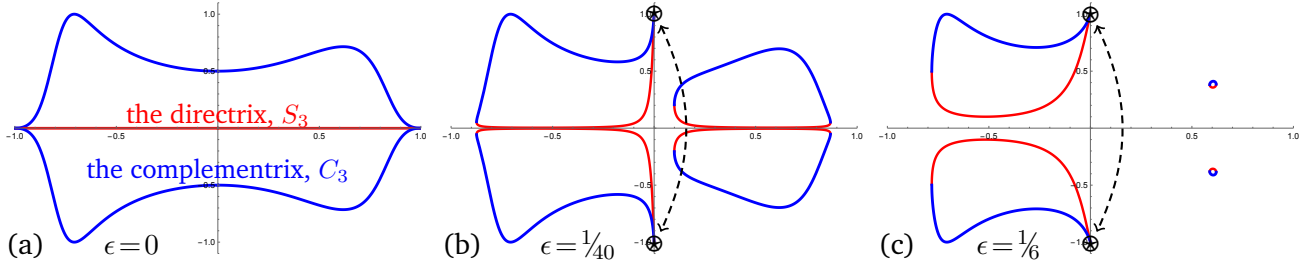

While $a_1a_4\!^5\<\neq a_2a_3\!^5$, the polynomial~\eqref{e:csp2F3} is transverse:
 The gradient $\vd_i f(x;a)$ cannot vanish without setting $x_1\<=0\<=x_2$ --- which cannot happen in the exceptional $\IP^1$ in $\FF[2]3$. In the toric specification~\eqref{e:nFmQnu}, the 1-cones $\n_1,\n_2$ do not form a 2-cone in $\fan{\FF[2]3}$, $\vev{x_1x_2}$ is in the Stanley-Reisner ideal, and the exceptional set
\begin{equation}
  Z(\fan{\FF[2]3})
  = \{x_1\<=0\<=x_2\}\<\times\IC^2_{x_3,x_4} \cup
    \IC^2_{x_1,x_2}\<\times\{x_3\<=0\<=x_4\}
\end{equation}
is excised from $\FF[2]3=\big(\IC^4\ssm Z(\fan{\FF[2]3})\big)/(\IC^*)^2$~\cite{rCLS-TV}. (Parts of this base locus {\em\/are included,} appropriately and self-consistently, in the Landau-Ginzburg and the so-called hybrid phases of the GLSM model~\cite{rBH-gB}.)
 For this same reason, $x_2\<\neq0$ and so $f(x;a)\<\neq0$ at $\{x_1\<=0\}$, so that
\begin{equation}
   \big(S_m\<=\{x_1\<=0\}\big) \cap \{ f(x;a)\<=0 \} = \varnothing.
\end{equation}
For any $a_3,a_4\<\neq0$, the zero-locus $\{f(x;a)\<=0\}$ is moved away from the directrix, and thus also from the singular set~\eqref{e:CpXred}.
 This effectively smoothes the Tyurin-degenerate model, as illustrated in the series of plots in Figure~\ref{f:smDrX}, where $a_3\<\to\e$ and $a_4\<\to0$, and an additional regular monomial ($x_1x_2x_3\!^2$) is added,
\begin{equation}
  f(x;a) \leadsto
  F(x;\e)=x_1\!^2 x_3\!^5 +x_1\!^2 x_4\!^5 -x_1 x_2 x_3\!^2 +\e\frac{x_2\!^2}{x_4},
 \label{e:F(x)}
\end{equation}
to allow for real solutions in the real ($S^1{\times}S^1$) ``slice'' within
 $\IP^1_{\sss\text{fibre}}\<\into\FF[2]3\<\onto\IP^1_{\sss\text{base}}$.
Already for $\e=\fRc1{40}$, the red-plotted slice segment is visibly deformed away from the directrix, $S_3\<=\{x_1\<=0\}$ (the horizontal mid-line in Figure~\ref{f:smDrX}), and this separation only increases as $\e$ does.
 In turn, the presence of the rational deformations~\eqref{e:csp2F3} includes a $\frac00$-like putative pole in the defining section, which requires special attention.

\paragraph{Intrinsic Limit:}
The main concern with a Laurent defining polynomial such as~\eqref{e:F(x)} is that the {\em\/unqualified\/} limits of the rational terms, $\lim_{x_3\to0}\frac{x_2\!^2}{x_3}$ and  $\lim_{x_4\to0}\frac{x_2\!^2}{x_4}$, are not well defined.
 The zero-locus of the defining equation at hand is of course well defined away from the putative pole-location, $\{x_4\<=0\}\<\cup\{x_3\<=0\}$, which then defines the required qualification: approach the putative pole-locations from {\em\/within\/} the desired zero-locus, thus defining the {\em\/intrinsic limit.} For the case at hand, $\{F(x;\e)\<=0\}$ with~\eqref{e:F(x)}, we solve:
\begin{equation}
  F(x;\e)\<=0,\quad x_4\<\neq0~\To\quad
  x_2 =
  x_1\frac{x_4 x_3^2 \pm \sqrt{x_4} \sqrt{x_4 x_3^4 -4\e(x_3^5+x_4^5)}}{2\e}.
\end{equation}
Substituting this in $F(x;\e)$ but keeping the summands separately produces
\begin{alignat}9
 0\<=F(x;\e)
 &=x_1\!^2 x_3\!^5 {+}x_1\!^2 x_4\!^5
 {-}x_1\!^2x_3\!^2\frac{x_4 x_3^2{\pm}
                        \sqrt{x_4}\sqrt{x_4 x_3^4{-}4\e(x_3^5{+}x_4^5)}}{2\e}
 {+}x_1\!^2\frac{\big(x_3^2\sqrt{x_4}
                 {\pm}\sqrt{x_4 x_3^4{-}4\e(x_3^5{+}x_4^5)}\big)^2}{4\e},
 \nn\\
 &\too{x_4\to0}\, x_1^2 x_3^5 +0 -0 +({-}x_1^2 x_3^5),
\end{alignat}
making it clear that each monomial is seprately well defined in the so qualified $x_4\<\to0$ limit.
 Effectively, the a priori independent (Cox) variable $x_2$ is replaced with the function $x_2\<=x_2(x_1,x_3,x_4)$ that guarantees the vanishing of $F(x;\e)$ everywhere, including the intersection of this subspace with the pole-locus of concern, $\{x_4\<=0\}$. In this sense, the evaluation of the limit of the rational summand
\begin{equation}
  \lim_{x_4\to0}
  \left[\e\,\frac{\big(x_2\<=x_2(x_1,x_3,x_4)\big)^2}{x_4}\right]
  ~=-x_1\!^2 x_3\!^5
\end{equation}
is an application of L'Hopital's rule, and extends straightforwardly to poles of higher order.

 With the so-resolved putative pole-locus and at least as real manifolds, we expect the transversal Laurent deformation~\eqref{e:csp2F3} of the Tyurin degeneration~\eqref{e:nXm} to be no different than other nonsingular models in the deformation family $\ssK[{r||c|c}{\IP^n&1&n\\ \IP^1&m&2{-}m}]$, built over the effectively discrete family $\ssK[{r||c}{\IP^n&1\\ \IP^1&m}]$ in Figure~\ref{f:disDef}. These Laurent-deformed Calabi-Yau models are then expected to have the same Betti and Euler numbers, $b_2\<=2$, $b_3\<=174$ and $\c\<={-}168$, and then also to admit a Hodge decomposition with $h^{11}\<=2$ and $h^{21}\<=86$. However, we are not aware of a rigorous proof --- or correction.

\paragraph{Alternative:}
The additional care required to specify the nature of the limit-points such as $x_4\<\to0$ in the zero-locus of the Laurent defining function~\eqref{e:F(x)} stems from the fact that the pole-locus of concern, $\{x_4\<=0\}$, intersects the zero-locus of interest,
 $\{F(x)=0\}$. This situation is amenable to the following sequence of standard algebro-geometric operations:
\begin{proc}\label{P:BuLBd}
For a Laurent polynomial $F(x)$ such as~\eqref{e:F(x)} over an ambient space $A$, let
 $P\<\subset A$ denote the pole-locus of $F(x)$,
 $Z\<\subset A$ the well-defined open (non-compact) zero-locus of $F(x)$, and let 
 $x_*$ be a common point of (limiting sequences in) $P$ and $Z$. Then:
\begin{enumerate}[itemsep=-1pt, topsep=0pt]
 \item Let $\Ht{A}$ be a blowup of $A$ at $x_*$, possibly iterated so the closure of
 $\Ht{Z}$, identified as the zero-locus $\{\Ht{F}(x)=0\}\subset\Ht{A}$, is well defined and separate from the proper transform of the pole-locus, $\Ht{P}$.

 \item The blowdown (along $\hat{x}_*$) of the zero-locus
   $\{\Ht{F}(x)=0\}\subset\Ht A$ is then a well-defined subspace of $A$.\qedhere
\end{enumerate}
\begin{equation}
  \vC{\TikZ{[scale=.9]\path[use as bounding box](-.3,-.3)--(16.3,4.3);
            \draw[ultra thick,](0,0)to[out=60,in=120]++(3,3)to[out=-60,in=210]++(1,1);
             \path(3.15,3.4)node{\large$Z$};
            \draw[very thick,red](0,3)to[out=10,in=160]++(4,-2.5);
             \path[red](3.5,1.1)node{\large$P$};
            \filldraw[fill=pink,very thick](1.51,2.6)circle(.9mm);
             \path(1.51,2.6)node[right]{\large$x_*$};
            \draw[blue,very thick, dashed,->]
                 (2.3,2.5)to[out=-15,in=180]++(4.3,-.5);
             \path[blue](4.45,2.3)node[rotate=-6]{\footnotesize blowup};
            \draw[ultra thick,](6,3)to[out=-20,in=120]++(3,0)to[out=-60,in=210]++(1,1);
             \path(9.15,3.5)node{\large$\Ht{Z}$};
            \draw[very thick,red](6,1)to[out=10,in=160]++(4,-.5);
             \path[red](9.5,1.1)node{\large$\Ht{P}$};
            \draw[blue,very thick](7,.5)to[out=100,in=255]++(0,3);
             \path[blue](6.8,2)node[right]{\large$\hat{x}_*$};
            \filldraw[fill=black,draw=blue,very thick](6.885,2.98)circle(.8mm);
             \draw[stealth-,thick](6.98,3.1)--++(.7,.23);
             \draw[stealth-,thick](6.75,3.05)to[out=190,in=-15]++(-.7,.03);
             \path(6.8,2.8)node[right]{$\hat{x}_*\<\cap\,\7{\!\Ht{Z}}$};
            \filldraw[fill=pink,draw=blue,very thick](6.9,1.115)circle(.8mm);
             \path(6.8,1.4)node[right]{$\hat{x}_*\<\cap\Ht{P}$};
             \path(6,4)node[right]{include $\hat{x}_*\<\cap\,\7{\!\Ht{Z}}$};
            \draw[blue,very thick,->>]
                 (7.5,2)--++(5.8,.6);
             \path[blue](10.45,2.5)node[rotate=6]{\footnotesize blowdown};
            \draw[ultra thick,](12,0)to[out=60,in=120]++(3,3)to[out=-60,in=210]++(1,1);
             \path(15.15,3.4)node{\large$\,\7{\!Z}$};
            \draw[very thick,red!33](12,3)to[out=10,in=160]++(4,-2.5);
             \path[red!33](15.5,1.1)node{\large$P$};
            \fill(13.51,2.6)circle(.9mm);
             \path(13.51,2.6)node[right]{\large$x_*$};
            }}
\end{equation}
\end{proc}
\noindent
This separates the limiting sequences within the zero-locus from those within the pole-locus, and so conceptually corroborates the above-defined {\em\/intrinsic limit.}
 It also seems to suggest a reformulation wherein coincident points are separated based on limiting sequences that lead to them, perhaps not too dissimilar from the framework of Ref.~\cite{rGG-gCI}.

\paragraph{An Overview:}
In the footsteps of \SS\,\ref{s:DscDef}, consider the triply deformed 4-fold
\begin{equation}
 \K[{r||c}{\IP^4&1\\ \IP^1&5}]\ni~~
 x_0\,y_0\!^5 \<+x_1\,y_1\!^5
  + x_2\,y_0\!^4\,y_1 \<+x_3\,y_0\!^3\,y_1\!^2 \<+x_4\,y_0\!^2\,y_1\!^3 =0,
\end{equation}
which admits a collection of one degree-$\pM{~~1\\-2}$ and {\em\/three\/} algebraically independent degree-$\pM{~~1\\-1}$ directrices. Via the analogous constant-Jacobian change of variables, this leads to the toric
 $\FF[4]{(2,1,1,1)}\<{\approx_{\sss\IR}}\FF[4]{(1,0,0,0)}$.
For each $n\<\geqslant2$, $\FF{(1,0,\cdots)}$ is Fano: both its spanning and its Newton polytope is convex and reflexive.

So,~\eqref{e:bPnFmE} is an explicitly constructed deformation family that includes both Fano and non-Fano Hirzebruch scrolls, all of which (for any given $n,m$) are diffeomorphic to each other. This then induces a deformation {\em\/connection\/} between the ({\em\/secondary\/} deformation families of) respective Calabi-Yau hypersurfaces, such as:
\begin{equation}
   \underbrace{\overbrace{\FF[4]{\sss(1,0,0,0)}[c_1]}^{\text{smooth}}
                \lhook\mkern-4mu\too{~q=0~}
                 \FF[4]{\sss(1,0,0,0)}}_{\text{generic}}
    ~\vC{\TikZ{\path[use as bounding box](0,0);
                \draw[blue,->](-2.5,.25)to[out=30,in=180]++(1.5,.3)--++(3.3,0)
                                  to[out=0,in=150]++(1.6,-.4);
            }}%
     \;\C3{\too{~\e\to0~}}\;~
   \underbrace{\FF[4]{\sss(5,0,0,0)} \fro{~q=0~}\joinrel\rhook
                \overbrace{\FF[4]{\sss(5,0,0,0)}[c_1]}^{\text{Tyurin-degenerate}}\!\!}_
              {\text{central}}
  \supset \overbrace{\underbrace{\Sing\!\big(\XX[3]{\sss(5,0,0,0)}\big)}_
                                {{}^\sharp\!\XX[2]{(5,0,0,0)}\,=\,\text{K3}}}^{\text{matryoshka}}
 \label{e:OV}
\end{equation}
The (irreducible) degree-$\pM{~~1\\-5}$ directrix in the central Hirzebruch scroll
 $\FF{5}$ thus serves as an {\em\/obstruction\/} to regular smoothing of the Tyurin-degenerate Calabi-Yau hypersurface, which disappears away from the central scroll. That is, we have the same real 8-dimensional manifold on the two sides of the $\e\<\to0$ arrow, equipped however with discretely different complex structures:
\begin{enumerate}[itemsep=-1pt, topsep=-1pt]
 \item The anticanonical sections that are holomorphic with respect to a generic choice of the complex structure are transverse and can define smooth Calabi-Yau hypersurfaces.
 \item The anticanonical sections that are holomorphic with respect to the central choice of the complex structure factorize and can define only Tyurin-degenerate Calabi-Yau hypersurfaces.
\end{enumerate}
That is, there always exist smooth defining equations of the correct degree to define a smooth and Ricci-flat zero locus, they are just not holomorphic with respect to the choice of the complex structure in which the smooth directrix is also holomorphic. It is then tempting to conclude:
\begin{conj}
The Laurent deformations of the Calabi-Yau hypersurface in the central Hirzebruch scroll are $\e\<\to0$ limit-images of the regular smoothing deformations in the Calabi-Yau hypersurface within the generic Hirzebruch scrolls.
\end{conj}

\subsection{Mirror Pairs}
\label{s:MTM}
We now turn to our primary motivation, the Laurent generalization of the transposition mirror model construction~\cite{rBH,rBH-LGO+EG,rMK-diss} and Batyrev's toric construction~\cite{rBaty01}; see also~\cite{rF+K-BHK} and references therein.

\begin{figure}[htb]
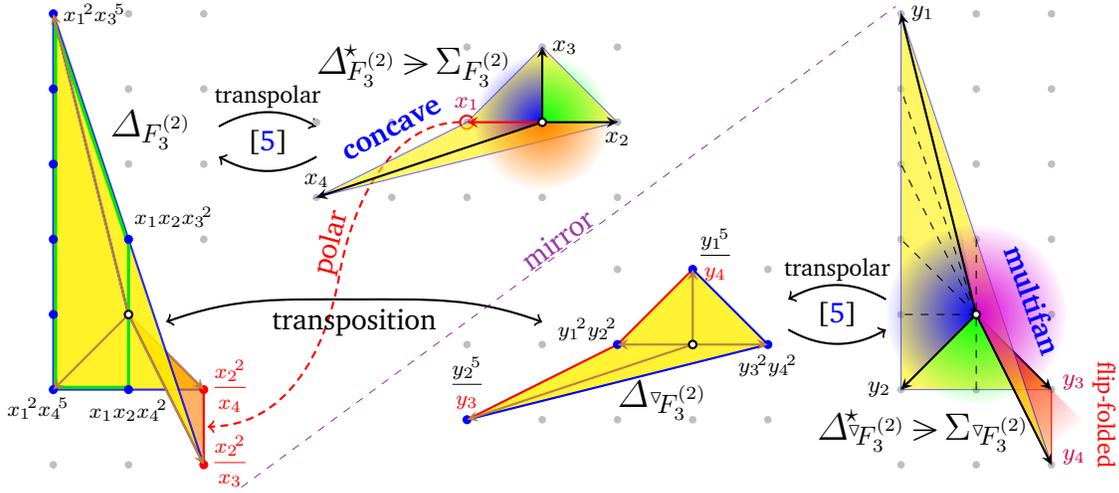

$$
 \vC{\TikZ{\path[use as bounding box](-1,-2)--(2.5,4);
            \foreach\y in{1,...,4} \fill[gray!50](1,\y)circle(.5mm);
            \foreach\y in{2,...,4} \fill[gray!50](0,\y)circle(.5mm);
            \foreach\x in{-1,...,0} \fill[gray!50](\x,-2)circle(.5mm);
           \fill[yellow,opacity=.85,draw=blue,thick,line join=round]
                (0,0)--(1,-1)--(-1,-1)--(-1,4)--(0,0);
           \fill[orange,opacity=.6](0,0)--(1,-1)--(1,-2)--cycle;
            \fill[red](1,-1)circle(.6mm); \fill[red](1,-2)circle(.6mm);
           \draw[thick,-stealth,brown](0,0)--(1,-1);
            \path[red](1.35,-.95)node{\scriptsize$\ddd\frac{x_2\!^2}{x_4}$};
           \fill[yellow,opacity=.85,draw=blue,thick]
                (0,0)--(-1,4)--(1,-2)--(0,0);
            \draw[thick,red](1,-2)--(1,-1);
           \draw[green!90!black,very thick]
                 (-.97,3.82)--(-.97,-.97)--(0,-.97)--(0,.92)--cycle;
            \path(0.6,1.3)node{\scriptsize$x_1x_2x_3\!^2$};
            \path(0,-1)node[below]{\scriptsize$x_1x_2x_4\!^2$};
           \foreach\y in{-1,...,4} \fill[blue](-1,\y)circle(.6mm);
            \fill[blue](0,-1)circle(.6mm); \fill[blue](0,1)circle(.6mm);
           \draw[thick,-stealth,brown](0,0)--(-1,-1);
            \path(-1.2,-1.25)node{\scriptsize$x_1\!^2x_4\!^5$};
           \draw[thick,-stealth,brown](0,0)--(-1,4);
            \path(-1,4)node[right]{\scriptsize$x_1\!^2x_3\!^5$};
           \draw[thick,-stealth,brown](0,0)--(1,-2);
            \path[red](1.35,-1.95)node{\scriptsize$\ddd\frac{x_2\!^2}{x_3}$};
           \filldraw[thick,fill=white](0,0)circle(.5mm);
           \draw[thick,->](1.2,2.5)to[out=30,in=150]++(1.3,0);
            \path(1.8,2.3)node{\cite{rBH-gB}};
            \path(1.85,2.9)node{\footnotesize transpolar};
           \draw[thick,<-](1.2,2.1)to[out=-30,in=210]++(1.3,0);
           \path(.3,2.5)node{\Large$\pDN{\FF[2]3}$};
            }}
 \MM{\TikZ{\path[use as bounding box](-3,-1)--(3,1);
           \foreach\x in{-1,...,1} \fill[gray!50](\x,1)circle(.5mm);
           \foreach\x in{-3,...,1}
            \foreach\y in{-1,...,0} \fill[gray!50](\x,\y)circle(.5mm);
           \draw[thick,red,densely dashed,o->](-.9,0)to[out=180,in=80]++(-1.8,-2.05)
                to[out=-100,in=0]++(-1.75,-2);
            \path(-2.8,-1.6)node[red,rotate=75]{polar};
           \filldraw[yellow,opacity=.6,draw=blue](1,0)--(0,1)--(-1,0)--(-3,-1)--cycle;
            \corner{(0,0)}{0}{90}{.9}{green};
            \corner{(0,0)}{90}{180}{.9}{blue};
            \corner{(0,0)}{180}{198}{.9}{violet};
            \corner{(0,0)}{198}{360}{.9}{orange};
           \draw[thick,-stealth](0,0)--(1,0);
            \path(1,0)node[below]{\footnotesize$x_2$};
           \draw[thick,-stealth](0,0)--(0,1);
            \path(0,1)node[right]{\footnotesize$x_3$};
           \draw[red,thick,-stealth](0,0)--(-1,0);
            \path(-1,.25)node[Rouge]{\footnotesize$x_1$};
           \draw[thick,-stealth](0,0)--(-3,-1);
            \path(-3,-1)node[above]{\footnotesize$x_4$};
           \filldraw[thick,fill=white](0,0)circle(.5mm);
           \path(-1.7,.8)node{\Large$\pDs{\FF[2]3}\<\lat\fan{\FF[2]3}$};
           \path(-2,-.1)node[blue, rotate=28]{\bf concave};
            }\\[8mm]
     \TikZ{\path[use as bounding box](-5,-1)--(1,1);
           \draw[Purple,dashed](-6,-1.9)--(2.7,4.5);
            \path(-1.8,1.4)node[Purple,rotate=36]{mirror};
           \draw[thick,<->](-7,.3)to[out=20,in=180]++(2.5,.3)to[out=0,in=160](-2,.3);
            \path(-4.5,.3)node{transposition};
           \foreach\x in{-1,...,1} \fill[gray!50](\x,1)circle(.5mm);
           \foreach\x in{-3,...,1}
            \foreach\y in{-1,...,0} \fill[gray!50](\x,\y)circle(.5mm);
           \fill[yellow,opacity=.8](1,0)--(0,1)--(-1,0)--(-3,-1)--cycle;
            \draw[blue,thick,line join=round](0,1)--(1,0)--(-3,-1);
            \draw[red,thick,line join=round](0,1)--(-1,0)--(-3,-1);
           \fill[blue](-1,0)circle(.6mm);
           \fill[blue](1,0)circle(.6mm);
           \fill[blue](0,1)circle(.6mm);
           \fill[blue](-3,-1)circle(.6mm);
           \draw[thick,-stealth,brown](0,0)--(1,0);
            \path(1,0)node[below]{\scriptsize$y_3\!^2y_4\!^2$};
           \draw[thick,-stealth,brown](0,0)--(0,1);
            \path(.3,1.2)node{\scriptsize$\ddd\frac{y_1\!^5}{\C1{y_4}}$};
           \draw[thick,-stealth,brown](0,0)--(-1,0);
            \path(-1.4,.2)node{\scriptsize$y_1\!^2y_2\!^2$};
           \draw[thick,-stealth,brown](0,0)--(-3,-1);
            \path(-3,-1)node[above]{\scriptsize$\ddd\frac{y_2\!^5}{\C1{y_3}}$};
           \filldraw[thick,fill=white](0,0)circle(.5mm);
           \path(-.4,-.7)node{\Large$\pDN{\MF[2]3}$};
            }}
\qquad
 \vC{\TikZ{\path[use as bounding box](-2,-2)--(2,4);
           \foreach\x in{-1,...,1}
            \foreach\y in{-2,...,4} \fill[gray!50](\x,\y)circle(.5mm);
           \fill[yellow,opacity=.6,draw=blue]
                (0,0)--(1,-1)--(-1,-1)--(-1,4)--(0,0);
            \corner{(0,0)}{-45}{-135}{1.2}{green};
            \corner{(0,0)}{225}{100}{1.2}{blue};
           \fill[orange,opacity=.4](0,0)--(1,-1)--(1,-2)--cycle;
            \corner{(0,0)}{-63}{-45}{2}{red};
           \draw[thick,-stealth](0,0)--(1,-1);
            \path[Rouge](1.3,-.9)node{\footnotesize$y_3$};
           \fill[yellow,opacity=.6,draw=blue]
                (0,0)--(-1,4)--(1,-2)--(0,0);
            \corner{(0,0)}{104}{-63}{1.2}{Magenta};
            \draw[red](1,-2)--(1,-1);
           \draw[thick,-stealth](0,0)--(-1,-1);
            \path(-1,-1)node[left]{\footnotesize$y_2$};
           \draw[thick,-stealth](0,0)--(-1,4);
            \path(-1,4)node[right]{\footnotesize$y_1$};
           \draw[thick,-stealth](0,0)--(1,-2);
            \path[Rouge](1.3,-1.9)node{\footnotesize$y_4$};
           \foreach\y in{-1,...,4} \draw[dashed](0,0)--(-1,\y);
            \draw[dashed](0,-1)--(0,1);
           \filldraw[thick,fill=white](0,0)circle(.5mm);
           \draw[thick,->](-1.2,.2)to[out=150,in=30]++(-1.3,0);
            \path(-1.9,0)node{\cite{rBH-gB}};
            \path(-1.85,.6)node{\footnotesize transpolar};
           \draw[thick,<-](-1.2,-.2)to[out=210,in=-30]++(-1.3,0);
           \path(-.7,-1.5)node{\Large$\pDs{\MF[2]3}\<\lat\fan{\MF[2]3}$};
           \path(.75,.2)node[blue, rotate=-70]{\bf multifan};
           \path(1.7,-1.33)node[red, rotate=-90]{\footnotesize flip-folded};
            }}
$$
 \caption{The transpolar pair of VEX polygons used in transposition-mirror fashion: one to define the Cox variables, the other to define anticanonical monomials --- and then the other way around}
 \label{f:NS2F3NS}
\end{figure}

\paragraph{Transpose Mirror:}
The standard anticanonical {\em\/cornerstone\/} polynomial of $\FF[2]{3}$ (green outline in Figure~\ref {f:NS2F3NS}, below),
\begin{equation}
   x_1\!^2x_3\!^5\<+x_1\!^2x_4\!^5\<+x_1x_2x_3\!^2\<+x_1x_2x_4\!^2
   =x_1\big( x_1x_3\!^5{+}x_1x_4\!^5{+}x_2x_3\!^2{+}x_2x_4\!^2 \big),
 \label{e:stdCS}
\end{equation}
is not transverse, but its Laurent analogue~\cite{rBH-gB},
\begin{equation}
   f(x)= a_1 x_1\!^2x_3\!^5 +a_2x_1\!^2x_4\!^5
        +a_3\frac{x_2\!^2}{x_4} +a_4\frac{x_2\!^2}{x_3}, \qquad
   a_1\,a_4\!^5\<\neq a_2\,a_3\!^5,
 \tag{\ref{e:csp2F3}$'$}
\end{equation}
is transverse away from the indicated discriminant locus. The matrix of exponents of~\eqref{e:csp2F3} is
\begin{equation}
  \IE[f(x)]=
  \AR{@{\;}c@{~~}c@{~~}c@{~~}c@{\;}}
  {2&0&~~5&~~0\\[-2pt] 2&0&~~0&~~5\\[-2pt] 0&2&~~0&-1\\[-2pt] 0&2&-1&~~0},
  \qquad
  \det\IE[f(x)]=0,~~\rank\IE[f(x)]=3,
 \label{e:EE}
\end{equation}
so that~\eqref{e:csp2F3} is {\em\/not invertible\/} in the sense defined in~\cite{rMK-diss} (see also~\cite{Kreuzer:1992bi,rKreSka95}), which would seem to prevent constructing the mirror model.
 Nevertheless, the transpose~\cite{rBH,rBH-gB} of the defining equation~\eqref{e:csp2F3} is straightforward:
\begin{equation}
  f(x)^\sfT=
  g(y)=b_1y_1\!^2y_2\!^2 +b_2y_3\!^2y_4\!^2
      +b_3\frac{y_1\!^5}{y_4} +b_4\frac{y_2\!^5}{y_3}, \qquad
   b_1\!^5\<\neq b_2\,b_3\!^2\,b_4\!^2,
 \label{e:csp*2F3}
\end{equation}
and is homogeneous for continuously\ftn{This continuousness of scaling symmetry choices correlates with the reduced rank of the matrix of exponents~\eqref{e:EE}.} many choices of $y_i$-degrees:
\begin{equation}
   \deg[g(y)]=1~\To~~
   q(y_1)\<=\frc15{+}\frc15q(y_4),~~
   q(y_2)\<=\frc3{10}{-}\frc15q(y_4),~~
   q(y_3)\<=\frc12{-}q(y_4),
 \label{e:csp2F3*}
\end{equation}
all of which automatically satisfy the Calabi-Yau condition, $\sum_{j=1}^4q(y_j)=1$.
Choosing a rational value for $q(y_4)$ and clearing denominators, one finds suitable linearly independent 4-vectors $Q(y_i)$, reconstructs the fan of the toric space for which $g(y)$ in~\eqref{e:csp*2F3} is an anticanonical section, and then refines\ftn{The procedure in~\cite{rBKK-tvMirr} yields more than two 4-vectors. Among these, we selects the two of which integral linear combinations reproduce all others. For brevity, we display only this final choice. In turn, the columns formed from the components of this pair of 4-vectors $Q(y_i)$ are 2-vectors with co-prime components; see~\eqref{e:*2F3Qmu}.} the choice of $Q(y_i)$ to proper Mori vectors~\cite{rBKK-tvMirr}. To this end:
\begin{equation}
 \begin{array}{r@{\,}l@{}}
 q(y_4)&=-1\\
 q(y_4)&=\frc32
\end{array}\bigg\}\To~~
\BM{\Tw{Q}^1(y_i)\\\Tw{Q}^2(y_i)}\!\<=\!\AR{@{}r@{~}r@{~}r@{~}r@{}}{1&0&-2&3\\0&1&3&-2}
\To\quad
  \AR{@{}r@{~}r@{~}r@{~}r@{}}{1&0&-2&3\\0&1&3&-2}\!{\cdot}
   \C3{\bM{-3&\3-2\\[2pt] \3-2&-3\\[2pt] \3-0&\3-1\\[2pt] \3-1&\3-0}}
    \<{\isBy{\sss\eqref{e:Qnu=0}}}0
   ~~\6\star\leadsto~~
   \C5{\bM{-1&-1\\[2pt] -1&\3-4\\[2pt] \3-1&-2\\[2pt] \3-1&-0}}
 \label{e:TQ2F3}
\end{equation}
where
 $\Tw{Q}^1(y_j)=2\,q(y_j)|_{q(y_4)=-1}$ and
 $\Tw{Q}^2(y_j)=2\,q(y_j)|_{q(y_4)=\frac32}$ are linearly independent integral choices.
The null-space of their matrix-stack is spanned by
$\Tw\m^1_j=(-3,2,0,1)$ and $\Tw\m^2_j=(2,-3,1,0)$, given in the two columns of the $2{\times}4$ right-hand side (blue) matrix in the middle.
These two 4-vectors, $\Tw\m_j^\k$ for $\k\<=1,2$, define via the $\star$-labeled arrow~\eqref{e:TQ2F3} the final 4-vectors, $\m^1_j=\Tw\m^1_j{+}\Tw\m^2_j$ and $\m^2_j={-}\Tw\m^1_j{-}2\Tw\m^2_j$, given here by the columns of the right-most matrix in~\eqref{e:TQ2F3}.
 Each of these pairs of 4-vectors, $\Tw\m^\k_j$ and $\m^\k_j$, defines a 4-tuple of 2-vectors, for which the $\star$-labeled transformation~\eqref{e:TQ2F3} is simply a $\GL(2;\ZZ)$ basis change:
\begin{equation}
  \ARR.{r@{\,=\,}l@{}}{\m^1_j&\Tw\m^1_j{+}\Tw\m^2_j\\ \m^2_j&{-}\Tw\m^1_j{-}2\Tw\m^2_j}.
  \bigg\}\To~~
  \AR{@{}c@{~}c@{}}{\3-1&\3-1\\[-4pt]-1&-2}\!{\cdot}\!
  \AR{@{}c|c|c|c@{}}{-3&\3-2&0&1\\[-4pt]\3-2&-3&1&0} \6*=
  \AR{@{}c|c|c|c@{}}{-1&-1&\3-1&\3-1\\[-4pt]-1&\3-4&-2&-1}
 \label{e:N2F3}
\end{equation}
Using this last 4-tuple of 2-vectors as generators of a fan we have, akin to~\eqref{e:nFmQnu}:
\begin{equation}
 \vC{\TikZ{[scale=.6]
           \path[use as bounding box](-2,-2)--(3,3);
           \foreach\x in{-2,...,3}
            \foreach\y in{-2,...,3} \fill[gray!50](\x,\y)circle(.7mm);
            \corner{(0,0)}{0}{90}{1.3}{green};
             \corner{(0,0)}{90}{123}{1.3}{yellow};
            \corner{(0,0)}{123}{225}{1.3}{red};
            \corner{(0,0)}{225}{327}{1.3}{Purple};
            \corner{(0,0)}{327}{360}{1.3}{blue};
           \draw[thick,-stealth](0,0)--++(1,0);
            \path(1.8,.4)node{\footnotesize$Q(y_1)$};
           \draw[thick,-stealth](0,0)--++(0,1);
            \path(.4,1.4)node{\footnotesize$Q(y_2)$};
           \draw[thick,-stealth](0,0)--++(-2,3);
            \path(-1.2,2.6)node[rotate=-52]{\footnotesize$Q(y_3)$};
           \draw[thick,-stealth](0,0)--++(3,-2);
            \path(2.8,-1.3)node[rotate=-30]{\footnotesize$Q(y_4)$};
           \draw[blue,very thick,-stealth](0,0)--++(-2,-2);
            \path[blue](-1.5,-.8)node[rotate=45]{\footnotesize$-\!\sum_{i=0}^4Q(y_i)$};
           \filldraw[thick,fill=white,thick](0,0)circle(.9mm);
            }}
 \qquad
  \begin{array}{@{}c@{~}|c@{~}c@{~}c@{~}c@{~}c@{\;}|@{\;}c@{~}c@{}}
    &~y_1 &~y_2 &~y_3 &~y_4 \\ \toprule\nGlu{-2pt}
    & -1  & -1  &~~1  &~~1  
 \TikZ{\path[use as bounding box](0,0);
        \path(.2,-.1)node{$\Big\}$};
        \draw[thick,->](.3,-.1)--++(.4,0);
            }%
                            \\[-3pt] 
    & -1  &~~4  & -2  & -1  \\[-1pt] \midrule\nGlu{-2pt}
 \TikZ{\path[use as bounding box](0,0);
        \path(-.15,-.1)node{$\Big\{$};
        \draw[thick,->](-.15,-.1)--++(-.45,0);
            }%
 Q^1&~~1  &~~0  & -2  &~~3  \\[-3pt]
 Q^2&~~0  &~~1  &~~3  & -2  \\
  \end{array}
  \qquad
 \vC{\TikZ{[scale=.67]\path[use as bounding box](-3,-3)--(2,2);
               \foreach\x in{-3,...,2}
                \foreach\y in{-3,...,2} \fill[gray!50](\x,\y)circle(.5mm);
               \draw[blue!50](-3,2)--(2,-3)--(0,1)--(1,0)--cycle;
               \draw[thick,-stealth](0,0)--++(-3,2);
                \path(-3,2)node[below]{\footnotesize$\Tw\m_1$};
               \draw[thick,-stealth](0,0)--++(2,-3);
                \path(2,-3)node[left]{\footnotesize$\Tw\m_2$};
               \draw[thick,-stealth](0,0)--++(0,1);
                \path(0,1)node[above]{\footnotesize$\Tw\m_3$};
               \draw[thick,-stealth](0,0)--++(1,0);
                \path(1,0)node[right]{\footnotesize$\Tw\m_4$};
               \filldraw[thick,fill=white,thick](0,0)circle(.67mm);
              }}
 ~~\6{\sss\GL(2;\ZZ)}{\approx}~~
 \vC{\TikZ{[scale=.67]\path[use as bounding box](-1,-2)--(1.5,4);
               \foreach\x in{-1,...,1}
                \foreach\y in{-1,...,4} \fill[gray!50](\x,\y)circle(.5mm);
               \draw[blue!50](-1,-1)--(-1,4)--(1,-2)--(1,-1)--cycle;
               \draw[thick,-stealth](0,0)--++(-1,-1);
                \path(-1,-1)node[below]{\footnotesize$\m_1$};
               \draw[thick,-stealth](0,0)--++(-1,4);
                \path(-1,4)node[right]{\footnotesize$\m_2$};
               \draw[thick,-stealth](0,0)--++(1,-2);
                \path(1,-2)node[left]{\footnotesize$\m_3$};
               \draw[thick,-stealth](0,0)--++(1,-1);
                \path(1,-1)node[right]{\footnotesize$\m_4$};
               \foreach\y in{-1,...,4} \draw[dashed](0,0)--(-1,\y);
                \draw[dashed](0,-1)--(0,1);
               \filldraw[thick,fill=white,thick](0,0)circle(.67mm);
              }}
 \label{e:*2F3Qmu}
\end{equation}
The secondary fan (far left) was read from the columns of the $\Tw{Q}$-matrix in~\eqref{e:TQ2F3}; the corresponding tabulated toric specification and fan generators on the right are read from the indicated columns in~\eqref{e:N2F3}.
The blue line links the $\Tw\m$- and $\m$-vertices in their order determined by~\eqref{e:TQ2F3} and~\eqref{e:N2F3}, and the result outlines precisely the self-crossing polygon that is the {\em\/transpolar\/}~\cite{rBH-gB} (roughly, iteratively face-wise polar~\eqref{e:StdP}) of the non-convex VEX polygon that spans the $\fan{\FF[2]3}$ fan:
\begin{equation}
 \vC{\TikZ{[scale=.6]\path[use as bounding box](-2,-3)--(2,1);
           \foreach\x in{-2,...,1}
            \foreach\y in{-3,...,1} \fill[gray!50](\x,\y)circle(.5mm);
            \corner{(0,0)}{0}{90}{1}{green};
            \corner{(0,0)}{90}{153}{1}{red};
            \corner{(0,0)}{153}{288}{1}{blue};
            \corner{(0,0)}{288}{360}{1}{yellow};
           \draw[thick,-stealth](0,0)--(1,0);
            \path(1,0)node[right]{\footnotesize$Q(x_2)$};
           \draw[thick,-stealth](0,0)--(1,-3);
            \path(1,-3)node[left]{\footnotesize$Q(x_1)$};
           \draw[blue,thick,-stealth](0,0)--(-2,1);
            \path[blue](-1.8,.2)node[rotate=-30]{\footnotesize$-\!\sum_{i=1}^4Q(x_i)$};
           \draw[thick,-stealth](0,0)--(0,1);
            \path(0,1)node[above]{\footnotesize~~~$Q(x_3)\<=Q(x_4)$};
           \filldraw[thick,fill=white](0,0)circle(.8mm);
            }}
\qquad
  \begin{array}{@{}c@{~}|c@{~}c@{~}c@{~}c@{~}c@{\;}|@{\;}c@{~}c@{}}
   & x_1  & x_2  & x_3  & x_4 \\ \toprule\nGlu{-2pt}
\MR2*{\rotatebox{90}{$\fan{\FF[2]{m}}$}}
   &  -1  &  1   &  0   &  -3 
 \TikZ{\path[use as bounding box](0,0);
        \path(.2,-.1)node{$\Big\}$};
        \draw[thick,->](.3,-.1)--++(.4,0);
            }%
                            \\[-3pt] 
   &\3-0  &  0   &  1   &  -1 \\[-1pt] \midrule\nGlu{-2pt}
 \TikZ{\path[use as bounding box](0,0);
        \path(-.15,-.1)node{$\Big\{$};
        \draw[thick,->](-.15,-.1)--++(-.45,0);
            }%
Q^1&\3-1  &  1   &  0   &\3-0\\[-3pt]
Q^2&  -3  &  0   &  1   &\3-1\\
  \end{array}
\qquad
 \vC{\TikZ{\path[use as bounding box](-3,-1)--(1,1);
           \foreach\x in{-3,...,1}
            \foreach\y in{-1,...,1} \fill[gray!50](\x,\y)circle(.5mm);
           \filldraw[yellow,opacity=.5,draw=blue](1,0)--(0,1)--(-1,0)--(-3,-1)--cycle;
            \corner{(0,0)}{0}{90}{.9}{green};
            \corner{(0,0)}{90}{180}{.9}{blue};
            \corner{(0,0)}{180}{198}{.9}{violet};
            \corner{(0,0)}{198}{360}{.9}{orange};
           \draw[thick,-stealth](0,0)--(1,0);
            \path(1,0)node[below]{\footnotesize$x_2$};
           \draw[thick,-stealth](0,0)--(0,1);
            \path(0,1)node[right]{\footnotesize$x_3$};
           \draw[red,thick,-stealth](0,0)--(-1,0);
            \path(-1,0)node[above left]{\footnotesize$x_1$};
           \draw[thick,-stealth](0,0)--(-3,-1);
            \path(-3,-1)node[above]{\footnotesize$x_4$};
           \filldraw[thick,fill=white](0,0)circle(.5mm);
           \path(.2,-.7)node{\Large$\fan{\FF[2]3}$};
            }}
\qquad
 \vC{\TikZ{[scale=.75]\path[use as bounding box](-2,-2)--(1,4);
           \foreach\x in{-1,...,1}
            \foreach\y in{-2,...,4} \fill[gray!50](\x,\y)circle(.5mm);
           \fill[yellow,opacity=.85,draw=blue]
                (0,0)--(1,-1)--(-1,-1)--(-1,4)--(0,0);
           \fill[orange,opacity=.6](0,0)--(1,-1)--(1,-2)--cycle;
           \draw[thick,-stealth](0,0)--(1,-1);
            \path(1.35,-.65)node{\scriptsize$\frac{x_2\!^2}{x_4}$};
           \fill[yellow,opacity=.85,draw=blue]
                (0,0)--(-1,4)--(1,-2)--(0,0);
            \draw[red](1,-2)--(1,-1);
           \draw[thick,-stealth](0,0)--(-1,-1);
            \path(-1,-1)node[below]{\scriptsize$x_1\!^2x_4\!^5$};
           \draw[thick,-stealth](0,0)--(-1,4);
            \path(-1,4)node[right]{\scriptsize$x_1\!^2x_3\!^5$};
           \draw[thick,-stealth](0,0)--(1,-2);
            \path(1.4,-1.8)node{\scriptsize$\frac{x_2\!^2}{x_3}$};
           \filldraw[thick,fill=white](0,0)circle(.6mm);
           \draw[thick,->](-1.2,1.8)to[out=150,in=30]++(-1.7,0);
            \path(-2.1,1.5)node{\cite{rBH-gB}};
           \draw[thick,<-](-1.2,1.2)to[out=210,in=-30]++(-1.7,0);
           \path(.5,2.5)node{\Large$\pDN{\FF[2]3}$};
            }}
 \label{e:2F3SN}
\end{equation}
The flip-folded polygon at the right-hand side of both~\eqref{e:*2F3Qmu} and~\eqref{e:2F3SN} spans a {\em\/multifan\/} that, with the dashed unit-degree subdivisions, specifies a smooth toric space in much the same way as does the fan $\fan{\FF[2]3}$, depicted mid-right in~\eqref{e:2F3SN}\cite{rHM-MFs,Nishimura:2006vs,rBH-gB}.

\paragraph{Toric Pr\'ecis:}
The foregoing then demonstrates that the straightforward generalization~\cite{rBH-gB} of the transposition mirror models~\cite{rBH,rBH-LGO+EG,rMK-diss} illustrated in Figure~\ref{f:NS2F3NS} for $\FF[2]3[c_1]$ and its mirror,
where ``$\pDs{V}\<\lat\S_{V}$'' means that the polytope (multitope) $\pDs{V}$ spans the (multi)fan $\S_{V}$, i.e., $\S_{V}$ star-subdivides $\pDs{V}$. The evident relations
\begin{equation}
  \fan{\FF[2]3}\smt\pDs{\FF[2]3}=\pDN{\MF[2]3}
   ~\fif[\!\text{\cite{rBH-gB}}]{~\wtd~}~
  \pDN{\FF[2]3}=\pDs{\MF[2]3}\lat\fan{\MF[2]3}
\end{equation}
provide the unit-subdivided (multi)fans, $\fan{\FF[2]3}$ and $\fan{\MF[2]3}$, which specify the atlas of smooth local charts for the {\em\/ambient\/} toric spaces, $\FF[2]3$ and $\MF[2]3$, respectively, as each others' {\em\/transpolar\/} toric space.
 Also, the Cox (homogeneous) variables, displayed in the column-heading rows in~\eqref{e:2F3SN} and~\eqref{e:*2F3Qmu}, were used to express the polynomials~\eqref{e:csp2F3} and~\eqref{e:csp*2F3}, respectively:
\begin{subequations}
 \label{e:2F3cs}
\begin{alignat}9
 g(y)^\sfT=f(x)
 &=a_1 x_1\!^2x_3\!^5 +a_2x_1\!^2x_4\!^5
        +a_3\frac{x_2\!^2}{x_4} +a_4\frac{x_2\!^2}{x_3}&
 &=\sum_{\m_j\smt\pDN{\FF[2]3}} a_j
    \prod_{\n_i\smt\pDs{\FF[2]3}} x_i^{\vev{\m_j,\n_i}+1}; \label{e:cs2F3}\\
 f(x)^\sfT=g(y)
 &=b_1y_1\!^2y_2\!^2 +b_2y_3\!^2y_4\!^2
      +b_3\frac{y_1\!^5}{y_4} +b_4\frac{y_2\!^5}{y_3}&
 &=\sum_{\n_i\smt\pDs{\FF[2]3}} b_i
    \prod_{\m_j\smt\pDN{\FF[2]3}} y_j^{\vev{\m_j,\n_i}+1}, \label{e:cs*2F3}
\end{alignat}
\end{subequations}
where the symbol ``$\smt$'' stands for ``is a vertex of'' the polytope, i.e., a 1-cone generator of the fan spanned by that polytope. This generalizes the transposition prescription~\cite{rBH,rBH-LGO+EG,rMK-diss} and~\cite{rBaty01}, in Cox variables~\cite{rCox}, to transpolar pairs of VEX polytopes~\cite{rBH-gB}.

For all $n\<\geqslant2$ and $m\<\geqslant3$,
the fan $\fan{\FF{m}}$ is spanned by a non-convex polytope, $\pDs{\FF{m}}$, reflecting that $\FF{m}$ is not Fano.
In turn, the transpolar polytope $\pDN{\FF{m}}$ is flip-folded and spans a multifan $\fan{\MF{m}}$, which when unit-subdivided encodes $\MF{m}$ as a (smooth) {\em\/toric manifold\/}~\cite{rHM-MFs,Nishimura:2006vs}.

\paragraph{Laurent Deformations Rationale:}
The specific choice of the rational monomials included as Laurent deformations of the anticanonical sections, such as in~\eqref{e:csp2F3}, is then specified by the following:
\begin{itemize}[itemsep=-1pt, topsep=-1pt]
 \item As indicated (red-ink dashed arrow) in Figure~\ref{f:NS2F3NS}, the rational monomials, $\big\{\frac{x_2\!^2}{x_4},\frac{x_2\!^2}{x_3}\big\}$, form the flip-folded edge in $\pDN{\FF[2]3}$, which is polar to the concave (MPCP-desingularizing~\cite{rBaty01}) vertex $\n_1\<\in\pDs{\FF[2]3}$. The Cox variables in the denominators are defined by the vertices delimiting the concavity in $\pDs{\FF[2]3}$.

 \item Through the looking glass, the rational monomials $\big\{\frac{y_1\!^5}{y_4},\frac{y_2\!^5}{y_3}\big\}$ correspond to vertices that delimit the concavity in $\pDN{\MF[2]3}$, and the Cox variables in their denominators are defined by vertices that form the flip-folded edge in $\pDs{\MF[2]3}$, the one that is polar to the non-convex vertex in $\pDN{\MF[2]3}$.

\end{itemize}
Thus, the rational monomials included both in~\eqref{e:cs2F3} and in~\eqref{e:cs*2F3} are precisely those that correspond to the {\em\/concave subset\/} in
 $\pDs{\FF[2]3}\<=\pDN{\MF[2]3}$ and {\em\/flip-folded subset\/} in
 $\pDN{\FF[2]3}\<=\pDs{\MF[2]3}$ ---
precisely the features by which VEX polytopes~\cite{rBH-gB} generalize the by now familiar reflexive polytopes~\cite{rBaty01,rKreSka98}.
 Such self-crossing polygons have been used to encode {\em\/(pre)symplectic\/} spaces~\cite{rK+T-pSympTM}, which correlates with mirror symmetry relating complex and symplectic structures~\cite{Kontsevich:1995wk,rSYZ-Mirr}; of course, each Calabi-Yau space admits both structures.

 Thus, the Newton polygons $\pDN{\FF[2]3}$ and $\pDN{\MF[2]3}$ (Figure~\ref{f:NS2F3NS} far left and mid-bottom, respectively), both non-convex but VEX~\cite{rBH-gB}, specify the (Laurent-deformed) sections for the defining equation of the Calabi-Yau hypersurfaces:
\begin{equation}
  \FF[2]3[c_1]\ni\{f(x)\<=0\}
  \quad\fIf[1pt]{\sss\text{pair}}{\sss\text{mirror}}\quad
  \{g(y)\<=0\}\in\MF[2]3[c_1].
 \label{e:MM}
\end{equation}
 Incidentally, the {\em\/standard\/} Newton polytope of $\FF[2]3$ is highlighted in green in the leftmost illustration in Figure~\ref {f:NS2F3NS} and specifies the non-transverse polynomial~\eqref{e:stdCS}; for a 3-dimensional example, see~\cite{rBH-gB}.

The above computations~(\ref{e:csp2F3}$'$)--\eqref{e:MM} are fairly routine for (convex) reflexive polytopes~\cite{rBaty01,rKreSka98}.
 With the standard {\em\/polar\/} operation generalized to the transpolar --- they continue to also hold for VEX polytopes~\cite{rBH-gB}. In fact, suffice it here to mention that numerous combinatorial formulae generalizing the ``12-theorem''~\cite[Theorem~10.5.10]{rCLS-TV} and various results about Chern and other characteristic classes~\cite{rD-TV} continue to hold provided all polytopes and (multi)fans are taken with orientation-dependent multiplicity. In turn, the so-called A-discriminants for both the complex structure and the K\"ahler class moduli, as well as the Yukawa couplings continue to be computable and conform to general mirror symmetry expectations. Details of these results will be reported separately.

\begin{figure}[htb]
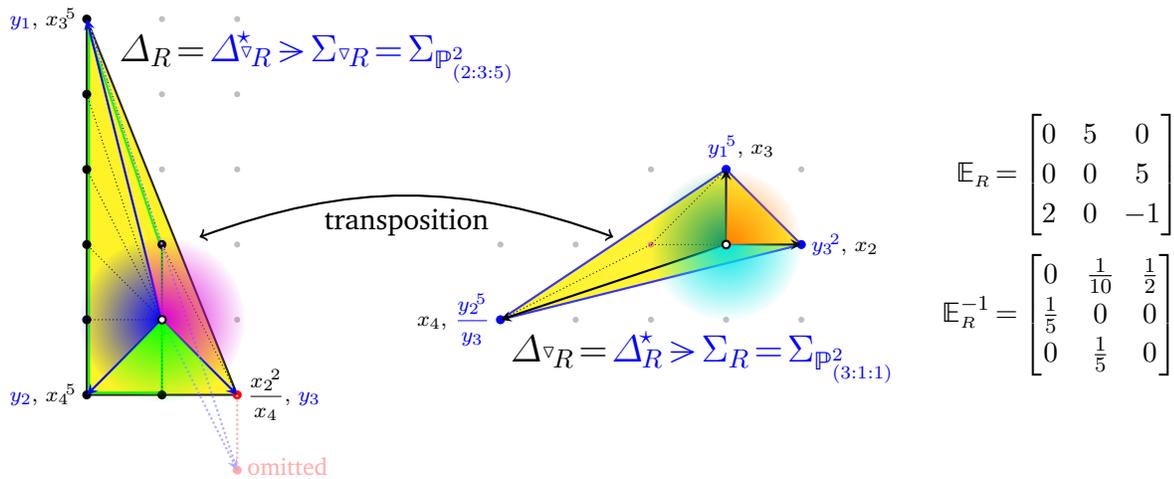

 $$
 \vC{\TikZ{\path[use as bounding box](-1,-2.2)--(2.5,4.2);
            \foreach\y in{0,...,4} \fill[gray!50](1,\y)circle(.4mm);
            \foreach\y in{2,...,4} \fill[gray!50](0,\y)circle(.4mm);
           \fill[yellow,opacity=.85,draw=black,thick,line join=round]
                (1,-1)--(-1,-1)--(-1,4)--cycle;
            \fill[red](1,-1)circle(.6mm);
             \fill[red!33](1,-2)circle(.6mm);
             \path[red!33](1,-1.95)node[right]{\footnotesize omitted};
           \draw[green,thick]
                 (-.97,3.82)--(-.97,-.97)--(0,-.97)--(0,.92)--cycle;
           \foreach\y in{-1,...,4}
             { \fill(-1,\y)circle(.6mm); \draw[densely dotted](0,0)--(-1,\y); }
               \draw[densely dotted](0,-1)--(0,1)--(1,-1);
               \draw[densely dotted](0,1)--(-1,4);
            \fill(0,-1)circle(.6mm); \fill(0,1)circle(.6mm);
           \draw[blue,thick,-stealth](0,0)--(1,-1);
            \path(1,-1)node[right]{\scriptsize$\ddd\frac{x_2\!^2}{x_4},\,\C3{y_3}$};
           \draw[blue!33,thick, densely dotted,-stealth](0,0)--(1,-2);
            \draw[red!33,thick, densely dotted](1,-1)--(1,-2);
            \draw[blue!33,thick, densely dotted](1,-2)--(-1,4);
           \draw[blue,thick,-stealth](0,0)--(-1,-1);
            \path(-1,-1)node[left]{\scriptsize$\C3{y_2},\,x_4\!^5$};
           \draw[blue,thick,-stealth](0,0)--(-1,4);
            \path(-1,4)node[left]{\scriptsize$\C3{y_1},\,x_3\!^5$};
            \corner{(0,0)}{-45}{-135}{1.1}{green};
            \corner{(0,0)}{225}{105}{1.1}{blue};
            \corner{(0,0)}{105}{-45}{1.1}{Magenta};
           \filldraw[thick,fill=white](0,0)circle(.5mm);
           \path(-.7,3.5)node[right]
           {\Large$\pDN{R}\<=\C3{\pDs{{}^\wtd\!R}\<\lat
                    \S_{{}^\wtd\!R}\<=\S_{\IP^2_{(2{\mathchar"3A}3{\mathchar"3A}5)}}}$};
            }}
     \vC{\TikZ{\path[use as bounding box](-5,-1)--(2,1);
           \draw[thick,<->](-7,.1)to[out=20,in=160](-1.5,.1);
            \path(-4.25,.3)node{transposition};
           \foreach\x in{-1,...,1} \fill[gray!50](\x,1)circle(.4mm);
           \foreach\x in{-3,...,1}
            \foreach\y in{-1,...,0} \fill[gray!50](\x,\y)circle(.4mm);
           \filldraw[yellow,opacity=.8,draw=blue,thick,line join=round]
                 (1,0)--(0,1)--(-3,-1)--cycle;
           \fill[red!50](-1,0)circle(.4mm);
           \fill[blue](1,0)circle(.6mm);
           \fill[blue](0,1)circle(.6mm);
           \fill[blue](-3,-1)circle(.6mm);
             \draw[densely dotted](0,1)--(-1,0)--(-3,-1);
             \draw[densely dotted](0,0)--(-1,0);
           \draw[thick,-stealth](0,0)--(1,0);
            \path(1,0)node[right]{\scriptsize$\C3{y_3\!^2},\,x_2$};
           \draw[thick,-stealth](0,0)--(0,1);
            \path(.2,1.3)node{\scriptsize$\C3{y_1\!^5},\,x_3$};
           \draw[thick,-stealth](0,0)--(-3,-1);
            \path(-3,-1)node[left]{\scriptsize$x_4,\,\ddd\C3{\frac{y_2\!^5}{y_3}}$};
            \corner{(0,0)}{0}{90}{1}{orange};
            \corner{(0,0)}{90}{198}{1}{Sage};
            \corner{(0,0)}{198}{360}{1}{Turque};
           \filldraw[thick,fill=white](0,0)circle(.5mm);
           \path(-3,-1.5)node[right]
           {\Large$\pDN{{}^\wtd\!R}\<=\C3{\pDs{R}\<\lat\S_R\<=
                    \S_{\IP^2_{(3{\mathchar"3A}1{\mathchar"3A}1)}}}$};
            }}
\qquad
 \ARR.{@{}r@{\,=\,}l@{}}
      {\IE_{\sss R}&\BM{0&5&0\\ 0&0&5\\ 2&0&-1\\}\\[8mm]
     \IE_{\sss R}^{-1}&\BM{0&\frc1{10}&\frc12\\ \frc15&0&0\\ 0&\frc15&0\\}}.
 $$
 \caption{The toric data of the reduced mirror pair~\eqref{e:csR2F3}--\eqref{e:csR*2F3}}
 \label{f:Red2F3}
\end{figure}

\paragraph{Specific Mirror Models:}
Since the polynomials~\eqref{e:csp2F3} and~\eqref{e:csp*2F3} and the matrix of exponents~\eqref{e:EE} are not invertible, a direct relation to the original transposition prescription~\cite{rBH,rBH-LGO+EG,rMK-diss} is provided by particular complementary reducing deformations of~\eqref{e:cs2F3} and~\eqref{e:cs*2F3}, such as:
\begin{alignat}9
 x_1&\to1 &~~\&~~& a_4&\to0&~~\To&~~
 f_{\sss\!R}(x)
 &=a_1 x_3\!^5 +a_2x_4\!^5 +a_3\frac{x_2\!^2}{x_4}; \label{e:csR2F3}\\
 b_1&\to0 &~~\&~~& y_4&\to1&~~\To&~~
 g_{\sss R}(y)
 &=b_2y_3\!^2 +b_3y_1\!^5 +b_4\frac{y_2\!^5}{y_3}. \label{e:csR*2F3}
\end{alignat}
This reduces the matrix~\eqref{e:EE} to an invertible $3{\times}3$ matrix, and may be depicted in terms of the toric data in Figure~\ref {f:NS2F3NS} as shown in Figure~\ref{f:Red2F3}.
Both diagrams here serve both as Newton polygons ($\pDN{R}$ and $\pDN{{}^\wtd\!R}$, respectively) with the cornerstone (extreme) monomials indicated, as well as spanning the fans ($\pDs{{}^\wtd\!R}\<\lat\S_{{}^\wtd\!R}$ and $\pDs{R}\<\lat\S_R$, respectively) that identify the Cox variables, and specify the indicated weighted projective spaces identified by the degrees (lattice areas) of the three major cones. Here, $R\<=\IP^2_{(3{:}1{:}1)}$ is identified as the blowdown (by un-subdividing at the MPCP-smoothing $\n_1\<=(-1,0)$ 1-cone generator) of $\FF[2]3$. Analogously, ${}^\wtd\!R\<=\IP^2_{(2{:}3{:}5)}$ is identifiable as a blowdown of $\MF[2]3$ by omitting the $\m_4\<=(1,-2)$ 1-cone generator of the multifan.

Finally, the columns of $\IE^{-1}$ specify the discrete symmetries of $f_{\sss R}(x)$ in~\eqref{e:csR2F3}, while its rows analogously pertain to $g_{\sss R}(y)$ in~\eqref{e:csR*2F3}~\cite{rMK-diss}:
\begin{alignat}9
 f_{\sss\!R}(x)
 &=a_1 x_3\!^5 +a_2x_4\!^5 +a_3\frac{x_2\!^2}{x_4},&&\quad
 \bigg\{\ARR.{@{}r@{\,=\,}l}
             {\cQ&\zZ5{\frc35,\frc15,\frc15},\\
              \cG&\zZ{10}{\frc1{10},0,\frc15},}.
 &(x_2,x_3,x_4)&\in\IP^2_{(3{:}1{:}1)}; \label{e:P3115} \\
 g_{\sss R}(y)
 &=b_2y_3\!^2 +b_3y_1\!^5 +b_4\frac{y_2\!^5}{y_3},&&\quad
 \bigg\{\ARR.{@{}r@{\,=\,}l@{~}l}
             {\Tw\cQ&\zZ{10}{\frc2{10},\frc3{10},\frc5{10}},\\
              \Tw\cG&\zZ5{\frc15,\frc15,0},}.
 &(y_1,y_2,y_3)&\in\IP^2_{(2{:}3{:}5)}.  \label{e:P23510}
\end{alignat}
The action of the ``quantum symmetry'' $\cQ$ (resp., $\Tw\cQ$) on the indicated homogeneous coordinates is specified by the sum of the columns (resp., rows) of $\IE^{-1}$, and of course coincides with the (rescaled) weights of $\IP^2_{(3{:}1{:}1)}$ (resp., $\IP^2_{(2{:}3{:}5)}$).
The action of the ``geometric'' symmetry is specified by linear combinations of the columns (resp., rows) independent of $\cQ$ (resp., $\Tw\cQ$), and can here be chosen to be generated by the 2nd column (resp., 2nd+3rd row) of $\IE^{-1}$. The total degree of discrete symmetries being
\begin{equation}
  |\cQ||\cG|=\det[\IE]\<=50=|\Tw\cQ||\Tw\cG|
\end{equation}
verifies that the $(\ZZ_5,\ZZ_{10})$ pair exhausts the options.

\paragraph{The General Case:}
The above computations are straightforward to follow through for $\FF[2]m$, for all $m$:
\begin{subequations}
\begin{alignat}9
 f_{\sss\!R}(x;m)
 &=a_1 x_3\!^{m+2} +a_2x_4\!^{m+2} +a_3\frac{x_2\!^2}{x_4\!^{m-2}},\quad
 &(x_2,x_3,x_4)&\in\IP^2_{(m{:}1{:}1)}; \label{e:2FmRx}\\
 g_{\sss R}(y;m)
 &=b_2y_3\!^2 +b_3y_1\!^{m+2} +b_4\frac{y_2\!^{m+2}}{y_3\!^{m-2}},\quad
 &(y_1,y_2,y_3)&\in\IP^2_{(2{:}m{:}m{+}2)}. \label{e:2FmRy}
\end{alignat}
\end{subequations}
The symmetries depend on the parity of $m$, but are as straightforward to find from $\IE^{-1}$:
\begin{equation}
 \cQ(\FF[2]m) \<=\ttt\zZ{m{+}2}{\frac{m}{m{+}2},\frac1{m{+}2},\frac1{m{+}2}},\quad
 \cG(\FF[2]m)\<=
 \ARR\{{@{}l@{~~}l@{}}
      {\zZ{2(m{+}2)}{\frac{m{-}2}{2(m{+}2)},0,\frac1{m{+}2}},
        &m\,\text{odd};\\[2mm]
       \zZ{m{+}2}{\frac{m/2{-}1}{(m{+}2)},0,\frac1{m{+}2}}
                  \times\zZ2{\frc12,0,0},
        &m\,\text{even}}.
\end{equation}
The symmetries of $g_{\sss R}(y;m)$ are of course flipped:
\begin{equation}
 \cQ(\MF[2]m)\<=
 \ARR\{{@{}l@{~~}l@{}}
      {\zZ{2(m{+}2)}{\frac1{m{+}2},\frac{m}{2(m{+}2)},\frc12},
        &m\,\text{odd};\\[2mm]
       \zZ{m{+}2}{\frac1{m{+}2},\frac{m/2}{m{+}2},\frac12}
                  \times\zZ2{0,\frc12,\frc12},
        &m\,\text{even};}.\quad
 \cG(\MF[2]m)\<=\ttt\zZ{m{+}2}{\frac1{m{+}2},\frac1{m{+}2},0}.
\end{equation}

The total Hilbert space in a Landau-Ginzburg orbifold\cite{rLGO0,rLGO} or the corresponding {\em\/phase\/} of the gauged linear sigma model~\cite{rPhases} is a direct sum of the ``untwisted'' and several ``twisted'' sectors, which span representations of the ``geometric'' and ``quantum'' symmetries, respectively.
The hallmark flipped identifications $\cQ\<\approx\Tw\cG$ and $\Tw\cQ\<\approx\cG$ therefore insure~\cite{rBH} that the ``untwisted'' and ``twisted'' sectors of the $f_{\sss R}(x)$-model match those of the  ``twisted'' and ``untwisted'' sectors of the $g_{\sss R}(y)$-model --- as required by mirror symmetry.

\subsection{Multiple Mirrors}
\label{s:MTMs}
\begin{figure}[htb]
 \begin{center}
  \begin{picture}(160,115)(0,-2)
       \put(0,90){$
  \begin{array}{@{}c@{~}|@{~}r@{~}r|@{~}r@{~~}r@{~~}r@{~~}r@{}}
 & y_1   &y_2   &y_3     &y_4     &y_5     &y_6 \\ [2pt]\toprule
\MR3*{\rotatebox{90}{$\fan{\MF{m}}$}}
 & -1     &-1     & 2       & 2       &-1       &-1   \\[-1pt]
 & -1     &-1     &-1       &-1       & 2       & 2   \\[-1pt]
 & 2m{+}1 &-1     &1{-}m    &-1       &1{-}m    &-1   \\ \midrule
\Tw{Q}^1&  1  & 1 & 2 &\C30    &\C30    & 2 \\ 
\Tw{Q}^2&  2m{-}1 & 3     & 2(m{+}1) &\C30    & 2(m{+}1) &\C30\\ 
\Tw{Q}^3& -(m{-}2)  & m{-}2 &-2(m{+}1) & 2(m{+}1) &\C30    &\C30\\ \bottomrule
  \end{array}$}
       \put(0,48){$
  \begin{array}{@{}r@{~}|@{~}r@{~}r@{~}r@{\;}|@{\;}r@{~~}r@{}}
   & x_1  & x_2  & x_3  & x_4  & x_5 \\ \toprule\nGlu{-2pt}
\MR3*{\rotatebox{90}{$\fan{\FF{m}}$}}
   &\!-1  &  1   &  0   &  0   &-m \\[-3pt] 
   &\!-1  &  0   &  1   &  0   &-m \\[-3pt]
   &   0  &  0   &  0   &  1   &-1 \\[-1pt] \midrule\nGlu{-2pt}
Q^1&   1  &  1   &  1   &  0   &\3-0\\[-3pt]
Q^2&\!-m  &  0   &  0   &  1   &\3-1\\ \bottomrule
  \end{array}$}
       \put(0,10){$\5{\sss(m=3)}{~\IE\<=}
                    \BM{3 & 0 & 0 & 8 & 0\\[-2pt]
                        3 & 0 & 0 & 0 & 8\\[-2pt]
                        0 & 3 & 0 &-1 & 0\\[-2pt]
                        0 & 3 & 0 & 0 &-1\\[-2pt]
                        0 & 0 & 3 &-1 & 0\\[-2pt]
                        0 & 0 & 3 & 0 &-1\\ }$}
       \put(95,70){\reflectbox{\includegraphics[width=40mm]{K3N4.pdf}}}
       \put(105,87.7){\TikZ{\path[use as bounding box](0,0);
                       \path(1.2,1.2)node[right]
                           {\Large$\pDs{\FF[3]3}\<\lat\fan{\FF[3]3}$};
                       \path[red](.85,.4)node{$\n_1$};
                       \path(2.25,-.2)node{$\n_2$};
                       \path(-1,-.2)node{$\n_3$};
                       \path(.4,1.8)node{$\n_4$};
                       \path(2.9,-1.25)node{$\n_5$};
                       \path(.75,-1.1)node[rotate=-25]{see~\eqref{e:nFmQnu}};
              }}
       \put(85,37){\reflectbox{\includegraphics[width=40mm]{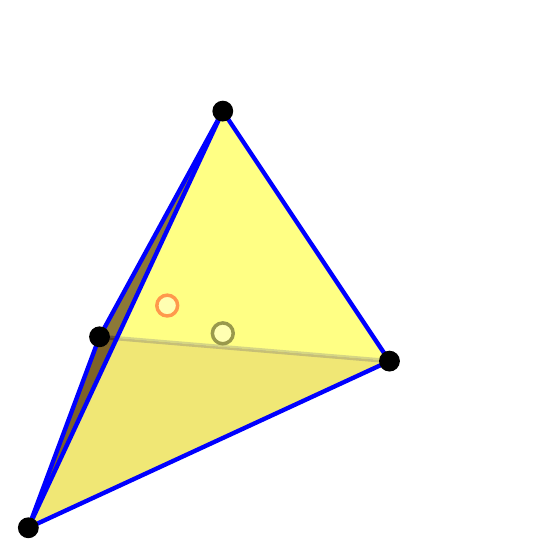}}}
       \put(98,45){\rotatebox{-25}{$\Conv(\pDs{\FF[3]3})$}}
       \put(105,-8){\reflectbox{\includegraphics[width=50mm]{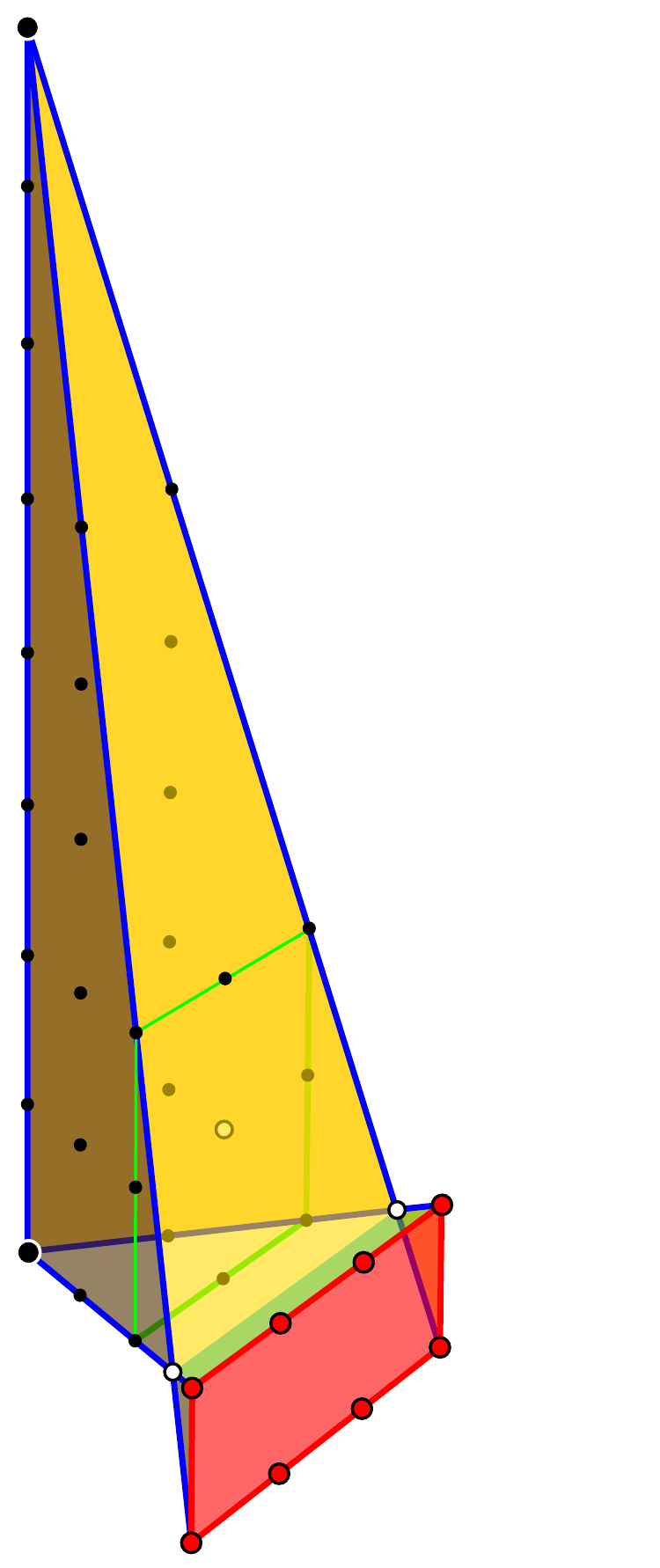}}}
       \put(148,0){\Large$\pDN{\FF[3]{3}}$}
       \lput(155,107){$\m_1$}
       \lput(155,16){$\m_2$}
       \lput(115,8){\C1{$\m_3$}}
       \lput(115,19){\C1{$\m_4$}}
       \put(143,4){\C1{$\m_6$}}
       \put(142,-7){\C1{$\m_5$}}
       \put(155,30){\rotatebox{90}{the standard, incomplete part of $\pDN{\FF[3]{3}}$}}
       \rput(128,-1){\color{Rouge}the ``extension,''}
       \rput(133,-5){\color{Rouge}included in $\pDN{\FF[3]{3}}$}
   \put(46,-7){\includegraphics[width=22mm]{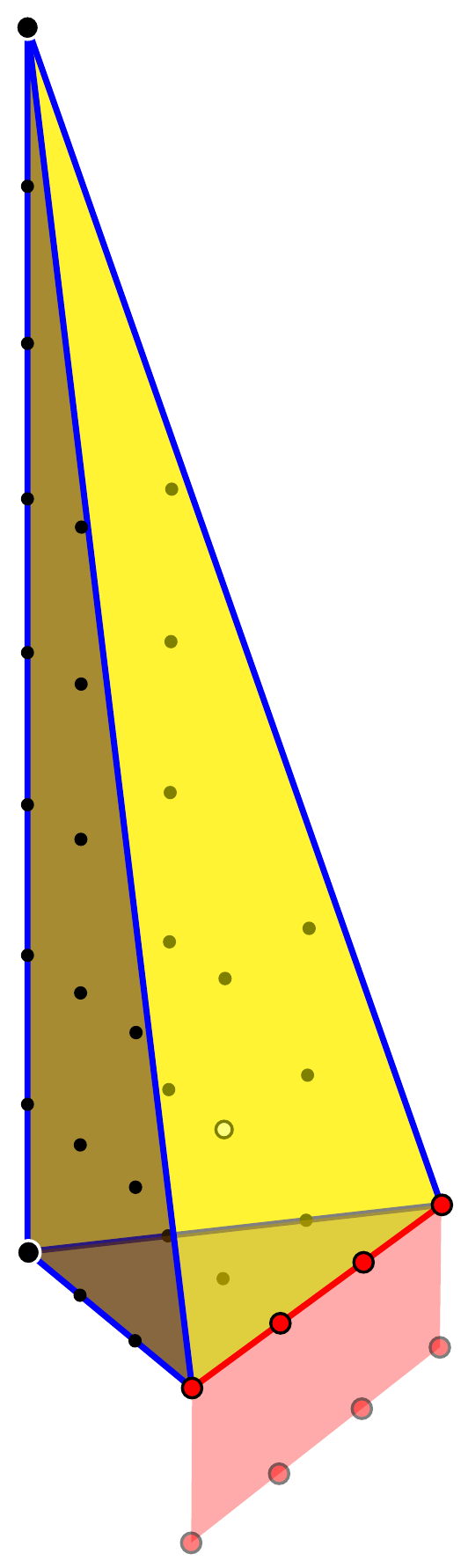}}
    \put(55,45){\rotatebox{-67}{$\Red_{3,5}[\pDN{\FF[3]{3}}]$}}
   \put(73,-7){\includegraphics[width=22mm]{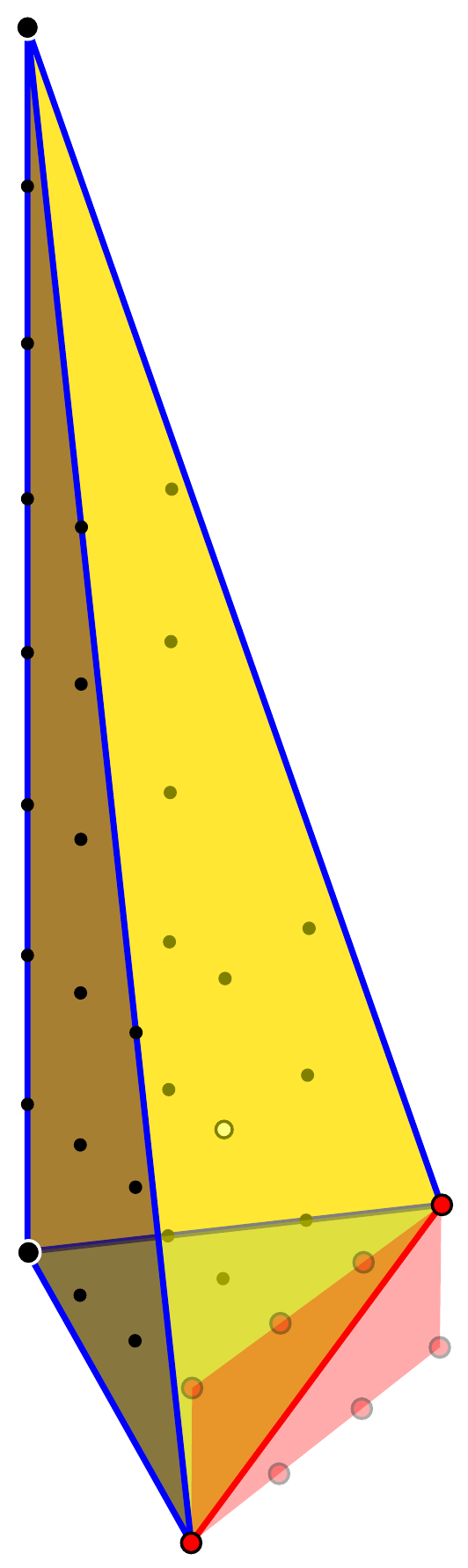}}
    \put(82,45){\rotatebox{-67}{$\Red_{4,5}[\pDN{\FF[3]{3}}]$}}
  \end{picture}
 \end{center}
 \caption{The Newton polytope $\pDN{\FF[3]m}$ specification (top, left), with the $m\<=3$ case depicted at right; the reduced polytopes are the convex hulls of a minimal subset of the vertices of the original polytope indicated}
 \label{f:*nFmQmu}
\end{figure}

The 3-dimensional case was analyzed in some detail in~\cite{rBH-gB} and we adapt some of that specification in Figure~\ref{f:*nFmQmu}; see also the display~\eqref{e:nFmQnu}.
Defining the Cox variables by the vertices of the non-convex polytope $\pDs{\FF[3]m}$ that spans the fan of $\FF[3]m$ and limiting to the vertices of the (extended) Newton polytope, $\pDN{\FF[3]m}$, to specify the cornerstone (extremal) monomials --- and then the other way around, produces the 3-dimensional analogue of the transposition mirror pair~\eqref{e:2F3cs}:
\begin{subequations}
 \label{e:3Fmcs}
\begin{alignat}9
 g(y)^\sfT\<=f(x)
 &= a_1\,x_1\!^3\,x_4\!^{2m+2} +a_2\,x_1\!^3\,x_5\!^{2m+2}
   +a_3\,\frac{x_2\!^3}{x_4\!^{m-2}} +a_4\,\frac{x_2\!^3}{x_5\!^{m-2}} 
   +a_5\,\frac{x_3\!^3}{x_4\!^{m-2}} +a_6\,\frac{x_3\!^3}{x_5\!^{m-2}},
 \label{e:cs3F3x}\\
 f(x)^\sfT\<=g(y)
 &=b_1\,y_1\!^3\,y_2\!^3 +b_2\,y_3\!^3\,y_4\!^3
  +b_3\,y_5\!^3\,y_6\!^3 
  +b_4\,\frac{y_1^{\,2m+2}}{(y_3\,y_5)^{m-2}} +b_5\,\frac{y_2^{\,2m+2}}{(y_4\,y_6)^{m-2}}.
 \label{e:cs3F3y}
\end{alignat}
\end{subequations}
For simplicity, we focus on the $m\<=3$ case.
 The $5{\times}6$ matrix of exponents is shown in Figure~\ref{f:*nFmQmu} and has rank 4. The bottom-central two diagrams illustrate two inequivalent cornerstone (extremal) reductions of the Newton polytope, whereas only $\n_1$ may be omitted from $\pDs{\FF[3]3}$ while retaining the origin inside the polytope. This results in two inequivalent, though still extremal, reductions of the matrix of exponents to an invertible $4{\times}4$ submatrix. With its many lattice points, the Newton polytope can be reduced in many other ways, leading to a web of mirror models --- all ``generated'' from the transpolar pair
 $(\pDs{\FF[3]m},\pDN{\FF[3]m})$.

\paragraph{Mirror-Pair \#1:} The first of these two pairs
\begin{subequations}
\label{e:nK3min1}
\begin{alignat}9
 \big(\Red_{1;3,5} g(y)\big)^\sfT=\Red_{1;3,5} f(x)
 &=&\,
 &a_1\,x_4\!^8 +a_2\,x_5\!^8 
   +a_4\,\frac{x_2\!^3}{x_5} 
   +a_6\,\frac{x_3\!^3}{x_5}\quad
 &\in&~\IP^3_{(3:3:1:1)}[8],
 \label{e:3F3min1x}\\
 \big(\Red_{1;3,5} f(x)\big)^\sfT=\Red_{1;3,5} g(y)
 &=&
 &b_2\,y_4\!^3
  +b_3\,y_6\!^3 
  +b_4\,y_1\!^8 +b_5\,\frac{y_2\!^8}{y_4\,y_6}\quad
 &\in&~\IP^3_{(3:5:8:8)}[24],
 \label{e:3F3min1y}
\end{alignat}
\end{subequations}
corresponds to
 $\Conv\big(\pDs{\FF[3]{3}}\ssm\n_1\big)$ and 
 $\Conv\big(\pDN{\FF[3]{3}}\ssm\{\m_3,\m_5\}\big)$.
For a generic choice of the coefficients, the polynomials~\eqref{e:nK3min1} are $\D$-regular, and are each other's transpose. The so-reduced matrix of exponents is regular:
\begin{equation}
  \eqref{e:nK3min1}\To\quad
  \Red_{1;3,5}\IE(\FF[3]{3})
   =\BM{ 0& 0& 8&~\>0\\ 0& 0& 0&~\>8\\ 3& 0& 0&-1\\ 0& 3& 0&-1\\},\quad
  \big(\Red_{1;3,5}\IE(\FF[3]{3})\big)^{-1}
   =\BM{ 0& \frac1{24}& \frac13&0\\[0pt] 0& \frac1{24}& 0&\frac13\\[0pt]
         \frac18& 0& 0&0\\[0pt] 0& \frac18& 0&0\\ },
 \label{e:E3F3min1}
\end{equation}
and the discrete symmetries of the polynomials~\eqref{e:nK3min1} are read off from the inverse matrix:
\begin{subequations}
 \label{e:MM3F31}
\begin{alignat}9
  a_1\,x_4\!^8 +a_2\,x_5\!^8 
   +a_4\,\frac{x_2\!^3}{x_5} 
   +a_6\,\frac{x_3\!^3}{x_5}:
 &~~\ARR\{{@{}r@{}}
         {\zZ3{\frac13,\frac23,0,0}\\
          \zZ{24}{\frac1{24},\frac1{24},0,\frac18}\\ \midrule\nGlu{-3pt}
          \zZ8{\frac38,\frac38,\frac18,\frac18}}.
     \bM{\ttt x_2\\[2pt]\ttt x_3\\[2pt]\ttt x_4\\[2pt]\ttt x_5}:&
 &~~\ARR\{{@{}r@{\,=\,}l}
          {\cG&\ZZ_3\<\times\ZZ_{24},\\[-2pt] \midrule\nGlu{-3pt}
           \cQ&\ZZ_8.}. \label{e:3F3-1x}\\
  b_2\,y_4\!^3
  +b_3\,y_6\!^3 
  +b_4\,y_1\!^8 +b_5\,\frac{y_2\!^8}{y_4\,y_6}:
 &~~\ARR\{{@{}r@{}}
         {\zZ8{\frac18,0,0,0}\\
          \zZ3{0,0,\frac13,\frac23}\\ \midrule\nGlu{-3pt}
          \zZ8{\frac5{24},\frac3{24},\frac13,\frac13}}.
     \bM{\ttt y_1\\[2pt]\ttt y_2\\[2pt]\ttt y_4\\[2pt]\ttt y_6}:&
 &~~\ARR\{{@{}r@{\,=\,}l}
          {\cG^\wtd&\ZZ_8\times\ZZ_3,\\[-2pt] \midrule\nGlu{-3pt}
           \cQ^\wtd&\ZZ_{24}.}. \label{e:3F3-1y}
\end{alignat}
\end{subequations}
To insure the geometric and quantum symmetry swap, we may consider the models
\begin{equation}
 \big(\,\eqref{e:3F3-1x}/\ZZ_3\,,\,\eqref{e:3F3-1y}\,\big)
  \qquad\text{and}\qquad
 \big(\,\eqref{e:3F3-1x}\,,\,\eqref{e:3F3-1y}/\ZZ_3\,\big)
 \label{e:3F3mm1}
\end{equation}
for two possible mirror pairs, in each case using the traceless $\ZZ_3$-action indicated in~\eqref{e:3F3-1x} and~\eqref{e:3F3-1y}, respectively.
Finally, notice the factor ``3'' in the relation
$d(\Red_{3,5}[\pDN{\FF[3]{3}}])=3d(\pDs{\IP^3_{(8:8:5:3)}})$, correlating to the order of the $\ZZ_3$ group in~\eqref{e:3F3mm1}, which was called $H$ in Ref.~\cite{rBH}.

\paragraph{Mirror-Pair \#2:}
On the other hand,
\begin{subequations}
\label{e:nK3min2}
\begin{alignat}9
 \big(\Red_{1;4,5} g(y)\big)^\sfT=\Red_{1;4,5} f(x;\FF[3]{3})
 &=&\,
 &a_1\,x_4\!^8 +a_2\,x_5\!^8 
   +a_3\,\frac{x_2\!^3}{x_4}
   +a_6\,\frac{x_3\!^3}{x_5}\quad
 &\in&~\IP^3_{(3:3:1:1)}[8],
 \label{e:3F3min2x}\\
 \big(\Red_{1;4,5} f(x)\big)^\sfT=\Red_{1;4,5} g(y;\MF[3]{3})
 &=&
 &b_2\,y_3\!^3
  +b_3\,y_6\!^3 
  +b_4\,\frac{y_1\!^8}{y_3} +b_5\,\frac{y_2\!^8}{y_6}\quad
 &\in&~\IP^3_{(1:1:2:2)}[6],
 \label{e:3F3min2y}
\end{alignat}
\end{subequations}
corresponds to
 $\Conv\big(\pDs{\FF[3]{3}}\ssm\{\n_1\}\big)$ and
 $\Conv\big(\pDN{\FF[3]{3}}\ssm\{\m_4,\m_5\}\big)$.
For a generic choice of the coefficients, the polynomials~\eqref{e:nK3min2} are again $\D$-regular, and are each other's transpose. The so-reduced matrix of exponents is again regular:
\begin{equation}
  \eqref{e:3F3min2x}~\&~\eqref{e:3F3min2y}\To\quad
  \Red_{1;4,5}\IE(\FF[3]{3})
   =\BM{ 0& 0&~\>8&~\>0\\ 0& 0&~\>0&~\>8\\ 3& 0&-1&~\>0\\ 0& 3&~\>0&-1\\},\quad
  \big(\Red_{1;4,5}\IE(\FF[3]{3})\big)^{-1}
   =\BM{\frac1{24}& 0&\frac13&0\\[0pt] 0&\frac1{24}& 0&\frac13\\[0pt]
         \frac18& 0& 0& 0\\[0pt] 0&\frac18& 0& 0\\},
 \label{e:E3F3min2}
\end{equation}
and the discrete symmetries of the polynomials~\eqref{e:nK3min1} are read off from the inverse matrix:
\begin{alignat}9
  a_1\,x_4\!^8 +a_2\,x_5\!^8 
   +a_4\,\frac{x_2\!^3}{x_5} +a_5\,\frac{x_3\!^3}{x_4}:
 &~~\ARR\{{@{}r@{}}
         {\zZ3{\frac13,\frac13,0,0}\\
          \zZ{24}{\frac1{24},\frac{23}{24},\frac18,\frac78}\\ \midrule\nGlu{-3pt}
          \zZ8{\frac38,\frac38,\frac18,\frac18}}.
  \bM{\ttt x_2\\[2pt]\ttt x_3\\[2pt]\ttt x_4\\[2pt]\ttt x_5}:&
 &~~\ARR\{{@{}r@{\,=\,}l}
          {\cG&\ZZ_3\<\times\ZZ_{24},\\[-1pt] \midrule\nGlu{-2pt}
          \cQ&\ZZ_8.}. \label{e:3F3-2x}\\
  b_2\,y_4\!^3 +b_3\,y_5\!^3 
   +b_4\,\frac{y_1\!^8}{y_5} +b_5\,\frac{y_2\!^8}{y_4}:
 &~~\ARR\{{@{}r@{}}
         {\zZ4{\frac14,\frac14,0,0}\\
          \zZ{24}{\frac1{24},\frac{23}{24},\frac13,\frac23}\\ \midrule\nGlu{-3pt}
          \zZ6{\frac16,\frac16,\frac13,\frac13}}.
  \bM{\ttt y_1\\[2pt]\ttt y_2\\[2pt]\ttt y_3\\[2pt]\ttt y_6}:&
 &~~\ARR\{{@{}r@{\,=\,}l}
          {\cG^\wtd&\ZZ_4\times\ZZ_{24},\\[-1pt] \midrule\nGlu{-2pt}
          \cQ^\wtd&\ZZ_6.}. \label{e:3F3-2y}
\end{alignat}
The desired swap of ``geometric'' and ``quantum'' symmetries can be achieved following~\cite{rBH}: we should consider instead the quotient models
$\eqref{e:3F3-2x}/\ZZ_4$ and $\eqref{e:3F3-2y}/\ZZ_3$ for a mirror pair. To this end, we may use the $\zZ4{\frc14,\frc34,\frc14,\frc34}$ generated by the 6-fold difference between the two leftmost columns,
 and the $\zZ3{\frc13,\frc23,\frc23,\frc13}$ generated by the 8-fold difference between the two topmost rows in $(\Red_{1;4,5}\IE(\FF[3]{3}))^{-1}$. For the so-defined models,
\begin{equation}
  \eqref{e:3F3-2x}/\ZZ_4
  ~\bigg\{\begin{array}{r@{\,=\,}l}
            \Tw{\cG}&\ZZ_3\<\times\ZZ_6,\\[2mm]
            \Tw{\cQ}&\ZZ_8\<\times\ZZ_4;
          \end{array}
  \quad\text{vs.}\quad
          \begin{array}{r@{\,=\,}l}
            \Tw{\cG}^\wtd&\ZZ_4\<\times\ZZ_8,\\[2mm]
            \Tw{\cQ}^\wtd&\ZZ_6\<\times\ZZ_3;
          \end{array}\bigg\}~\eqref{e:3F3-2y}/\ZZ_3.
 \label{e:3F3mm2}
\end{equation}

\paragraph{Fractional Relation:}
Finally,~\eqref{e:3F3min1y} and~\eqref{e:3F3min2y} are, respectively, transposes of~\eqref{e:3F3min1x} and~\eqref{e:3F3min2x}, which are evidently related by deformation --- a variation in the coefficient space of the $a_\tI$'s in~\eqref{e:cs3F3x}. It then follows that~\eqref{e:3F3min1y} and~\eqref{e:3F3min2y} should be related by a corresponding, {\em\/dual\/} transformation, in the $y_\tI$-space. 
Indeed, the requisite ({\em\/constant-Jacobian\/}) fractional change of variables (\`a la~\cite{Lynker:1990vw,rMirr00,Greene:1991iv,rRS-FracTr,rRS-TheMS,rLS-Cfld+MS}) is
\begin{equation}
 \eqref{e:3F3min1y}:~
 \IP^3_{(3{:}5{:}8{:}8)}\ni(y_1,y_2,y_4,y_6)
                   \to \Big(\frac{y_1}{\sqrt[8]{y_6}}, y_2\sqrt[8]{y_6}, y_4, y_6\Big)
                   \mapsto (y_1,y_2,y_4,y_5)\in\IP^3_{(1{:}1{:}2{:}2)}
 ~:\eqref{e:3F3min2y},
 \label{e:FrTr}
\end{equation}
which also turns the $\ZZ_{24}$ $\cQ^\wtd$-action from~\eqref{e:3F3-1y} into the $\ZZ_6$ $\cQ^\wtd$-action in~\eqref{e:3F3-2y}:
\begin{equation}
  \begin{array}{r|cccc}
 {\ddd\Red_{\sss1;3,5}}
  &y_1&y_2&y_4&y_6\\[1pt] \hline\rule{0pt}{2.67ex}
  \ZZ_{24}&\frac5{24}&\frac3{24}&\frac13&\frac13\\
  \end{array} ~~\mapsto~~
  \begin{array}{r|cccc}
  &y_1/\sqrt[8]{y_6}&y_2\sqrt[8]{y_6}&y_4&y_6\\[1pt] \hline\rule{0pt}{2.67ex}
  \ZZ_{24}&\frac{5-1}{24}&\frac{3+1}{24}&\frac13&\frac13\\
  \end{array} ~~\simeq~~
  \begin{array}{r|cccc}
 {\ddd\Red_{\sss1;4,5}}
  &y_1&y_2&y_4&y_5\\[1pt] \hline\rule{0pt}{2.67ex}
  \ZZ_6&\frac16&\frac16&\frac13&\frac13\\
  \end{array}
\end{equation}
Since this assignment involves the $8^\text{th}$ root, the mapping also involves a $\ZZ_8$-orbifold quotient, indicating that the models~\eqref{e:3F3min1y} and~\eqref{e:3F3min2y} are birational to each other: they are so-called ``multiple mirrors''~\cite{rS-BirBHK,rK-BHK,rPC-birBHK,rPC-birLG}.

 We take this as further evidence that the wide selection of $K3$ surfaces one can define with the pair of polynomials~\eqref{e:3Fmcs}, their deformations and after requisite complementary finite quotients as in~\eqref{e:3F3mm2}, and so ultimately with the transpolar pair of polytopes
 $\pDs{\FF[3]{3}}$ and $\pDN{\FF[3]{3}}\<=\pDs{\MF[3]{3}}$,
indeed form mirror pairs --- after appropriate MPCP-desingularization encoded by the unit star-subdivisions.
 In turn, as shown in \SS\,\ref{s:HH}, this collection stems merely from the central member of the deformation family $\ssK[{r||c}{\IP^3&1\\ \IP^1&3}]$, which leaves many other open avenues for further exploration; see Figure~\ref{f:disDef}.

\paragraph{Tyurin Degenerations, Again:}
The straightforward transposition mirror pair of polynomials, such as explicitly given in~\eqref{e:3Fmcs} for $\FF[3]m$, include rational monomials for $m\<\geq3$ and are both transverse.

Consider now the limit in which the rational monomials are omitted. For the anticanonical system of $\FF{m}$~\eqref{e:CsK*}, this amounts to omitting the extension of the Newton polytope, $\pDN{\FF{m}}$ (e.g., Figure~\ref{f:*nFmQmu}) reducing it at a (green-outlined) new facet that includes the origin. This regular part of the Newton polytope, $\reg[\pDN{\FF{m}}]$, spans an {\em\/incomplete fan\/} that covers only a half-space; each anticanonical section,
\begin{alignat}9
  \reg[f(x)]
  &=\!\!\!\sum_{\m\in M\cap\,\reg[\pDN{\FF{m}}]}\!\!\!
          a_\mu \Big(\prod_{\n_i\smt\pDs{\FF{m}}} x_i^{~\vev{\n_i,\m}+1}\Big),
 \qquad \reg[\pDN{\FF{m}}]\<=\big(\!\Conv[\pDs{\FF{m}}]\big){}^\circ,
\iText{factorizes, and has a Tyurin degenerate zero locus.
 In turn, direct computation shows that the regular part of the transpose of the full Laurent polynomial,}
  \reg\Big[f(x)^\sfT&=\sum_{\n\in N\cap\,\pDs{\FF{m}}}
          b_\nu \Big(\prod_{\m_i\smt\pDN{\FF{m}}} y_i^{~\vev{\m_i,\n}+1}\Big)
          =g(y)\Big],
\iText{does not factorize but fails to be transverse only at isolated points.
 Finally, the transpose of $\reg[f(x)]$,}
  \big(\!\reg[f(x)]\big){}^\sfT
  &=\sum_{\n\in N\cap\,\pDs{\FF{m}}}
          b_\nu \Big(\prod_{\m'_i\smt\reg[\pDN{\FF{m}}]}\h_i^{~\vev{\m'_i,\n}+1}\Big)
          =g(\h),
\end{alignat}
consists of the same sections as the complete Laurent $g(y)$, but is re-expressed in terms of the Cox variables $\h$, defined by the vertices of $\reg[\pDN{\FF{m}}]$. This $g(\h)$ does not factorize either, but fails to be transverse at a 1-dimensional curve. The increase in the singularity,
\begin{equation}
   \dim\big[\,\big\{\rd\reg\big[f(x)^\sfT\<=g(y)\big]\<=0\big\}\,\big]\<=0  ~~\leadsto~~
   \dim\big[\,\big\{\rd\big(\reg[f(x)]^\sfT\<=g(\h)\big)\<=0\big\}\,\big]\<=1,
 \label{e:0>1}
\end{equation}
thus seems to stem from the use of Cox variables that correspond to the incomplete fan
 $\S'\<\smt\reg[\pDN{\FF{m}}]$.

For the various deformations of the central Hirzebruch scroll, such as discussed in \SS\,\ref{s:DscDef}, the regular part of the anticanonical systems is less singular. For example,
both the generic $\reg[\pDN{\FF{\!\sss(3,2,\cdots)}}]$-sections and their transposes are transverse. Here, Tyurin degenerations both in the ``original'' and in the transpose mirror model are smoothable by regular sections.
 In turn, the generic $\reg[\pDN{\FF{\!\sss(4,1,\cdots)}}]$-sections have isolated singular points, while their transposes are transverse. Here, Tyurin degenerations of the transpose mirror model is fully smoothable by regular sections, while the singularity of the ``original'' may be reduced by regular sections to isolated singular points, but needs rational sections for full smoothing.

\subsection{Algebro-Geometric Avenues}
\label{s:Cox}
The discussion of \SS\,\ref{s:LdL'H} involved the notion of the {\em\/intrinsic limit,} which is ostensibly not part of the standard tool-set in algebraic geometry. Although Procedure~\ref{P:BuLBd} provides an alternative formulation that involves decidedly more familiar algebro-geometric operations, it seems worth indicating two more alternative formulations of these Laurent models.

\paragraph{Weil Divisors:}
Consider the $a_i\<\to1$ special case of the particular Laurent polynomial~\eqref{e:P3115}:
\begin{equation}
 f_{\sss\!R}(x) =x_3\!^5 +x_4\!^5 +\frac{x_2\!^2}{x_4}
                =\frac{x_3\!^5x_4 +x_4\!^6 +x_2\!^2}{x_4},
 \qquad \in\IP^2_{(3{:}1{:}1)}[5].
 \label{e:3115}
\end{equation}
The factorization of the regular anticanonical sections~\eqref{e:q=cs} of $\FF{m}$ for $m\<\geqslant3$ implies that their zero-locus~\eqref{e:nXm} reduces to a union of two hypersurfaces, i.e., to a sum, $[\Fc^{-1}(0)]+[\Fs^{-1}(0)]$, of the corresponding divisors.
The formal factorization~\eqref{e:3115} analogously corresponds to\ftn{We thank Amin Gholampour for alerting us to this formulation and avenue for further study.}:
\begin{subequations}
 \label{e:DivRef}
\begin{equation}
 \big[f_{\sss R}^{-1}(0)\big]
  \<=\big[\big\{\Fn(x)/\Fd(x)\<=0\big\}\big]
   \<=\big[\Fn^{-1}(0)\big]{-}\big[\Fd^{-1}(0)\big],
 \label{e:5=6-1}
\end{equation}
where\vspace*{-3mm}
\begin{alignat}9
 \Fn(x)      &\coeq x_3\!^5x_4 +x_4\!^6 +x_2\!^2,&\qquad
 \Fn^{-1}(0) &\in\IP^2_{(3{:}1{:}1)}[6], \label{e:N6}\\
 \Fd(x)      &\coeq x_4,&\qquad
 \Fd^{-1}(0) &\in\IP^2_{(3{:}1{:}1)}[1] \label{e:D1}
\end{alignat}
\end{subequations}
are the (sextic) numerator and (linear) denominator divisors in $\IP^2_{(3{:}1{:}1)}$.

Formal integral {\em\/differences\/} of divisors (in this case, zero-loci of otherwise regular sections) such as~\eqref{e:5=6-1} are {\em\/Weil divisors\/}~\cite{rF-TV,rGE-CCAG,rCLS-TV}, introduced a century ago as {\em\/virtual varieties\/} by F.~Severi~\cite{rSeveri-VirtVar}. These then provide a standard algebro-geometric framework for the Calabi-Yau zero-locus $f_{\sss R}^{-1}(0)$ --- and in fact all the Laurent-deformed codimension-1 Calabi-Yau models of in non-Fano varieties~\cite{rBH-gB}.

Owing to their respective degrees, $\Fn(x)$ and $\Fd(x)$ differ significantly:
 Regular sextic polynomials such as $\Fn(x)$ on $\IP^2_{(3{:}1{:}1)}$ freely involve all three quasi-homogeneous coordinates and are sections of the {\em\/line bundle\/} $\cO(6)$.
 By contrast, regular linear polynomials such as the denominator, $\Fd(x)$, on
 $\IP^2_{(3{:}1{:}1)}$ can involve only the latter two quasi-homogeneous coordinates, $x_3,x_4$, and are sections of the {\em\/sheaf\/} $\cO(1)$. This then implies that $f_{\sss R}(x)$ also is a section of the $\cO(5)$ {\em\/sheaf\/} on $\IP^2_{(3{:}1{:}1)}$.
 For most of the physics-motivated computations the precise distinction between sheaves and bundles does not seem to matter in the intended physics applications~\cite{rBeast}, but the distinction exists and may well be worth a formally more rigorous analysis.

 This reformulation~\eqref{e:5=6-1} of the Laurent model~\eqref{e:3115} as the formal {\em\/difference\/} of two regular divisors~\eqref{e:DivRef} then opens another avenue of studying such subspaces of well-understood ``ambient'' spaces, $X\<\subset A$. Of particular interest are of course methods for computing their numerical characteristics such as the Euler number, Hodge numbers, and then also (topological) intersection numbers of their own divisors and other subspaces, i.e., various Yukawa couplings. It is worth noticing that the contributions of the denominator divisor, $[\Fd^{-1}(0)]$, to such numerical characteristics typically {\em\/subtract\/} from those of the numerator divisor, $[\Fn^{-1}(0)]$ --- and this resonates with the overall structure of results in~\cite{rBH-gB}.

\paragraph{Fractional Mapping:}
Having already seen fractional coordinate changes such as~\eqref{e:FrTr} above, it is perhaps no surprise that another fractional coordinate change may simplify --- indeed, {\em\/regularize\/}\ftn{We are grateful to David Cox for providing this proof-of-concept example and Hal Schenck for communicating it to us.} --- the Laurent defining equation~\eqref{e:P3115}:
\begin{subequations}
 \label{e:DC}
\begin{alignat}9
  f(x)&= x_3\!^5+x_4\!^5+\frac{x_2\!^2}{x_4}
  &(x_2,x_3,x_4)&\in\IP^2_{(3{:}1{:}1)}[5] \tag{\ref{e:P3115}}\\*
  \vC{\TikZ{\path[use as bounding box](0,-.3)--(0,.2);
            \draw[thick, densely dashed,|->](0,.4)--++(0,-.9);
            }}~~&(x_2,x_3,x_4)\<\mapsto(z_3\sqrt{z_2},z_1\!^2,z_2)&
  &\vC{\TikZ{\path[use as bounding box](0,-.3)--(0,.2);
            \draw[thick, densely dashed,<-](.6,.4)--++(0,-.9);
            }} \label{e:DCmap}\\*[3pt]
  h(z)&= z_1\!^{10}+z_2\!^5+z_3\!^2
  &\qquad(z_1,z_2,z_3)&\in\IP^2_{(1{:}2{:}5)}[10]. \label{e:P12510}
 \end{alignat}
\end{subequations}
This effectively re-renders the Laurent model as a regular algebraic variety --- at a price: The indicated mapping\ftn{\label{fn:bck}The direction of the left-hand side dashed arrows in~\eqref{e:DC} and~\eqref{e:DC'} follows the coordinate assignment, which is dual --- and so opposite of the direction of the mapping between the underlying spaces, shown on the right-hand side.} involves $\sqrt{z_2}$ and so a double cover that is branched over the $z_2\<=0$ locus; it also maps $x_1\mapsto z_1\!^2$ and so involves a $\ZZ_2$ quotient (with respect to $z_1\<\to{-}z_1$) with the $z_1\<=0$ fixed-point set. 
 This then relates the Calabi-Yau Laurent model $\IP^2_{(1{:}1{:}3)}[5]$ with
 the decic hypersurface $\IP^2_{(1{:}2{:}5)}[10]$ --- a variety of {\em\/the general-type\/} ($c_1\<<0$) --- via this $\ZZ_2$-quotient of a branched double covering map; the reverse mapping is no simpler. While this makes the analysis fairly convoluted (cf.~\cite[\SS\,5.4--5.5]{rBeast}) it does show that the original Laurent model $\IP^2_{(1{:}1{:}3)}[5]$ is closely related to a regular algebraic variety, and in a way that involves standard and more familiar algebro-geometric operations.

The mapping~\eqref{e:DC} might appear to be a fortuitous fluke. However, it does have a generalization that is applicable not only to the (rationally extended) anticanonical system of $\IP^2_{(3{:}1{:}1)}$, but in fact to the entire anticanonical system of $\FF[2]3$ from whence~\eqref{e:P3115} originally stems, and also extends straightforwardly to higher-twisted Hirzebruch scrolls. Using the toric rendition of these monomials, we have:
\begin{subequations}
 \label{e:DC'}
\begin{alignat}9
  f(x)&= x_1\!^2 (x_3\oplus x_4)^5 ~\oplus~ x_1\,x_2(x_3\oplus x_4)^2 ~\oplus~
         x_2\!^2 \Big(\frac1{x_4}\<\oplus \frac1{x_3}\Big)
  &&~~\FF[2]3\big[\mM{~\,2\\\!-1}\big]\<=
       \ssK[{r||c|c}{\IP^2&1&2\\ \IP^1&3&-1}]\\*
  \vC{\TikZ{\path[use as bounding box](0,-.3)--(0,.2);
            \draw[thick, densely dashed,|->](0,.4)--++(0,-.9);
            }}~~&\!\! (x_1,\, x_2,\, x_3,\, x_4)  \mapsto
       (z_1\sqrt{z_3},\, z_2\sqrt{z_3}\,z_4^3,\, z_3,\, z_4)
  &&\qquad\vC{\TikZ{\path[use as bounding box](0,-.3)--(0,.2);
            \draw[thick, densely dashed,<-](0,.4)--++(0,-.9);
            }} \label{e:DC'map}\\*[3pt]
  h(z)&= z_1\!^2\,\big(z_3(z_3\oplus z_4)^5\big) ~\oplus~
         z_1\,z_2\,\big(z_3(z_3\oplus z_4)^2\,z_4\!^3\big) ~\oplus~
         z_2\!^2 \,\big((z_3\oplus z_4)\,z_4\!^5\big)
  &&~~\FF[2]0\big[\mM{2\\6}\big]\<=\ssK[{r||c}{\IP^1&2\\ \IP^1&6}], \label{e:F026}\\*
  &\deg(x_1,x_2,x_3,x_4)\<=\big(\mM{~~1\\-3},\mM{1\\0},\mM{0\\1},\mM{0\\1}\big),\quad
   \deg(z_1,z_2,z_3,z_4)\<=\big(\mM{1\\0},\mM{1\\0},\mM{0\\1},\mM{0\\1}\big)
\end{alignat}
\end{subequations}
While it looks a little more involved than~\eqref{e:DC}, the (dash-arrow) mapping~\eqref{e:DC'} now also includes 
 ({\small\bf1})~the hallmark directrix $\{x_1\<=0\}\<\subset\FF[2]3$, and
 ({\small\bf2})~the ``untwisted'' Hirzebruch scroll, $\FF[2]0$ --- all while staying at the same level of conceptual complexity in mapping the Calabi-Yau subspace:
 The (presumably desingularized) finite quotient of a branched multiple cover mapping~\eqref{e:DC'} relates
  the Calabi-Yau hypersurface (2-torus) $\XX[1]3\<\subset\FF[2]3$ in the non-Fano Hirzebruch scroll $\FF[2]3$ (a 3-twisted $\IP^1$-bundle over $\IP^1$) to
  a regular degree-$\pM{2\\6}$ hypersurface in $\FF[2]0\<=\IP^1{\times}\IP^1$ (the ``untwisted'' plain product), which is of (semi-)general type: $c_1\<=0$ over one $\IP^1$-factor, but $c_1\<<0$ over the other.
 
This ``un/twisting'' fractional mapping (see footnote~\ref{fn:bck}) between $\FF[2]0$ and $\FF[2]3$ is most definitely {\em\/not\/} the classical diffeomorphism, $\FF{m}\<{\approx_{\sss\IR}}\FF{m\smash{\pMod{n}}}$. Also, we note that this (dash-arrow) mapping of defining polynomials is injective but most definitely not surjective: only 10 of the 21 monomials from the full deformation family $\ssK[{r||c}{\IP^1&2\\ \IP^1&6}]$ turn up in the image~\eqref{e:F026}. That is, this (dash-arrow) mapping exists only over the special subset in the full deformation family $\ssK[{r||c}{\IP^1&2\\ \IP^1&6}]$, specified by the particular degree-$\pM{2\\6}$ polynomials~\eqref{e:F026}. Qualitatively, this reminds of the situation illustrated in Figure~\ref{f:disDef}.
 While more precise details of such mappings are needed to effectively compute numerical characteristics of Laurent models in general, suffice it here to establish their existence and state their general nature.

\paragraph{Infinite Pools of Constructions:}
The numerator divisor, $\big[\Fn^{-1}(0)\<\in\IP^2_{(3{:}1{:}1)}[6]\big]$ in~\eqref{e:DivRef}, has a negative 1st Chern class, and so is a subvariety of {\em\/general type,} as is the regular variety, $\IP^2_{(1{:}2{:}5)}[10]$, in~\eqref{e:DC}. More generally, such ``regularizing'' mappings involve a variety with a {\em\/partially negative\/} 1st Chern class: for $\FF[2]0\big[\mM{2\\6}\big]\<=\ssK[{r||c}{\IP^1&2\\ \IP^1&6}]$ in~\eqref{e:F026}, $c_1$ vanishes over the first (upper) $\IP^1$-factor and is negative over the second (lower) factor.

 The a priori infinite number of algebraic varieties of general type to serve in such ``regularizing'' mappings correlates with the infinite number of VEX polytopes usable in encoding the Calabi-Yau models of~\cite{rBH-gB}.
 Also, this supports the possibility that the pool of (g)CICYs connects to {\em\/all\/} Calabi-Yau models --- including all toric models~\cite{rKreSka00b}, which resonates with the second part of~\cite{rGHC} (and closing paragraph of~\cite[Ch.\,D]{rBeast}) that discusses conifold transitions to branched multiple covers.

\section{Gauged Linear Sigma Model Aspects}
\label{s:GLSM}
Each Cox variable in a toric model is identified 1--1 with a {\em\/chiral\/} superfield of the the gauged linear sigma model (GLSM)~\cite{rPhases,rMP0}. In turn, each toric, i.e., projective space projectivization transformation corresponds to a {\em\/twisted-chiral,} $U(1;\IC)$-{\em\/gauge superfield.}
 Then, the constant-Jacobian changes of variables such as~\eqref{e:0psxy}, \eqref{e:41psxy}, \eqref{e:32psxy} and~\eqref{e:221psxy} correspond to those same superfield redefinitions. In fact, even the non-constant Jacobian changes of variables~\eqref{e:DC} and~\eqref{e:DC'} nevertheless turn out to provide supersymmetry-preserving mappings of superfields. This provides a direct ``translation'' of the toric computations discussed herein into the GLSM framework.

To this end, we focus on {\em\/worldsheet\/} supersymmetry as needed in the usual application of GLSMs, and note that all superfields are formal power-series such as $\bS\F=\f(\x){+}\q^\a\j_\a(\x){+}\q^2F(\x)\dots$~\cite{r1001,rWB,rBK}. Here, $\x$ denotes the ordinary (bosonic, commuting) coordinates on the worldsheet, and $\q^\a,\bar\q^{\dot\a}$ with $\a\<=1,2,\dots,p$ and $\dot\a\<=1,2,\dots,q$ denote Grassmann (fermionic, anticommuting) coordinates of the $(p,q)$-superspace extension of the worldsheet; routinely, $\q^2\<\coeq\frc12(\q^\a\q^\b{-}\q^\b\q^\a)$, etc. In this $\q,\bar\q$-expansion, coefficient functions of even (vs.\ odd) order, $\f(\x),F(\x),\dots$ (vs.\  $\j_\a(\x),\dots$), have the same (vs.\ opposite) boson/fermion parity as the superfield $\bS\F$ itself. Owing to the nilpotence and anticommutativity of $\q,\bar\q$, all superfields in fact terminate into order-$(p,q)$ polynomials in $\q,\bar\q$.

Focusing now on $(2,2)$-supersymmetry and using the customary labels $\a,\dot\a=-,+$~\cite{rHSS}, the particular class of chiral ($\bS\F$) and twisted-chiral ($\bS\S$) superfields are specified by the 1st-order superdifferential conditions
\begin{equation}
  \bar{D}_{\pm}\bS\F\<=0\qquad\text{and}\qquad \bar{D}_-\bS\S\<=0\<=D_+\bS\S.
\end{equation}
It is immediate that such superfields form a ring under ordinary multiplication and all analytic functions of superfields are also superfields. Moreover, even division by a superfield is well defined as long as division by its leading (``lowest-component'') coefficient function, $\f(\x)$, is:
\begin{equation}
  \frac{f(\bS\F_1,\bS\F_2,\dots)}{\bS\F}
  =f(\bS\F_1,\bS\F_2,\dots)\;
   \Big(\frac1{\f} -\big(\q^\pm\j_\pm +\q^2F\big)\frac1{\f^2} +\q^2\j_-\j_+\frac1{\f^3} +\dots\Big).
\end{equation}
Similar $\q,\bar\q$-expansions are just as well defined for fractional powers\ftn{In fact, logarithms of superfields are quite commonplace in this type of analysis also in the original works~\cite{rPhases,rMP0}, and are defined by analogous $\q,\bar\q$-expansions.}, such as needed in~\eqref{e:DC} and~\eqref{e:DC'}. Since chiral superfields in a GLSM are assigned to the Cox variables in the corresponding toric model, the various rational expressions involving chiral superfields are well defined as long as their Cox variable counterparts are. As discussed in \SS\,\ref{s:LdL'H}, this is true in all the cases of interest here.

Finally, it remains to ascertain that the Laurent polynomials of chiral superfields as used herein are themselves chiral superfields --- as required of the superpotential in the GLSM. In particular, the superpotentials of interest are all of the general form $W(\bS{X})=\bS{X}_0\,f(\bS{X}_i)$ for $i\<=1,2,\dots$, as we check:
\begin{subequations}
 \label{e:DbW=0}
\begin{alignat}9
 \bar{D}_\pm\big(\bS{X}_0\,f(\bS{X}_i)\big)
 &=\big(\underbrace{\bar{D}_\pm\bS{X}_0}_{=\,0}\big)\,f(\bS{X}_i) +
   \bS{X}_0\sum_{i>0}\pd{f}{\bS{X}_i}\,\big(\underbrace{\bar{D}_\pm\bS{X}_i}_{=\,0}\big),\\
 &=0 \quad\text{precisely if}\quad
   \big|f(\bS{X}_i)\big|,\, \Big|\bS{X}_0\pd{f}{\bS{X}_i}\Big|<\infty,
 \label{e:DbW=0c}
\end{alignat}
\end{subequations}
whatever the functional form of $f(\bS{X}_i)$. The analogous expansion checks that the so-called twisted superpotential itself remains a twisted chiral superfield.

Now, every {\em\/quantum\/} field can be expanded about a given {\em\/vacuum expectation value\/} (vev), which is also known as the {\em\/background field expansion.} The vev of every fermionic component field in every superfield must vanish to preserve Lorentz symmetry, and the vev of every auxiliary field must vanish to preserve supersymmetry~\cite{r1001,rWB,rBK}. It then follows that the vev of any chiral (and also twisted-chiral) superfield reduces to the vev of its lowest component field, $\vev{\bS\F}=\vev{\f}$ --- which for every ``$\q,\bar\q$-expandable'' function $f(\bS{X}_i)$ is the {\em\/value\/} of that function of the corresponding Cox variable, $f(X_i)$.

Thus, as long as the vevs of the lowest components of the superfield expressions appearing in the condition of~\eqref{e:DbW=0c} are finite, the superpotential is indeed a chiral superfield. Reduced to the lowest components in the $\q,\bar\q$-expansion, these expressions include precisely the defining (regular or Laurent) section, $f(X_i)$, its gradient components, $\pd{f}{X_i}$, and the Lagrange multiplier-like field, $X_0$, interpretable as the fibre coordinate of the canonical bundle. As discussed in \SS\,\ref{s:LdL'H}, all of these quantities are required to remain finite in all the toric models considered herein, thus verifying that the GLSM superpotential (as well as the twisted superpotential) remain chiral (twisted-chiral) superfields. This then guarantees the various by now standard {\em\/non-renormalization\/} arguments, insuring that the usual computational framework of supersymmetric GLSMs remains valid.

Finally, the various choices of the Mori vectors as discussed in \SS\,\ref{s:HH} correspond in the GLSM model to specific choices of generators for the $U(1;\IC)\<\times U(1;\IC)$ gauge symmetry. As discussed in~\cite{rMP0} and traced in full detail in~\cite{rBH-gB}, the particular choices are distinguished by leaving some of the Cox variables invariant, and so allowing them to acquire nonzero expectation values. The appearance of at least one neutral Cox variable in every such assignment of the $U(1;\IC)\<\times U(1;\IC)$ charges precisely reflects one of the requirements in the determination of candidate Mori vectors~\cite{rBKK-tvMirr}.
 This then reinterprets the secondary fan as encoding the {\em\/phases\/} of the GLSM and its possible phase transitions~\cite{rPhases,rMP0}. While illustrated in~\eqref{e:nFmQnu}, \eqref{e:*2F3Qmu} and~\eqref{e:2F3SN} for the simplest cases\ftn{The Calabi-Yau hypersurfaces in all $n\<=2$ cases are of course 2-tori, which exceptionally have a single K{\"a}hler class, and for which the shown 2-dimensional secondary fans collapse to a 1-dimensional one.}, this {\em\/semiclassical\/} characteristic of GLSMs is just as computable for all $n\<\geqslant2$ and all $m\<\in\ZZ$ and all their discrete deformations discussed in~\SS\,\ref{s:DscDef}. These {\em\/semiclassical\/} phase diagrams exhibit a detailed $m$-dependence --- unreduced by the Wall isomorphism, $m\<\simeq m\pMod{n}$, and so indicate an unreduced sequence of novel Calabi-Yau GLSM models.

\section{Concluding Remarks}
\label{s:Coda}
Every {\em\/configuration\/} of (generalized) complete intersections in products of projective spaces~\cite{rH-CY0,rCYCI1,rBeast} represents a continuous deformation family of multi-projective complete intersections. We have shown herein that even the very simplest~\eqref{e:bPnFmX} among their generalizations~\cite{rgCICY1,rBH-Fm,rJL-gCIShv,rGG-gCI} contain many discretely related toric models. 
 It is tempting and at least logically possible that the pool of gCICYs in fact includes mamy (if not all) toric constructions as variously sub-generic and even singular models as well as their various smoothings --- including the (infinitely many) VEX models with Laurent deformations~\cite{rBH-gB}, far exceeding the already immense database~\cite{rKreSka00b}.
 Of course, toric constructions tend to be computationally more approachable, partly because that is where much of the recent computer-aided technology has been developed. We should like to hope that the explicit relations of the kind explored herein will provide for a sinergy between these different approaches, for the benefit of all, including the somewhat more familiar albeit also more involved renditions discussed in \SS\,\ref{s:Cox}.

The inclusion of singular models such as the Tyurin degenerations (\SS\,\ref{s:CY-CY}) also raises an issue, a resolution of which will require further study:
 Standard methods of cohomology computations on $\FF{m}[c_1]$ for $m\<\geqslant3$ are ambiguous: singular spaces admit even different {\em\/notions\/} of cohomology, e.g.,~\cite{rFKJW,rB-IS+ST}, with no a priori obvious preference from string theory in these circumstances; see however~\cite{rFrGaZu86,Atiyah:1989ty,rBatyBor2,rSSC,rAR01,rD+G-DBrM}.
 The reduction $\XX{m}\<=C_m\<\cup S_m$ with $\Sing\XX{m}\<=C_m\<\cap S_m$ being very much akin to the so-called ``infinite complex structure'' limiting form of the Dwork pencil of quintics~\cite{rMirr00} suggests that Tyurin-degenerate models have an application in string compactifications, the details of which we defer to a future study.

\paragraph{Acknowledgments:}
 We would like to thank Lara Anderson, Charles Doran, Amin Gholampour, James Gray, Vishnu Jejjala, Ilarion Melnikov, Challenger Mishra, Dami\'an Kaloni Mayorga Pe{\~n}a, Hal Schenck, Weikun Wang and Richard Wentworth
 for helpful discussions on the topics discussed in this article.
 PB would like to thank the CERN Theory Group for their hospitality over the past several years. The work of PB is supported in part by the Department of Energy 
grant DE-SC0020220.
 TH is grateful to the Department of Physics, University of Maryland, College Park MD,  and the Physics Department of the Faculty of Natural Sciences of the University of Novi Sad, Serbia, for the recurring hospitality and resources.

\appendix
\section{Holomorphic Distinctions}
\label{s:holoD}
Besides the directrix, the Hirzebruch scrolls~\eqref{e:bPnFm} also exhibit both an $m$-dependent number of exceptional anticanonical sections, $H^0(\FF{m},\cK^*)$, and also an $m$-dependent number of exceptional local reparametrizations, $H^0(\FF{m},T)$ --- exactly matching the number of local deformations of the complex structure, $H^1(\FF{m},T)$. We discuss these in turn, and then also the quasi-Fano components in Tyurin degeneration.

\subsection{Exceptional Anticanonical Sections}
\label{s:xsK*}
While there exist exceptional anticanonical sections in $H^0\big(\FF{m;0},\cK^*\big)$ for all $n\<\geqslant2$ and $m\<\geqslant4$~\cite{rBH-Fm}, for notational simplicity we illustrate this here with the lowest-$n$, lowest-$m$ non-trivial example. Consider the simple deformation of Hirzebruch's original hypersurface~\cite{rH-Fm}:
\begin{subequations}
 \label{e:2F4P21e}
\begin{alignat}9
  \FF[2]{4;\e}&=\big\{(x,y)\in\IP^2\<\times\IP^1:~
                   p(x,y):=x_0\,y_0\!^4+x_1\,y_1\!^4+\e\,x_2\,y_0\!^2\,y_1\!^2=0\big\},
\iText{so that $p(x,y)\<=p_{a(ijkl)}x_a\,y_iy_jy_ky_l$ simplifies:}
  p(x,y)&:~~ p_{0(0000)}=1,\quad p_{1(1111)}=1,\quad p_{2(0011)}=\e
\end{alignat}
\end{subequations}
The anticanonical sections are determined by the Koszul resolution of
$\cK^*\<=\cO\pM{\3-1\\-2}\big|_{\FF[2]{4;\e}}$, where we stack the cohomology groups underneath the corresponding bundles and write the explicit tensor coefficients~\cite{rBeast}:
\begin{equation}
~~\begin{array}{@{}c@{~}c@{~}c@{~}c@{~}l@{}}
 \cO\pM{\3-1\\-6\\} &\hTc\too{p}
  &\cO\pM{\3-2\\-2\\} &\6{\r_\fF}\onto &\cK^*\<=\cO\pM{\3-1\\-2\\}\big|_{\FF[2]{4;\e}} \\
   \toprule
 0 & &0 & &H^0(\FF[2]{4;\e},\cK^*) \\
 \snake{.1}{.11}{6.9}{.075}{\rd}
  \{\vf^{i(jk_1\cdots k_4)}_a\} &\too{p}
  &\{\ve^{ij}\f_{(ab)}\} &\onto &H^1(\FF[2]{4;\e},\cK^*) \\
 0 & &0 & &H^2(\FF[2]{4;\e},\cK^*)\<=0 \\
 0 & &0 & &\text{---} \\
  \end{array}
 \quad
 \begin{array}{r@{\,\sim\,}r}
 H^0(\FF[2]{2;\e},\cK^*)
  &\ker\big[\{\underbrace{\vf^{i(jk_1\cdots k_4)}_a}_{\dim\,=\,15}\} \too{p}
              \{\underbrace{\ve^{ij}\f_{(ab)}}_{\dim\,=\,6}\}\big],\\[2mm]
 H^1(\FF[2]{2;\e},\cK^*)
  &\coker\big[\{\overbrace{\vf^{i(jk_1\cdots k_4)}_a}\} \too{p}
              \{\overbrace{\ve^{ij}\f_{(ab)}}\}\big],
 \end{array}
 \label{e:2F4K*}
\end{equation}
where $\vf^{i(jk_1\cdots k_4)}\<\approx \ve^{i(j}\vf^{k_1\cdots k_4)}$ is totally symmetric in $(jk_1\cdots k_4)$, but vanishes on total symmetrization of all indices, $\vf^{(i(jk_1\cdots k_4))}\<=0$.

With the choice~\eqref{e:2F4P21e}, the kernel of the $p$-mapping in~\eqref{e:2F4K*},
 $\vf^{i(jk_1\cdots k_4)}_{(a}\,p_{b)(k_1\cdots k_4)}\mapsto \ve^{ij}\f_{(ab)}$,
is spanned by the non-zero solutions of the system
\begin{equation}
 \MM{ \big\{\vf^{i(jk_1\cdots k_4)}_{(a}\,p^\9_{b)(k_1\cdots k_4)}=0\big\} \\
      \SSS \vf^{i(jk_1\cdots k_4)} \<\approx \ve^{i(j}\vf^{k_1\cdots k_4)} }~=~
 \left\{
 \begin{array}{r@{\,}lr@{\,}lr@{\,}l}
        \vf_1^{1111}             &=0, &     \vf_1^{0000}+\vf_0^{1111}&=0, &    \vf_0^{0000}&=0 \\
   6\e\,\vf_1^{0011}+\vf_2^{1111}&=0, &6\e\,\vf_0^{0011}+\vf_2^{0000}&=0, &\e\,\vf_2^{0011}&=0 \\
 \end{array}\right\}
 \label{e:vfijkl}
\end{equation}
 When $\e\<\neq0$, all six equations~\eqref{e:vfijkl} constrain, and $H^0(\FF[2]{2;\e},\cK^*)$ is spanned by the $\dim\ker(p)=15{-}6=9$ coefficients:
\begin{equation}
  \vf_1^{0000},~~ \vf_0^{0011},~~ \vf_1^{0011}, \quad\text{and}\quad 
  \vf_a^{0001},~~ \vf_a^{0111} ~~\text{for}~~a=0,\,1,\,2.
\end{equation}
However, on Hirzebruch's original hypersurface $\e\<=0$, the very last of the six equations~\eqref{e:vfijkl} is vacuous, and $\vf_2^{0011}$ is additionally left unconstrained. This leaves now a total of ten free coefficients
to parametrize anticanonical sections via the general formula~\cite{rBH-Fm},
\begin{equation}
  q(x,y)=\vf_{(a}^{i(j_1{\cdots}j_5)}\,p^{\9}_{b)(i\,j_3{\cdots}j_5)}\,
   \frac{x^a\,x^b}{h^{(j_1j_2)}(y)},
\end{equation}
and the exceptional contribution parametrized by $\vf_2^{0011}$ is no different.
 This not only confirms the counting based on $\FF[2]4\coeq\IP\big(\cO\<\oplus\cO(4)\big)$ provided in~\cite{rBH-Fm}, but explicitly constructs this exceptional anticanonical section.
 The by now standard argument~\cite{rgCICY1,rBH-Fm} as well as the general scheme-theoretic framework~\cite{rGG-gCI,rJL-gCIShv} verify that all so-constructed exceptional anticanonical sections are also holomorphic on $\FF[2]4$.

In turn, the cokernel of the $p$-mapping in~\eqref{e:2F4K*},
 $\{\ve^{ij}\f_{(ab)}
     \pMod{\ve^{i(j}_{\9}\vf^{k_1\cdots k_4)}_{(a}\,p^{\9}_{b)(k_1\cdots k_4)}}\}$
has all $\binom{2+2}2\<=6$ parameters $\f_{(ab)}$ gauged away if $\e\<\neq0$, leaving nothing to parametrize $H^1(\FF[2]{2;\e},\cK^*)=0$.
 In particular, $\ve^{ij}\f_{(22)}$ is gauged away by
 $\ve^{i(j}_{\9}\vf^{k_1\cdots k_4)}_{(2}\,p^{\9}_{2)(k_1\cdots k_4)}
  =\e\,\ve^{i(j}_{\9}\vf^{0011)}_2$, since $p_{2(k_1\cdots k_2)}=p_{2(0011)}=\e$.
However, on Hirzebruch's original hypersurface when $\e\<=0$, the contribution
 $\f_{(22)}\mapsto H^1(\FF[2]{2;\e},\cK^*)$ is not gauged away, leaving
 $\dim H^1(\FF[2]{2;\e},\cK^*)=1$.

These computations generalize straightforwardly to all $m\geqslant4$ and $n\geqslant2$, and verify the result~\cite{rBH-Fm} for these sub-generic hypersurfaces:
\begin{subequations}
 \label{e:XSK*nFm}
\begin{equation}
   \dim H^0(\FF{m;\e},\cK^*)
    =3{\ttt\binom{2n-1}{n}} +\hat\d^{\sss(n)}_{m;\e}
   \qquad\text{and}\qquad
   \dim H^1(\FF{m;\e},\cK^*)
    =\hat\d^{\sss(n)}_{m;\e},
 \label{e:ExcK*}
\end{equation}
where the number of exceptional contributions is
\begin{alignat}9
   \hat\d^{\sss(n)}_{m;0}&=\vq_3^m\,{\ttt\binom{2n-2}{2}}(m{-}3),&\quad&\text{for}~~
   \FF{m,0}=\{x_0\,y_0^{\,m}{+}x_1\,y_1^{\,m}\<=0\}
   \in\ssK[{r||c}{\IP^n&1\\\IP^1&m}], \label{e:XSK*nFm0}\\
   \hat\d^{\sss(n)}_{m;\e\neq0}&<\hat\d^{\sss(n)}_{m;0};&\quad&
    \text{for generic cases, }\hat\d^{\sss(n)}_{m;\e\neq0}=0.
\end{alignat}
\end{subequations}
Between Hirzebruch's central, maximally {\em\/non-generic\/} hypersurface~\eqref{e:bPnFm} and the maximally generic deformations~\eqref{e:bPnFmE}, there may well exist intermediately sub-generic hypersurfaces for which the number of exceptional anticanonical sections is also nonzero, depends on the $\e$'s, but does not reach the maximal value $\vq_3^m\,{\ttt\binom{2n-2}{2}}(m{-}3)$.

This ``jumping''~\eqref{e:ExcK*} in the dimensions of $H^*(\FF{m;\e},\cK^*)$ depending on the concrete choice of the defining equation~\eqref{e:bPnFm}--\eqref{e:bPnFmE} illustrates the fact that a deformation family of even simple hypersurfaces such as
 $\ssK[{r||c}{\IP^n&1\\ \IP^1&m}]$ easily contain {\em\/discretely different\/} complex manifolds.

\subsection{Exceptional Local Reparametrizations}
\label{s:xsT}
Another consequence of this complex structure subtlety is the existence of the exceptional local reparamet\-ri\-za\-tions of $\FF{m}$, parametrized by $H^0(\FF{m},T)$. As a holomorphic cohomology computation, this again showcases the subtle dependence on the complex structures.

Suffices it again to consider but the simplest, $n\<=2$ cases, and compute the cohomology groups $H^*(\FF[2]{m;\e},T)$ for deformed hypersurfaces such as~\eqref{e:bPnFmE} and~\eqref{e:2F4P21e}, parametrizing the complex structures of those hypersurfaces $\FF[2]{m;\e}\in\ssK[{r||c}{\IP^2&1\\\IP^1&m}]$. To this end, we use the adjunction formula combined with the Koszul resolution of the restriction to $\FF[2]{m;\e}$ of requisite bundles:
\begin{equation}
  \begin{array}{cc@{}c}
      & T_A\otimes\cO_A\pM{-1\\-m\\} & \cO_A \\*[-1mm]
      & \CW{\into}\crlap{p} & \CW{\into}\crlap{p} \\*
      & T_A & \cO_A\pM{1\\m\\} \\*[-1mm]
      & \CW{\onto}\crlap{\r} & \CW{\onto}\crlap{\r} \\*
    T_{\smash{\FF[2]{m;\e}}}~\into\mkern-20mu
      & T_A|_{\smash{\FF[2]{m;\e}}} &\onto\cO_{\smash{\FF[2]{m;\e}}}\pM{1\\m\\}\\ 
  \end{array}
\quad\text{i.e.}\quad
  \begin{array}{cc@{}c}
      & \pM{0\40&1\\m\40\\}\oplus\pM{~\quad1\40&0\\m-1\41\\}
                                       & \pM{0\40&0\\0\40\\} \\*
      & \CW{\into}\crlap{p} & \CW{\into}\crlap{p} \\*[-1pt]
      & \pM{-1\40&1\\\3-0\40\\}\oplus\pM{\3-0\40&0\\-1\41\\}
                                       & \pM{-1\40&0\\-m\40\\} \\*
      & \CW{\onto}\crlap{\r} & \CW{\onto}\crlap{\r} \\*[-1mm]
    T_{\smash{\FF[2]{m;\e}}}~\into\mkern-20mu
      & T_A|_{\smash{\FF[2]{m;\e}}} &\onto\cO_{\smash{\FF[2]{m;\e}}}\pM{1\\m\\}\\ 
  \end{array}
 \label{e:tTN}
\end{equation}
where ``$(a|b_1b_2)$'' encodes bundles on $\IP^2\<=\frac{U(3)}{U(1){\times}U(2)}$ in terms of $U(1){\times}U(2)$-representations: $a$ is the $U(1)$ charge, and $(b_1b_2)$ encodes the $U(2)$-representation by means of the Young tableau with $b_r$ boxes in the $r^{\text{th}}$ row. Analogously, $\IP^1\<=\frac{U(2)}{U(1){\times}U(1)}$, and ``$\mkern2mu\pM{a\4b_1&b_2\\c\4d\\}$'' encodes bundles on $\IP^2\<\times\IP^1$~\cite{rBE,rMETH,rBeast}.

The central column in~\eqref{e:tTN} produces
\begin{alignat}9
 &\begin{array}{c|c@{}c@{}c@{~}|@{~}c}
    & \pM{0\40&1\\m\40\\}{\oplus}\pM{~\quad\str81\40&0\\m-1\41\\}
    && \pM{-1\40&1\\\3-0\40\\}{\oplus}\pM{\3-0\40&0\\-1\41\\}
       & T_A|_{\FF[2]{m;\e}} \\[2mm]
   \toprule
   0. & \vq_m^0\{\vf^a\}^3_1 &\too{\,p\,}
             & \{\l_b\!^a\}^8_1\oplus\{\k_j\!^i\}^1_3 & H^0(\FF[2]{m;\e},T_A) \\[1mm]
   1. & \snake{.1}{.1}{9.05}{.1}{\rd}
        \vq_2^m\{\vf^{i(j_2\cdots i_m)a}\}^3_{m-1} &
             & 0 & 0 \\
   \vdots & \vdots && \vdots & \vdots \\ 
    \bottomrule
  \end{array}\\
 &H^0(\FF[2]{m;\e},T_A)\sim\{\l_b\!^a/(\vq_m^0\vf^ap_b)\}\oplus\{\k_j\!^i\}\sdpr\{\vq_2^m\vk_j\!^i\}
\end{alignat}
where ``$\SSS\str71|0\,0$'' indicates no cohomology,
 $\vq_a^b\<\coeq\{1~\text{if}~a\<\leqslant b,~0~\text{otherwise}\}$, the ``directed sum'' $A\<\sdpr B=C$ denotes the {\em\/extension of $B$ by $A$,} i.e., abbreviates the exact sequence $A\into C\onto B$, and where
\begin{equation}
  y^j\,\vk_j\!^i\,\pd{}{y^i},\qquad\text{with}\quad
  \vk_j\!^i:=\vf^{i(k_2\cdots k_m)a}\,p_{a(jk_2\cdots k_m)},
 \label{e:vk}
\end{equation}
is the minimal form of this contribution to reparametrizations, constructed by maximally contracting the tensor coefficients\ftn{Iteratively ``un-contracting'' this expression,
 $\vf^{\cdots i)a}p_{\cdots j)}\,\d_i{}^j
  \to\vf^{\cdots i)a}p_{\cdots j)} \frac{y^j}{y^i}\to\text{etc.}$, generalizes the representatives and enables a detailed and complete match with the direct Czech cohomology computations {\em\/\`a la\/} Ref.~\cite{rGG-gCI}.}.
 This gives $\dim H^0(\FF[2]{m;\e},T_A)=(8{-}\vq_m^03)+3+\vq_2^m3(m{-}1)=3m{+}8$.

The right-hand column in~\eqref{e:tTN} produces
\begin{alignat}9
 &\begin{array}{c|c@{}c@{}c@{~}|@{~}c}
    & \pM{0\40&0\\0\40\\}
    && \pM{-1\40&0\\-m\40\\}
       & \cO\pM{1\\m\\}|_{\FF[2]{m;\e}} \\[2mm] \toprule
   0. & \{\vq\}^1_1 &\too{\,p\,}
         & \{\f_{a(i_1,\cdots,i_m)}\}^3_{m+1} & H^0\big(\FF[2]{m;\e},\cO\pM{1\\m\\}\big) \\ 
   1. & 0 &  & 0 & 0 \\[-1mm]
   \vdots & \vdots && \vdots & \vdots \\ 
    \hline
  \end{array}\\
 &H^0\big(\FF[2]{m;\e},\cO\pM{1\\m\\}\big)\sim\{\f_{a(i_1,\cdots,i_m)}/\vq\,p_{a(i_1,\cdots,i_m)}\}_{3m+2}.
\end{alignat}

Finally, the long exact sequence from the bottom-row short exact sequence in~\eqref{e:tTN} reduces to:
\begin{equation}
  H^0(\FF[2]{m;\e},T)\into
  \underbrace{\{\l_b\!^a/(\vq_m^0\vf^ap_b)\}\oplus\{\k_j\!^i\}\sdpr\{\vq_2^m\vk_j\!^i\}}_{3m+8}\too{\rd p}
  \underbrace{\{\f_{a(i_1,\cdots,i_m)}/\vq\,p_{a(i_1,\cdots,i_m)}\}}_{3m+2}\onto
  H^1(\FF[2]{m;\e},T).
 \label{e:nFmeT}
\end{equation}
This leaves $\dim H^0(\FF[2]{m;\e},T)=6+\D_m$ reparametrizations and $\dim H^1(\FF[2]{m;\e},T)=\D_m$ (Kodaira-Spencer) deformations of the complex structure. The quantity $\D_m=\dim(\coker(\rd p))$ measures the corank of the $\rd p$-mapping, i.e., $\D_m=0$ if $\rd p$ is of maximal rank.

As the simplest concrete and non-trivial example, consider the $m\<=2\<=n$ family
\begin{equation}
  \FF[2]{2;\e}\coeq\{p_{\vec\e\,}(x,y)\<=0\} \in \K[{r||c}{\IP^2&1\\\IP^1&2}],\quad
  p_{\vec\e\,}(x,y)\coeq x_0\,y_0\!^2 +x_1\,y_1\!^2 +\e\,x_2\,y_0y_1
 \label{e:2F2pc}
\end{equation}
for which we have:
\begin{equation}
 \begin{array}{r@{~}c@{~}l}
  H^0(\FF[2]{m;\e},T)\into
  \{\l_b\!^a\}\oplus\{\k_j\!^i\}\sdpr\{\vk_j\!^i\}&\too{\rd p}&
  \{\f_{a(ij)}/\vf\,p_{a(ij)}\}\onto H^1(\FF[2]{m;\e},T),\\*[-1pt]
  \WC{\in}\qquad\qquad~&&\quad\WC{\in}\\*[-1pt]
  \big(\l_a\!^b\,p_{b(ij)}+p_{a\,k(i}\,\k_{j)}{}^k+p_{a\,k(i}\,\vk_{j)}{}^k\big)&=:&
   ~\Ht\f_{a(ij)}.\\
 \end{array}
 \label{e:dp}
\end{equation}
With~\eqref{e:2F2pc} and so $p_{0(00)}=1=p_{1(11)}$ and $p_{2(01)}\<=\e$,
 the definition~\eqref{e:vk} of $\vk_j\!^k$ specifies
\begin{equation}
   \vk_0\!^1=-\vf^0,\quad \vk_1\!^0=\vf^1,\quad \text{and}\quad \vk_0\!^0=\e\,\vf^2=-\vk_1\!^1.
\end{equation}
 $H^0(\FF[2]{m;\e},T)=\ker(\rd p)$ is spanned by the variables $\{\l_b\!^a,\k_j\!^i,\vk_j\!^i\}$ omitted in the assignment~\eqref{e:dp} in the target spanned by elements of the equivalence class $[\f_{a(ij)}\simeq\f_{a(ij)}+\vf\,p_{a(ij)}]$:
\begin{equation}
  \l_a\!^b\,p_{b(ij)}+p_{a\,k(i}\,\k_{j)}{}^k+p_{a\,k(i}\,\vk_{j)}{}^k
   ~~\mapsto~~ [\f_{a(ij)}\simeq\f_{a(ij)}+\vf\,p_{a(ij)}].
 \label{e:modpeq0}
\end{equation}
While $\f_{a(ij)}$ has $\binom{1+2}2\binom{2+1}1=9$ components, the $\mathrm{mod}~p_{\vec\e\,}(x,y)$-equivalence class on the right-hand side has $9{-}1\<=8$, so that~\eqref{e:modpeq0} encodes only 8 equations, not 9. With the choice~\eqref{e:2F2pc},
\begin{equation}
   [\f_{a(ij)}\simeq\f_{a(ij)}+\vf\,p_{a(ij)}] \sim
   \Bigr\{\begin{array}{l}
     \f_{0(01)},~~\f_{0(11)},~~\f_{1(00)},~~\f_{1(01)},~~\f_{2(00)},~~\f_{2(11)},\\
     \f_{0(00)}\simeq\f_{0(00)}+\vf,~~
     \f_{1(11)}\simeq\f_{1(11)}+\vf,~~
     \f_{2(01)}\simeq\f_{2(01)}+\e\,\vf\\
     \end{array}\Bigl\}.
\end{equation}
The $\vf\<={-}\f_{0(00)}$ ``gauge'' renders the $\Ht\f_{0(00)}=0$ equation vacuous, and replaces
\begin{equation}
  \Ht\f_{1(11)} \to (\Ht\f_{1(11)}-\Ht\f_{0(00)})\quad\text{and}\quad
  \Ht\f_{2(01)} \to (\Ht\f_{2(01)}-\e\,\Ht\f_{0(00)}),
\end{equation}
turning the system of assignments~\eqref{e:dp} into:
\begin{equation}
 \begin{array}{@{}r|r@{\;}l|r@{\;}l|r@{\;}l@{}}
 \bS{a} &\MC2{c|}{\bS{(ij)\<=(00)}} &\MC2{c|}{\bS{(ij)\<=(01)}} &\MC2c{\bS{(ij)\<=(11)}}\\ \toprule
 0 & \MC2{c|}{\hbox{---}\!\hbox{---}}&
    2\e\,\l_0\!^2{+}(\k_1\!^0{+}\vk_1\!^0)&\mapsto 2\f_{0(01)}&
    \l_0\!^1&\mapsto \f_{0(11)}\\
 1 &\l_1\!^0&\mapsto \f_{1(00)}&
    2\e\,\l_1\!^2{+}(\k_0\!^1{+}\vk_0\!^1)&\mapsto 2\f_{1(01)}&
    \l_0\!^0{-}\l_1\!^1{+}2(\k_0\!^0{+}\vk_0\!^0)&\mapsto \f_{1(11)}\\
 2 &\l_2\!^0{+}\e(\k_0\!^1{+}\vk_0\!^1)&\mapsto \f_{2(00)}&
    {-}\e\big(2\l_0\!^0{+}\l_1\!^1{+}(\k_0\!^0{+}\vk_0\!^0)\big)&\mapsto 2\f_{2(01)}&
    \l_2\!^1{+}\e(\k_1\!^0{+}\vk_1\!^0)&\mapsto \f_{2(11)} \\
 \end{array}
 \label{e:9-1}
\end{equation}
For $\e\<\neq0$, the system is solved by assigning, e.g.:
\begin{subequations}
 \label{e:H*Tsol}
\begin{alignat}9
 \vk_0\!^0&\mapsto\big(\frc1{2\e}\f_{2(01)}{+}\l_1\!^1\big)
           {-}\frc12\f_{1(11)} {-}\k_0\!^0,~~
 \l_0\!^0\mapsto {-}\big(\frc1\e\f_{2(01)}{-}\l_1\!^1\big),\quad~
 \l_0\!^1\mapsto \f _{0(11)},\quad~
 \l_1\!^0\mapsto \f _{1(00)},\\
 \vk_0\!^1&\mapsto2\f _{1(01)} {-}\k _0\!^1 {-}2\e\l_1\!^2,\quad~
 \l_2\!^0\mapsto\f_{2(00)} {+}2\e^2\l_1\!^2 {-}2\e\f_{1(01)},\\
 \vk_1\!^0&\mapsto2\f_{0(01)} {-}\k_1\!^0 {-}2\e\l_0\!^2,\quad~
 \l_2\!^1\mapsto\f_{2(11)} {-}\e(2\f_{0(01)}{+}\frc12\f_{1(11)})
     {+}\frc12\e(\l_0\!^0 {+}\l_1\!^1) {+}2\e^2\l_0\!^2,
\end{alignat}
\end{subequations}
which leaves six linearly independent local reparametrization generator representatives
\begin{equation}
  H^0(\FF[2]{m;\e},T)\ttt
  \sim\Big\{x^1\l_1\!^1\pd{}{x^1},~\>
            x^0\l_0\!^2\pd{}{x^2},~\>
            x^1\l_1\!^2\pd{}{x^2},~\>
            y^0\k_0\!^0\pd{}{y^0},~\>
            y^1\k_1\!^0\pd{}{y^0},~\>
            y^0\k_0\!^1\pd{}{y^1}\Big\}.
 \label{e:2F2eReps}
\end{equation}
At $\e\<\to0$, the (middle-column, bottom-row) $\f_{2(01)}$-assignment becomes vacuous, reducing the system from eight equations to seven, increasing
 $\dim H^0(\FF[2]{m;\e},T)\<=6{\to}7$ and $\dim H^1(\FF[2]{m;\e},T)\<=0{\to}1$.
The $\e\<\to0$ limit of the system~\eqref{e:9-1} is solved by the straightforward $\e\<\to0$ limit of the replacements~\eqref{e:H*Tsol} except that now
\begin{equation}
  \e\<\to0{:}\qquad
  \vk_0\!^0\to{-}\frc1{2}\f_{1(11)} {+}\frc12(\l_1\!^1{-}\l_0\!^0) {-}\k_0\!^0,\qquad
  \l_0\!^0~\text{free},
\end{equation}
adding $x^0\l_0\!^0\pd{}{x^0}$ to $H^0(\FF[2]{m;0},T)$,
and leaving $\f_{2(01)}\<\in H^1(\FF[2]{m;0},T)$ for the central, $\e\<=0$ member of the deformation family, the original Hirzebruch surface.

This explicit (if tedious) construction generalizes to all $m,n\geqslant2$, and produces the above-quoted results~\eqref{e:H*nFmT}, and is in full agreement with the \texttt{SAGE} result for the automorphism group of the toric realization of $\FF{m}$.
 As with $H^*(\FF{m},\cK^*)$, the deformation family~\eqref{e:bPnFmE} may well contain sub-generic hypersurfaces for which the number of exceptional reparametrizations is also nonzero, depends on the $\e$'s, but does not reach the maximal value $\vq_1^m\,(n{-}1)(m{-}1)$.
 
\subsection{Quasi-Fano Components}
\label{s:qFCS}
The cohomology of the components
 $\big(S_m\<=\Fs^{-1}(0)\big),~\big(C_m\<=\Fc^{-1}(0)\big)\<\subset\FF{m}$
is readily computed, as we show here using the bi-projective embedding~\eqref{e:bPnFm}. The results,
\begin{equation}
  \dim H^r(S_m,\cO)=\d_{r,0}=\dim H^r(C_m,\cO),\qquad
  \c(\cO_{S_m})=1=\c(\cO_{C_m})
\end{equation}
and the fact that $S_m\<\cap C_m\<={}^\sharp\!\XX[n-2]m$ is a smooth Calabi-Yau space for generic choices of $\Fc(x,y)$ satisfies the definition~\cite[Def.\,2.2]{tyurin2003fano} of a {\em\/quasi-Fano\/} space.

\paragraph{The Directrix:}
Consider first the hallmark directrix hypersurface $S_m\<\subset\FF{m}$. Its structure sheaf cohomology is computed from the network of Koszul resolutions:
\begin{equation}
  \begin{array}{c@{~}c@{~}c@{~}c@{~}c@{}c}
   \cO_A\pM{-2\\0} &\C3{\6[2pt]{\Fs}{\lhook\joinrel\dto}} &\cO_A\pM{-1\\-m} \\*
   \CW{\into}\crlap{ p_0} && \CW{\into}\crlap{ p_0} \\*[-1pt]
   \cO_A\pM{-1\\m} &\C3{\6[2pt]{\Fs}{\lhook\joinrel\dto}} & \cO_A \\*[-1pt]
   \CW{\onto} && \CW{\onto} \\*[-2pt]
   \cO_{\smash{\FF{m}}}\pM{-1\\m} &\lhook\joinrel\too{~\Fs~}
   &\cO_{\smash{\FF{m}}} &\onto&\cO_{S_m}\\ 
  \end{array}
 \label{e:OSm}
\end{equation}
where the mapping induced from multiplication by $\Fs(x,y)$ becomes regular on $\FF{m}$ (in the bottom row), while elsewhere on $A$ it involves multiplication by the equivalence class of Laurent polynomials~\eqref{e:DrX}. The vertical sequences are however induced from multiplication by the polynomial $ p_0(x,y)$, which is regular everywhere on $A$. We compute the associated cohomology from those first, using again the Young tableau notation as above~\cite{rBeast}:
\begin{equation}
 \begin{array}{c|c@{~}c@{~}c@{~}c@{~}l}
    & \pM{\str{10}2\40&\!\cdots\!&0\\0\40\\}
    &\lhook\joinrel\too{ p_0}& \pM{~~\str{11}1\40&\!\cdots\!&0\\-m\40\\}
       &\onto& \cO_{\FF{m}}\pM{-1\\m} \\[2mm]
   \toprule
   0. & 0 && 0 && H^0\<=0 \\
   1. & 0 && 0 && H^1\<=0 \\[-2mm]
   \vdots & \vdots && \vdots && ~~~\vdots \\ 
    \bottomrule
 \end{array}
 \qquad\text{and}\qquad
 \begin{array}{c|c@{~}c@{~}c@{~}c@{~}l}
    & \pM{\str{11}1\40&\!\cdots\!&0\\m\40\\}
    &\lhook\joinrel\too{ p_0}& \pM{0\40&\!\cdots\!&0\\0\40\\}
       &\onto& \cO_{\FF{m}} \\[2mm]
   \toprule
   0. & 0 && \pM{0&0&\!\cdots\!&0\\0&0\\} &\approx& H^0 \\
   1. & 0 && 0 && H^1\<=0 \\[-2mm]
   \vdots & \vdots && \vdots && ~~~\vdots \\ 
    \bottomrule
 \end{array}
\end{equation}
Combining these results for the bottom, horizontal sequence in~\eqref{e:OSm} yields
\begin{equation}
  \begin{array}{c@{~}c@{~}c@{~}c@{~}c@{}c}
    & \cO_{\smash{\FF{m}}}\pM{-1\\m} 
    &\lhook\joinrel\too{\Fs}& \cO_{\smash{\FF{m}}}
       &\onto& \cO_{S_m} \\[2mm]
   \toprule
   0. & 0 && \pM{0&0&\!\cdots\!&0\\0&0\\} &\approx& H^0\approx\IC \\
   1. & 0 && 0 && H^1\<=0 \\[-2mm]
   \vdots & \vdots && \vdots && ~~~\vdots \\ 
    \bottomrule
 \end{array}
 \qquad\qquad\text{so}\quad
 \c(\cO_{S_m})\<=1.
\end{equation}

\paragraph{The Complemetrix:}
The analogous computation for the complementrix,
 $(C_m\<\subset\FF{m})\<\in\ssK[{r||cc}{\IP^n&1&n{-}1\\ \IP^1&m&2}]$, is:
\begin{equation}
  \begin{array}{c@{~}c@{~}c@{~}c@{~}c@{}c}
   \cO_A\pM{-n\\-2-m} &\lhook\joinrel\too{~\Fc~} &\cO_A\pM{-1\\-m}
                       &\onto &\cO_{C'_m}\pM{-1\\-m} \\*[-2pt]
   \CW{\into}\crlap{ p_0} && \CW{\into}\crlap{ p_0} &&\mkern-20mu
                             \CW{\into}\crlap{ p_0} \\*
   \cO_A\pM{1-n\\-2} &\lhook\joinrel\too{~\Fc~} &\cO_A
                       &\onto &\cO_{C'_m} \\*[-3pt]
   \CW{\onto} && \CW{\onto} && \CW{\onto} \\*[-2pt]
   \cO_{\smash{\FF{m}}}\pM{1-n\\-2} &\lhook\joinrel\too{~\Fc~}
   &\cO_{\smash{\FF{m}}} &\onto&\cO_{C_m}\\ 
  \end{array}
 \qquad\text{where}\qquad
 \ARR\{{@{}r@{~}l}
       { C'_m &\in\ssK[{r||c}{\IP^n&n{-}1\\ \IP^1&2}], \\[2mm]
         C'_m &\coeq\big\{\Fc(x,y)\<=0\big\}\<\subset A. } .
 \label{e:OCm}
\end{equation}
Since both $ p_0(x,y)$ and $\Fc(x,y)$ are regular polynomials on $A\<=\IP^n{\times}\IP^1$, both the horizontal and the vertical mappings are well defined over all of $A$, and we short-cut the cohomology computation using the spectral sequence~\cite{rBeast}:
\begin{equation}
 \begin{array}{c|c@{~}c@{~}c@{~}c@{~}c@{~}c@{~}c}
    &\pM{\str{13}n\40&\!\cdots\!&0\\2{+}m\40\\}
    &\vC{\TikZ{[scale=.5]\path[use as bounding box](0,0)--(1,0);
               \draw[thick,->](0,.2)node[above right]{$\SSS\Fc$}--++(1,.3);
               \draw[thick,->](0,-.2)node[below right]{$\SSS p_0$}--++(1,-.3);
              }}
    &\MM{ \pM{\str{11}1\40&\!\cdots\!&0\\m\40\\} \\
          \pM{\str{15}n{-}1\40&\!\cdots\!&0\\2\40\\} }
    &\vC{\TikZ{[scale=.5]\path[use as bounding box](0,0)--(.8,0);
               \draw[thick,<-](.7,.2)node[above left]{$\SSS p_0$}--++(-1,.3);
               \draw[thick,<-](.7,-.2)node[below left]{$\SSS\Fc$}--++(-1,-.3);
              }}
    &\pM{0\40&\!\cdots\!&0\\0\40\\} &\onto&\cO_{C_m} \\[2mm]
   \toprule
   0. & 0 && 0 && \pM{0&0&\!\cdots\!&0\\0&0\\} &\approx& H^0\approx\IC \\
   1. & 0 && 0 && 0 &&H^1\<=0 \\[-2mm]
   \vdots & \vdots && \vdots && \vdots && ~~~\vdots \\ 
    \bottomrule
 \end{array}
 \qquad\text{so}\quad
 \c(\cO_{C_m})\<=1.
\end{equation}

\begingroup
\def\baselinestretch{1}
\small\addtolength{\baselineskip}{-1pt}
\raggedright

\endgroup

\end{document}